\documentclass[%
 aip,
rsi,%
 amsmath,amssymb,
preprint,%
]{revtex4-1}

\usepackage{float}
\usepackage{graphicx}
\usepackage{dcolumn}
\usepackage{bm}
\usepackage{color}
\usepackage[version=3]{mhchem} 
\usepackage{subcaption}

\begin{document}


\title[]{Investigating the influence of relativistic effects on absorption spectra for platinum complexes with light-activated activity against cancer cells.}

\author{Joel Creutzberg}
\affiliation{Division of Theoretical Chemistry, Lund University, Lund, Sweden}
\author{Erik Donovan Hedeg{\aa}rd}
\email{erik.hedegard@teokem.lu.se}
\affiliation{Division of Theoretical Chemistry, Lund University, Lund, Sweden}

\date{\today}

\begin{abstract}
\textbf{Abstract:} We report the first investigation of relativistic effects on the UV-vis spectra of two prototype  complexes for so-called photo-activated chemotherapy (PACT),  \textit{trans}-\textit{trans}-\textit{trans}-\ce{[Pt(N3)2(OH)2(NH3)2]} and \textit{cis}-\textit{trans}-\textit{cis}-\ce{[Pt(N3)2(OH)2(NH3)2]}. In PACT, design of new drugs requires in-depth understanding of the photo-activation mechanisms. A first step is usually to rationalize their UV-vis spectra for which time-dependent density functional theory (TD-DFT) is an indispensable tool.  

We carried out TD-DFT calculations with a systematic series of non-relativistic (NR), scalar-relativistic (SR), and four-component (4c) Hamiltonians. Large differences are found between spectra calculated within 4c and NR frameworks, while the most intense features (found at higher energies below 300 nm) can be reasonably well reproduced within a SR framework. Yet the underlying transitions can be strongly influenced  by spin-orbit coupling introduced in the 4c framework: while this can affect both intense and less intense transitions in the spectra, the effect is most pronounced for weaker transitions at lower energies, above 300 nm. Since  the investigated complexes are activated with light of wavelengths above 300 nm,  employing a method with explicit inclusion of spin-orbit coupling may be crucial to rationalize the activation  mechanism. 

All calculations were carried out with both the CAM-B3LYP and B3LYP functionals; we generally find the former to perform best in comparison with experimental spectra. 
\end{abstract}

\pacs{Valid PACS appear here}
\maketitle

\section{Introduction}
The use of platinum complexes in cancer therapy is among the most influential results of medicinal inorganic chemistry. Three simple platinum complexes are today approved worldwide for chemotherapy (cisplatin, carboplatin, and oxaliplatin), while three additional complexes (nedaplatin, lobaplatin, and heptaplatin) are approved in a few countries.\cite{johnstone2016} All these complexes contain Pt(II) centers and the activity against cancer cells is caused by the labile nature of Pt(II). Yet, this labile nature also causes severe side-effects (e.g., nausea and chronic kidney disease), limiting the use of  Pt(II)-based medicine. Another limitation is due to cancer cells with intrinsic or developed resistance. Complexes with fewer side-effects and different mechanisms to harm cancer cells are therefore sought for. 

 Recent investigations have a focus on  \textit{pro-drugs}, i.e., biologically inactive complexes that can be activated at the site of the tumor. 
For platinum, good candidates for pro-drugs are octahedral, low-spin $\text{d}^{6}$ Pt(IV) complexes, which  are kinetically stable (non-labile), compared to the square planar $\text{d}^{10}$  Pt(II) counterparts. Different activation mechanisms have been used: one strategy has been to  rely on bio-reducing reagents, such as ascorbic acid or  glutathione, reducing Pt(IV) to Pt(II).\cite{hall2002,graf2012} 
An alternative is to use light-activation in what have become known as \textit{photo-activated chemotherapy} (PACT) or \textit{photodynamic therapy} (PDT).\cite{butler2013,farrer2009,bonnet2018}  The PACT process involves an initial electronic excitation of the pro-drug, followed by a chemical transformation (e.g.~a reduction) into an active form. In this way, PACT differs from PDT, which employs a chemical substance (denoted \textit{photo-sensitizer}) to form an electronically excited state, which in turn induces generation of reactive oxygen species (ROS), harmful to cancer cells. The PACT process is not as far in development as PDT for clinical use\cite{brown2004}, but PACT offers advantages for cells with low oxygen levels\cite{chen2015}, which are among the most resistant to therapy.\cite{brown2004,wilson2011}   

For complexes involved in PACT (and similar for PDT), theoretical methods are often required to understand the (usually complex) activation mechanisms. Time-dependent density functional theory (TD-DFT) is a powerful tool to investigate the initial process of excitation and to analyze the character of the excited states generated upon radiation. In several studies, TD-DFT has been used with great insight to explain the experimentally observed photochemistry of platinum complexes used in PACT.\cite{salassa2009,mackay2009,farrer2010,westendorf2011,westendorf2012,zhao2013a,zhao2013b,shaili2019}   In one case, several DFT functionals were also benchmarked against complete active space second-order perturbation theory (CASPT2) and it was concluded that range-separated functionals reproduce both experiment and CASPT2 results best.\cite{sokolov2011} Despite this result, only a few investigations have employed range-separated functionals. Perhaps even more surprising is that none of the investigations have focused on the importance of relativistic effects, in particular concerning explicit inclusion of spin-orbit coupling.  
Instead, relativistic effects have usually been included indirectly  through  effective core potentials (ECPs) combined with a non-relativistic Hamiltonian. In one instance, relativistic effects have also been included through  Douglass-Kroll-Hess to second order (DKH2), focusing on scalar-relativistic (SR) parts.\cite{sokolov2011}
\begin{figure}[htb!]
\centering
\includegraphics[width=0.50\textwidth]{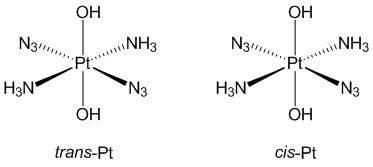}
\caption{Lewis structures of \textit{trans}-\textit{trans}-\textit{trans}-[Pt(N$_3$)$_2$(OH)$_2$(NH$_3$)$_2$] (\textit{trans}-Pt) and 
\textit{cis}-\textit{trans}-\textit{cis}-[Pt(N$_3$)$_2$(OH)$_2$(NH$_3$)$_2$] (\textit{cis}-Pt) investigated in this paper.}
\label{fig:lewis}
\end{figure}
The lack of systematic studies of relativistic effects can perhaps be attributed to that  TD-DFT implementations first relatively recently were developed for a large arsenal of relativistic Hamiltonians. Yet, several formulations of relativistic TD-DFT\cite{liu2018,saue2020,repisky2020} ranging from four-component (4c) Dirac-Kohn-Sham\cite{gao2004,gao2005,bast2009}   to various two-component frameworks\cite{wang2005,peng2005,bast2009,kuhn2013,egidi2016}  exist today.

 In this paper, we study the effect of employing TD-DFT based on relativistic Hamiltonians for complexes involved in PACT. We systematically investigate both non-relativistic (NR), scalar relativistic (SR) and four-component (4c) frameworks. In addition, we employ both the B3LYP functional and a range-separated variant (CAM-B3LYP). As targets, we employ two prototypical complexes (here denoted \textit{trans}-Pt and \textit{cis}-Pt), shown in Figure \ref{fig:lewis}. These two complexes are both active against tumors upon radiation with light\cite{muller2003,mackay2006,bednarski2006,imran2018} and are among the first reported platinum complexes for use in PACT. They are further the simplest among a number of related complexes which subsequently have shown similar photo-activity.\cite{kasparkova2003,mackay2007,mackay2009,farrer2010,pracharova2012,
 westendorf2011,westendorf2012,min2014,zhao2013a,zhao2013b,gandioso2015,shaili2019,imberti2020} The exact mechanism of photo-activation is not known in detail. However, the UV-vis spectra in combination with TD-DFT calculations\cite{mackay2006,salassa2009} have shown that both \textit{trans}- and \textit{cis}-Pt complexes display ligand-to-metal charge-transfer (LMCT) excitations. The population of the resulting states leads to decomposition, likely through multiple pathways\cite{imberti2020,ronconi2008,ronconi2011} , including dissociation of \ce{N3-} and/or \ce{NH3} ligands, generation of \ce{O2}, and reduction to Pt(II). 
 Possibly the photoreactions also involve triplet states which have been shown to be dissociative.\cite{salassa2009,westendorf2011}  
 
 Intriguingly, the complexes in Figure \ref{fig:lewis} only have intense transitions in high-energy part of the spectrum (285  and 256 nm for \textit{trans}- and \textit{cis}-Pt, respectively)\cite{muller2003,mackay2006,bednarski2006,salassa2009}. Yet, the light-induced reactivity has been achieved in regions without strong absorption. For instance \textit{trans}-Pt react with DNA or DNA models (guanosine 5'monophosphate) upon exposure of light at 365--366 nm and 647 nm\cite{mackay2006} (although the latter only slugishly), while \textit{cis}-Pt reacts after exposure to light at both 365--366 nm, 458 nm, and 647 nm.\cite{kasparkova2003,muller2003,bednarski2006} Our study therefore focuses on both high-energy parts of the spectra (approximately 250--300 nm) where the complexes absorb strongly, as well as lower-energy parts (above 300 nm) where the transitions are weaker, yet potentially important for anti-cancer activity.


\section{Computational details}
The structures of  \textit{trans}-\textit{trans}-\textit{trans}-[Pt(N$_3$)$_2$(OH)$_2$(NH$_3$)$_2$] (denoted \textit{trans}-Pt) and \textit{cis}-\textit{trans}-\textit{cis}-[Pt(N$_3$)$_2$(OH)$_2$(NH$_3$)$_2$] (denoted \textit{cis}-Pt) were optimized with Turbomole 7.1\cite{Ahlrichs1989}, employing the TPSS\cite{Tao2003}  functional  and a def2-SV(P) basis set\cite{Schafer1992}. Calculations in Turbomole  were sped up by expanding the Coulomb interactions in an auxiliary basis set (the resolution-of-identity approximation)\cite{eichkorn1995,eichkorn1997}, employing standard def2-SV(P) auxiliary basis sets.

All subsequent calculations of UV-vis spectra were performed with the DIRAC program\cite{DIRAC18,saue2020}, employing these optimized structures. The calculations were carried out with the def2-SV(P)\cite{Andrae1990a, Schafer1992, Weigend2005a} basis set for the ligands, and the dyall.v2z \cite{Dyall2009} basis set for the platinum atom. The basis set for Pt was employed uncontracted. We initially tested the effect of employing a completely uncontracted basis set, but since the obtained spectra were close to identical to calculations with contracted basis sets, all calculations reported here employed contracted basis sets for the ligands.  

The calculated spectra were obtained with TD-DFT within a four component (4c) Dirac-Kohn-Sham framework, using a Dirac--Coulomb Hamiltonian with (SS$|$SS) integrals replaced by interatomic SS energy corrections.\cite{Visscher1997} For the exchange-correlation functional, we 
employed the range-separated CAM-B3LYP\cite{Takeshi2004} functional as well as the B3LYP\cite{Becke1988,Lee1988,Becke1993} functional. The NR and SR calculations employed the same basis set and functional but invoked the L{\'e}vy-Leblond,\cite{Levy-leblond1967}
and Dyall's spin-free\cite{Dyall1994,Visscher2000} Hamiltonians, respectively. All  TD-DFT calculations (relativistic and non-relativistic) were performed with 80 roots. The spectra obtained were then broadened using a gaussian convolution with a broadening factor of 0.3 eV. Assignments of transitions were based on analyses of the response vectors for each transition, in combination with visual inspection. Even if not all 80 transitions are relevant, the density of states with non-zero intensities is still too high to warrant discussion of each transition (particular for the 4c calculations). Therefore, we have divided the spectra into  regions (\textbf{1}--\textbf{6} for \textit{trans}-Pt and \textbf{1}--\textbf{9} for \textit{cis}-Pt) from which we discuss the most intense transitions. We have as far as possible attempted to group the transitions in the regions according to their character so that the regions are comparable across different Hamiltonians and functionals. However, one-to-one correspondence could not always be achieved (especially when comparing functionals).   
Selected transition energies, oscillator strengths, and assignments are given (in terms of contributing orbitals) for each region in the supporting information (SI), Tables S1--S12. Many of the discussed transitions are of similar character and to avoid tedious repetition, we will use to short-hand notation $\pi\rightarrow d$ for the ligand-to-metal charge transfer (LMCT) transitions between $\pi$-orbitals on \ce{N3-} to orbitals of platinum d-character. Transitions between orbitals of d-character are denoted $d\rightarrow d$, while transitions from orbitals with p-character (lone pairs) on the \ce{OH-} oxygen are denoted $p \rightarrow d$. This nomenclature has also been employed in the Tables in the SI. 
In addition to Tables S1--S12, selected orbital densities are displayed in the SI, Figures S1--S12. The orbitals in the 4c calculations differ from orbitals in SR and NR calculations by (due to spin-orbit coupling) having both $\alpha$- or $\beta$-spin parts. Yet, most orbitals (the orbital densities are shown in Figures S1--S12) have a Mulliken population over 0.9 of either the $\alpha$- or the $\beta$-part. We denote orbitals with a mixing degree lower than this as "spin-mixed" ($\alpha$ and $\beta$ populations of the orbitals are also reported in the Figures in the SI).     

\section{Results}    

\subsection{Optimized structures}

\begin{figure}[htb!]
\centering
\includegraphics[width=0.75\textwidth]{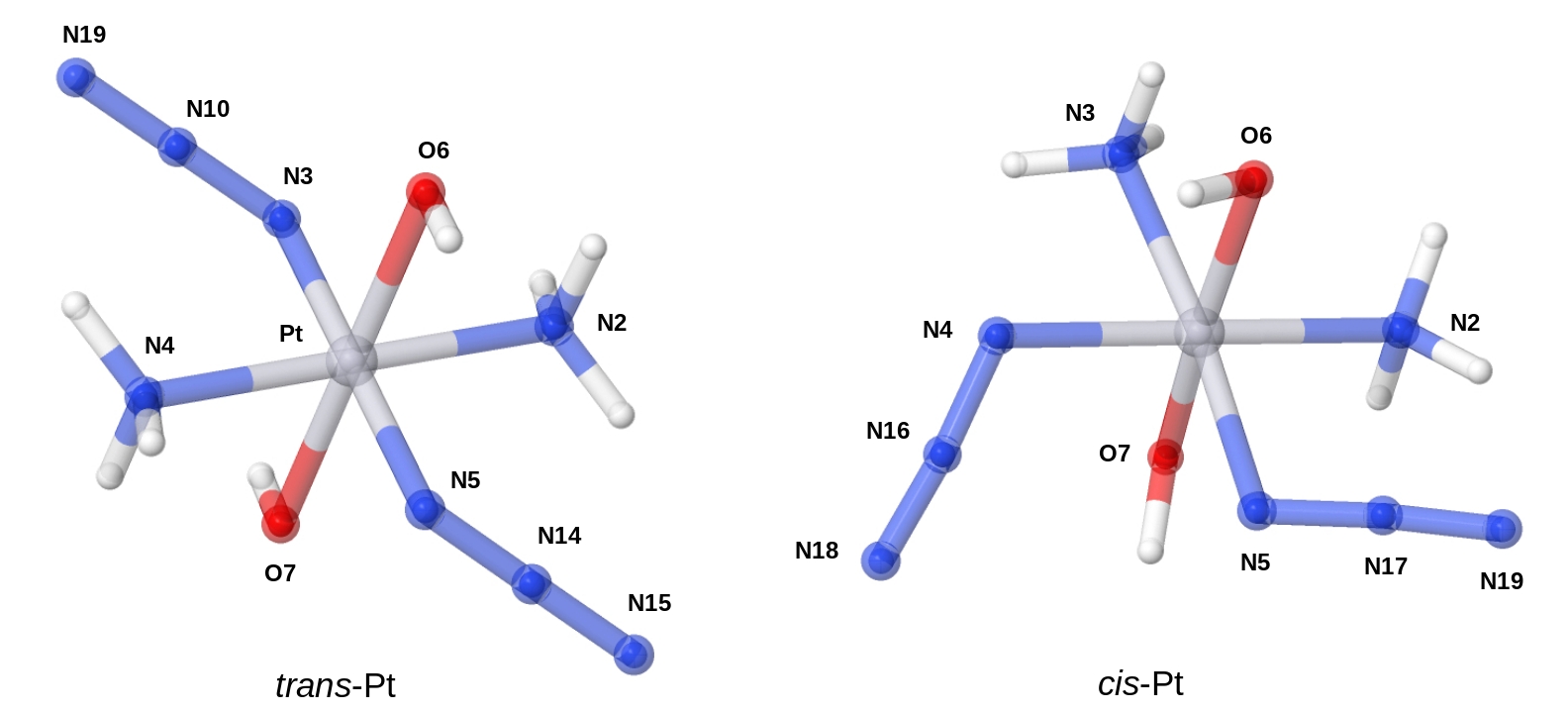}
\caption{Optimized structures of the two studied complexes. Bond distances and angles are shown in Table \ref{opt_struc}.}
\label{fig:complexes}
\end{figure}
The optimized structures of \textit{trans}-Pt and  \textit{cis}-Pt complexes are shown in Figure \ref{fig:complexes} and selected structural parameters are reported together with experimental values\cite{muller2003,mackay2006} in Table \ref{opt_struc}. The results are overall in good agreement with both experimental and earlier computational studies \cite{sokolov2011} (where several other functionals were  employed). This good correspondence is illustrated by the fact that the main difference is the \ce{Pt-N-N} and \ce{N-N-N} angles for  \textit{trans}-Pt only  deviate $\approx 3^{\circ}$ from the experimental results (previous DFT results have deviations around $\approx 0-2^{\circ}$). The differences for \textit{cis}-Pt complex are even smaller.
\begin{table*}[htb!]
\centering
\caption{Selected structural parameters for the optimized structures of \textit{trans}- Pt and \textit{cis}-Pt (numbers refer to Figure \ref{fig:complexes}). We show only one of the Pt and \ce{OH-}/\ce{N3-}/\ce{NH3} bonds and angles (the other is close to identical). Bonds
 are in {\AA} and angles in degrees ($^{\circ}$).   \label{opt_struc} }
\begin{tabular}{lcccccccccc}
 \hline\hline
\multicolumn{6}{c}{\textit{trans}-Pt} \\ [0.5ex] 
\hline
         & ~~~~\ce{Pt-N$_2$}~~~~  &     ~~~~\ce{Pt-N$_3$}~~~~ &    ~~~~\ce{Pt-O$_6$}~~~~  &  ~~~~\ce{Pt-N$_{3}$-N$_{10}$}~~~~ &   ~~~~\ce{N$_3$-N$_{10}$-N$_{19}$}~~~~      \\[0.5ex]
Calc.      & 2.072        &       2.101  &    2.053     & 114.0  &  177.05  \\[0.5ex]
Exp.       & 2.036        &       2.046  &    2.006     & 117.2  &  174.5   \\[0.5ex]
\hline 
\multicolumn{6}{c}{\textit{cis}-Pt} \\[0.5ex]
\hline
         &  ~~~~\ce{Pt-N$_2$}~~~~ &  ~~~~\ce{Pt-N$_4$}~~~~ &   ~~~~\ce{Pt-O$_6$}~~~~ &     ~~~~\ce{Pt-N$_4$-N$_{16}$}~~~~  &   ~~~~\ce{N$_4$-N$_{16}$-N$_{18}$}~~~~ &  \\[0.5ex] 
Calc.               & 2.136   &   2.057    &    2.047   &    116.4  &     174.5  \\[0.5ex]
Exp.                & 2.022   &   2.036    &    2.005   &    117.3  &     172.9  \\[0.5ex]
                    & 2.043   &   2.038    &    2.007   &    115.2  &            \\[0.5ex]
\hline \hline 
\end{tabular}
\end{table*} 
Further, the bonds from the platinum center to \ce{NH3} ligands are also slightly elongated, compared to previous results\cite{sokolov2011}.  This elongation is seen for both isomers; for instance the B3LYP result from Ref.~\citenum{sokolov2011} for the \textit{trans}-Pt complex is 2.087 {\AA} versus  2.101 {\AA} for TPSS (the corresponding numbers for \textit{cis}-Pt are 2.046 {\AA} versus 2.057/2.063 \AA). Yet, these elongations can still be considered minor.

\subsection{UV-vis spectra of the \textit{trans}-Pt complex}

\begin{figure}[htb!]
\centering
\includegraphics[width=1\textwidth,trim={1cm 6cm 2cm 10cm},clip]{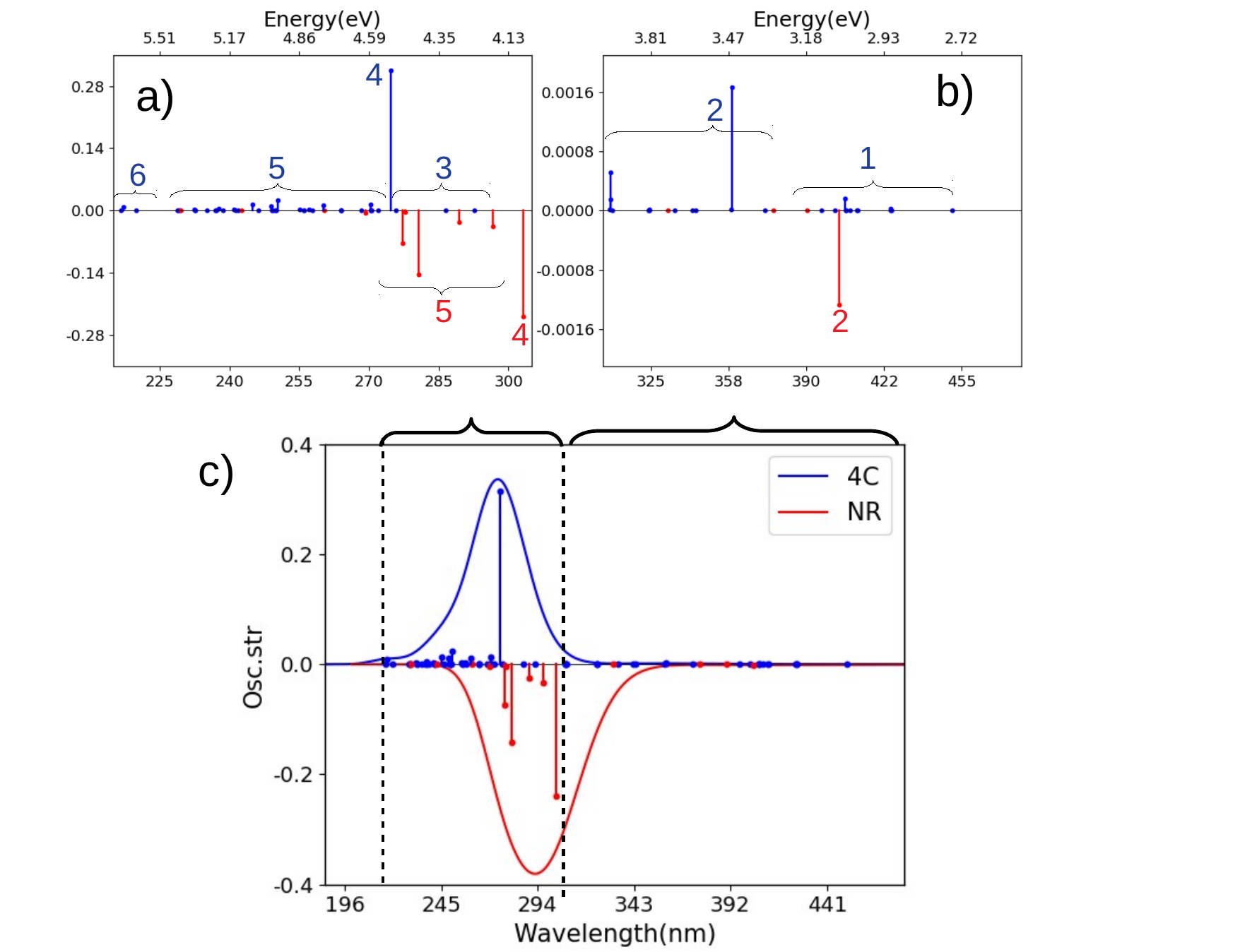}
\caption{Spectra for \textit{trans}-Pt calculated with CAM-B3LYP and NR or 4c  Hamiltonians. (a) and (b) are magnified for 215--305 nm and 305--480 nm. (c) shows the full spectrum. Assignments of main transitions within regions \textbf{1}--\textbf{6} are provided in Tables S1 and S2. } 
\label{fig:trans_camb3lyp_ll}
\end{figure}

\textbf{CAM-B3LYP results:} The spectra calculated in 4c and NR frameworks  are compared in Figure \ref{fig:trans_camb3lyp_ll}. Note that transitions of higher energy  are significantly more intense than transitions of lower energy and accordingly, Figures  \ref{fig:trans_camb3lyp_ll}(a) and \ref{fig:trans_camb3lyp_ll}(b) display high- and low-energy parts of the spectrum separately (with regions \textbf{1}--\textbf{6} marked), while the full spectrum is shown in Figure \ref{fig:trans_camb3lyp_ll}(c). From the latter figure, we immediately see that spectra from NR and 4c frameworks  are rather different.    
\begin{figure}[htb!]
\centering
\includegraphics[width=1\textwidth,trim={2cm 4cm 3cm 11cm},clip]{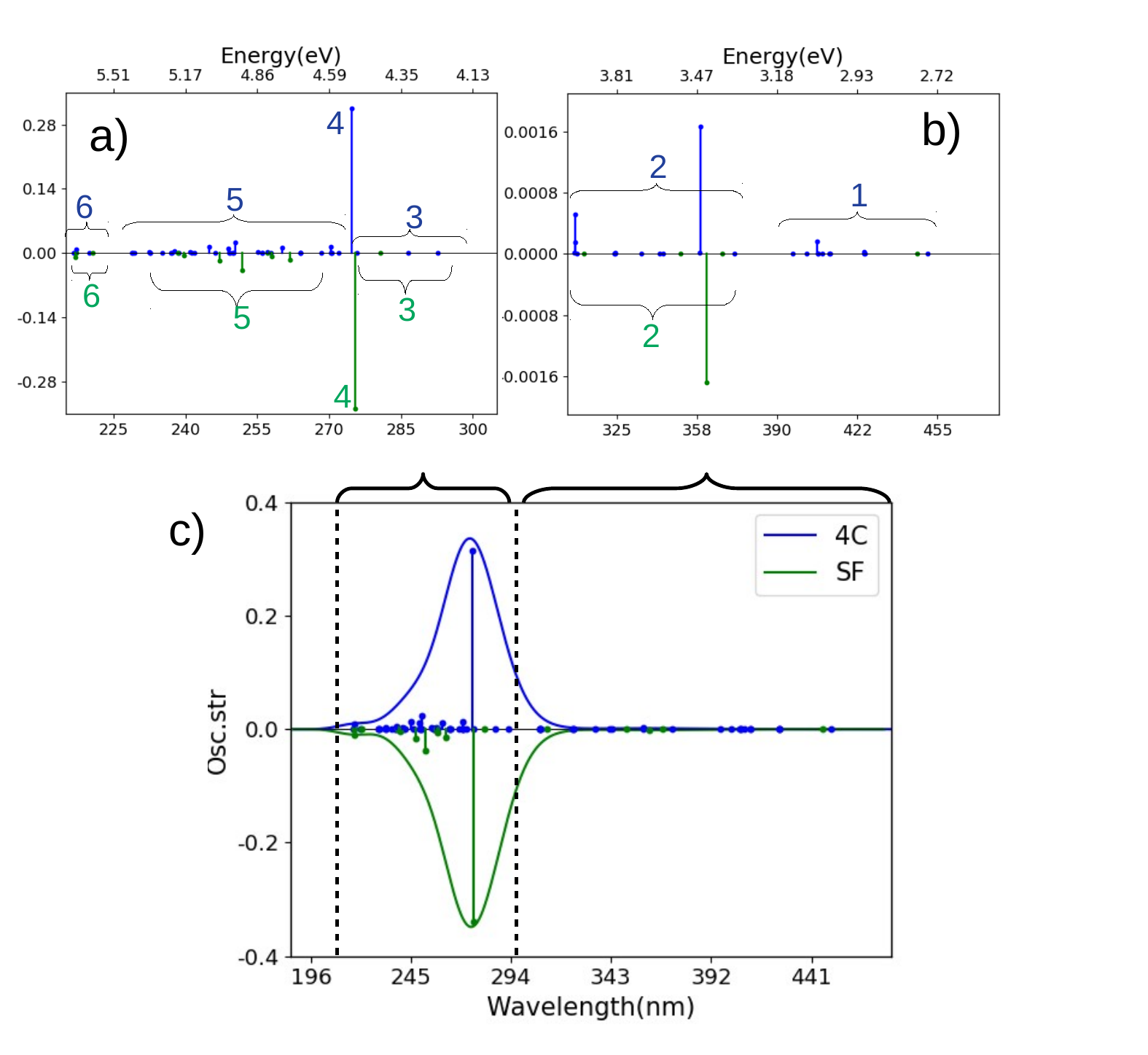}
\caption{Spectra for \textit{trans}-Pt calculated with CAM-B3LYP and SR or 4c  Hamiltonians. (a) and (b) are magnified for 215--305 nm and 305--480 nm. (c) shows the full spectrum. Assignments of main transitions within regions \textbf{1}--\textbf{6} are provided in Tables S1 and S3.  }
\label{fig:trans_camb3lyp_sf}
\end{figure}
At high energies (5.71--4.59 eV or 217--270 nm) a number of transitions of low intensity are found for the 4c calculation, labeled \textbf{6} and \textbf{5} in Figure \ref{fig:trans_camb3lyp_ll}(a). These transitions are mainly of LMCT character and contain a mixture of transitions from  $p$- and $\pi$-orbitals (on \ce{OH- } and \ce{N3-}, respectively) to a  $d$-orbital on platinum. The most intense transition (found in region   \textbf{5} at 4.95 eV or 250 nm) is mainly of $\pi \rightarrow d$ character (cf.~Table S1). We note that all transitions in region \textbf{5} also involve (to varying degree) transitions from the oxygen lone-pairs to the metal center ($p\rightarrow d$). 
Due to the neglect of all excitations to triplet states, the density of states is generally lower in the NR calculation and few transitions occur between  5.7--4.6 eV (i.e.~in what corresponds to  regions \textbf{6} and \textbf{5} in the 4c calculation). Rather, the transitions labeled \textbf{5} for the NR calculation appear somewhat red-shifted at  4.47--4.18 eV (277--297 nm) and with much higher intensity. In the NR calculation, the transition with highest intensity in region \textbf{5} (4.42 eV or 281 nm) primarily  involves orbitals on oxygen and the metal center ($p\rightarrow d$), but otherwise the transitions in \textbf{5} have compositions similar to those in the 4c calculation (i.e.~mixture of $\pi\rightarrow d$ and $p\rightarrow d$, cf.~Table S2).

The transitions labeled \textbf{4} are intense in both 4c and NR frameworks and corresponds in both cases to a transition of $\pi\rightarrow d$ character  (cf.~Tables  S1 and S2). As seen from Figure \ref{fig:trans_camb3lyp_ll}(a) the NR calculation is again  red-shifted (4.09 eV or 303 nm, compared to 4.51 eV or 275 nm in the 4c case). No intense transitions are found immediately after region \textbf{4} in the 4c-CAM-B3LYP calculation, although a few transitions with intensities close to zero can be seen in this region (denoted \textbf{3} in Figure \ref{fig:trans_camb3lyp_ll}a). 
 
 Moving to the lower-energy transitions in Figure \ref{fig:trans_camb3lyp_ll}(b), no transitions in the NR calculation match the (mainly $\pi\rightarrow d $) transitions in region  \textbf{1} around 3.08--2.75 eV (402--451 nm) in the 4c calculation. Region \textbf{2} is also quite different in 4c and NR frameworks: in the former, the region is a  mixture of $\pi\rightarrow d$ and $p\rightarrow d$ transitions of which the most intense (at 3.46 eV or 358 nm) is of $\pi\rightarrow d $ character. Only one transition is seen in the NR framework (also of $\pi\rightarrow d$ character), 
 but is  considerable red shifted compared to the 4c counterpart (3.07 eV or 404 nm, compared to 4.03--3.46 eV or 308--359 nm).
\begin{figure}[htb!]
\centering
\includegraphics[width=1\textwidth,trim={1cm 4cm 2cm 11cm},clip]{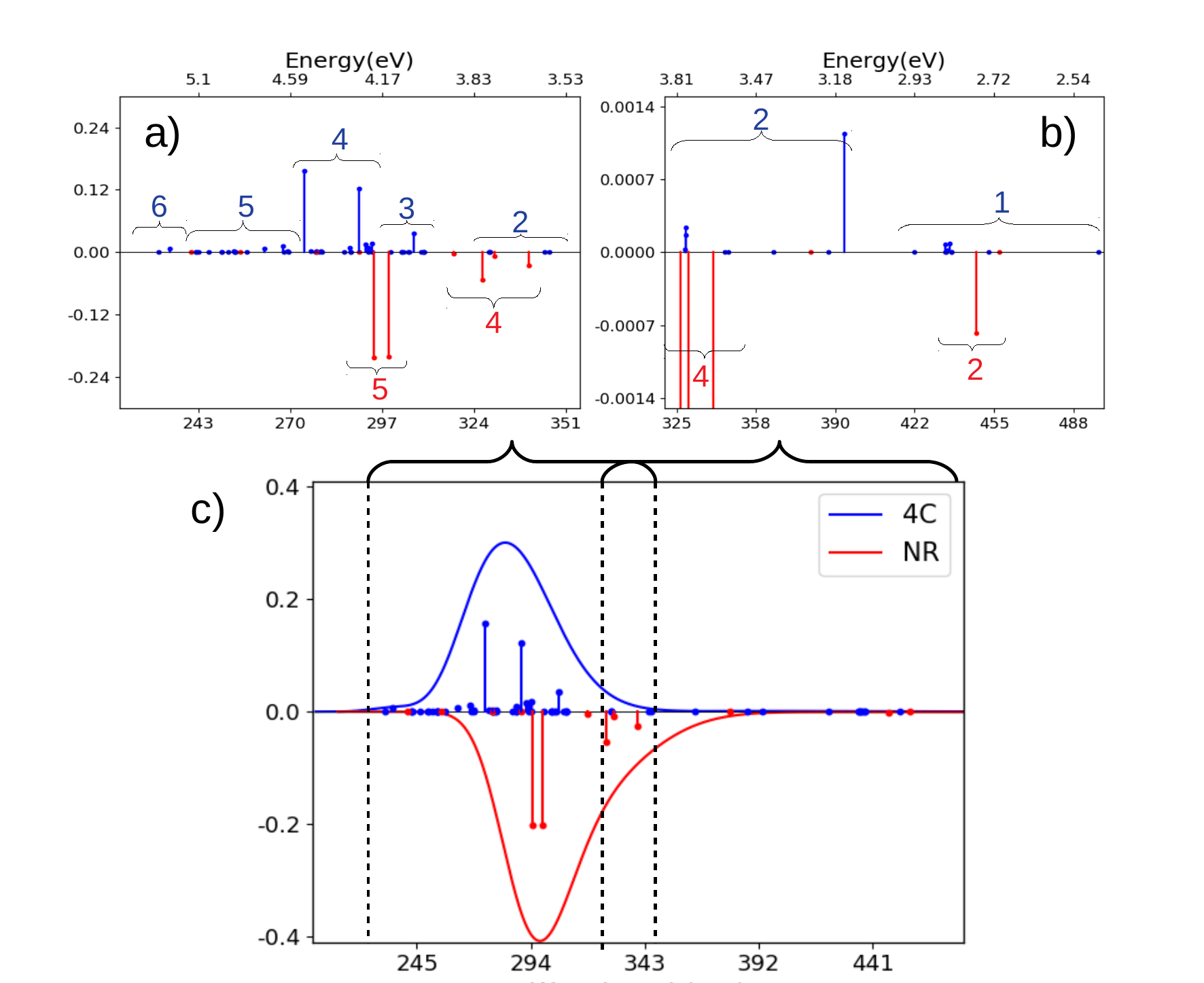}
\caption{Spectra for \textit{trans}-Pt calculated with B3LYP and NR or 4c  Hamiltonians. (a) and (b) are magnified for 220--355 nm and 328--500 nm. (c) shows the full spectrum. Assignments of main transitions within regions \textbf{1}--\textbf{6} are provided in Tables S4 and S5.}
\label{fig:trans_b3lyp_ll}
\end{figure}
 
While the NR and 4c calculations are both qualitatively and quantitatively different, we also investigated whether including only SR effects lead to improvement. A comparison of spectra obtained with 4c and SR Hamiltonians are given in Figure \ref{fig:trans_camb3lyp_sf}, employing the same regions \textbf{1}--\textbf{6} as above (Figure \ref{fig:trans_camb3lyp_sf}a and \ref{fig:trans_camb3lyp_sf}b). From Figures \ref{fig:trans_camb3lyp_sf}(a) and \ref{fig:trans_camb3lyp_sf}(c) we see that the SR calculation qualitatively reproduce the most intense features of the high-energy region, although the density of states (as expected) is higher in the 4c calculation. The characters of the transitions labeled \textbf{6}--\textbf{3} are  to large degree similar in 4c and SR calculations, respectively (cf. Tables S1 and S3). For instance, the most intense transition (region \textbf{4})  is in both cases of $\pi\rightarrow d$ LMCT character. 

Reproducing the most intense feature in the spectrum is sufficient to provide a seemingly correct description of the full spectrum in Figure \ref{fig:trans_camb3lyp_sf}(c). However, important differences occurs for the lower-energy part in Figure \ref{fig:trans_camb3lyp_sf}(b): while the most intense transitions in region \textbf{2} occur at similar energies in 4c and SR calculations  (3.46 and 3.43 eV or 358 and 361 nm) --  and both  have $\pi\rightarrow d$ LMCT character -- a closer investigation shows that the involved transitions are different: in the 4c calculation: they either involve  spin-mixed orbitals (cf.~Table S1 and Figure S1) or have dominantly triplet character, none of which can be reproduced by the SR calculation due to lack of spin-orbit coupling.  
The same is also true for the transitions labeled \textbf{1}, which do not even occur in the SR calculation. This illustrates that many transitions (due to spin-orbit coupling) have low, but non-zero intensity in the 4c calculation as opposed to the SR (and NR) calculation(s). These differences will be more pronounced in the \textit{cis}-Pt complex discussed below.  
\begin{figure}[htb!]
\centering
\includegraphics[width=1\textwidth,trim={1cm 4.5cm 2cm 10.5cm},clip]{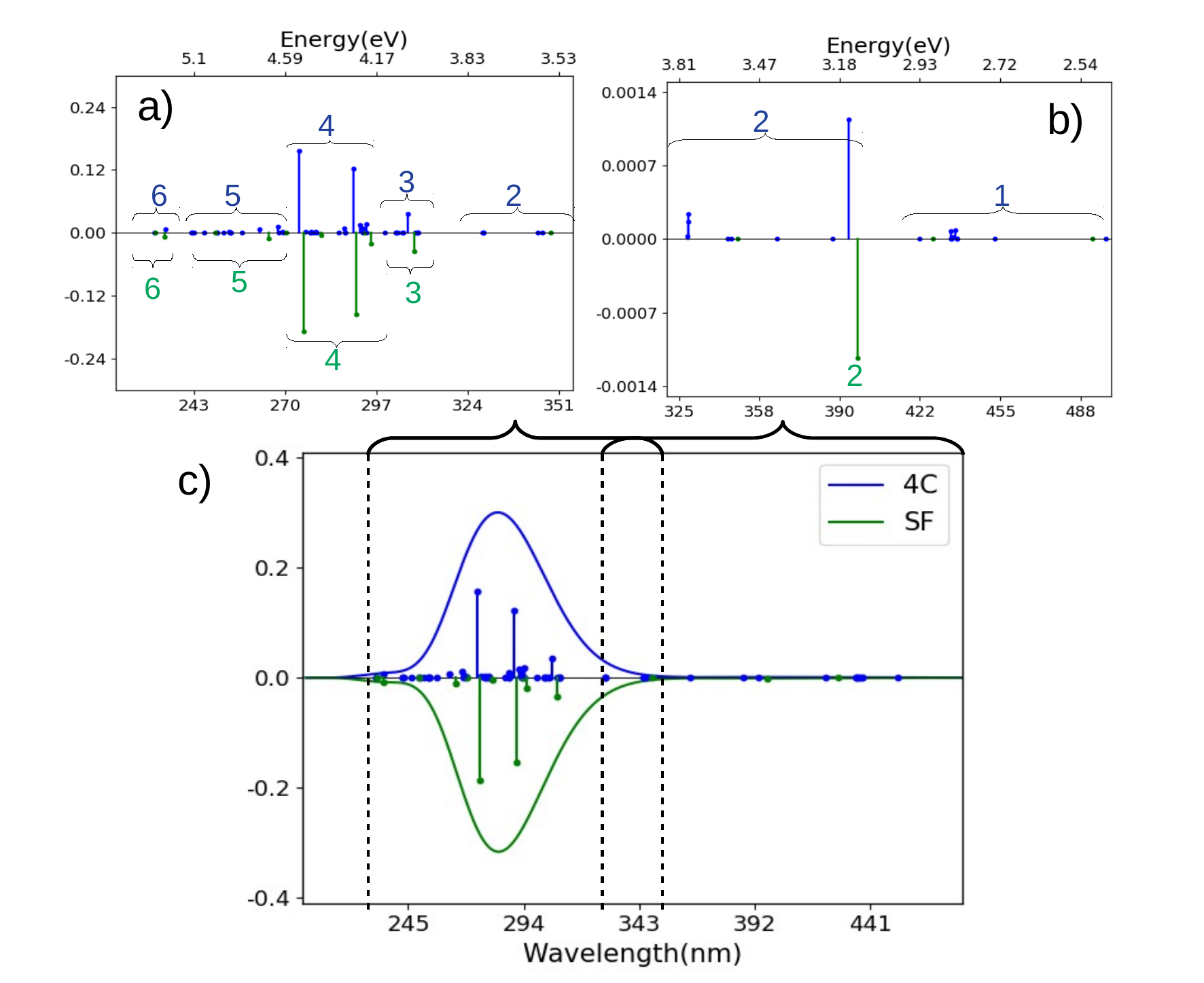}
\caption{Spectra for \textit{trans}-Pt calculated with B3LYP and SR or 4c  Hamiltonians. (a) and (b) are magnified for 220--355 nm and 328--500 nm. (c) shows the full spectrum. Assignments of main transitions within regions \textbf{1}--\textbf{6} are provided in Tables S4 and S5.}
\label{fig:trans_b3lyp_sf}
\end{figure}

\textbf{B3LYP results:} Since B3LYP has been employed extensively for complexes involved in PACT\cite{salassa2009,farrer2009,mackay2009,garino2013}, we repeated the above calculations with the B3LYP functional. The resulting spectra are shown in Figures \ref{fig:trans_b3lyp_ll} and \ref{fig:trans_b3lyp_sf}. We start by comparing 4c-B3LYP and 4c-CAM-B3LYP calculations (see e.g.~Figures  \ref{fig:trans_camb3lyp_ll} and \ref{fig:trans_b3lyp_ll}): from this comparison, we see that B3LYP generally predicts a larger number of intense transitions, particular in the high-energy region (this is most evident in regions \textbf{4} and \textbf{3}). 
A similar observation was made by Solokov and Schaefer\cite{sokolov2011}, comparing the range-separated $\omega$B97 and $\omega$B97X functionals with a number of hybrid and pure GGA functionals (with ECPs accounting for relativistic effects). As far as it is possible to compare the regions between the two functionals, B3LYP generally red-shifts the most intense transitions about 0.2--0.4 eV in the 4c calculations and 0.3--0.4 eV in the NR calculations. 

Despite the differences between the B3LYP and the range-separated variant, the changes  between 4c-B3LYP and NR-B3LYP calculations are qualitatively the same as seen for CAM-B3LYP. Thus, the density of states is  (as expected) much lower for the spectra calculated in the NR and SR frameworks. Further, for  region \textbf{5} the transitions in NR-B3LYP are much more intense than predicted by the corresponding 4c calculation and also considerably red-shifted (4.21--4.14 vs.~4.60--4.89 eV). Yet, the character of the transitions in region \textbf{5} (a mix of $\pi\rightarrow d$ and $p\rightarrow d$) is the same in both 4c and NR frameworks. Meanwhile, in region \textbf{4} the intense transitions are red-shifted (as for CAM-B3LYP); they appear at 4.52--4.21 eV (274--294 nm) and 3.90--3.65 eV (318--340 nm) in 4c- and NR-B3LYP calculations, respectively. Also here, the character of the transition is the same, i.e.~comprised  mostly of $\pi\rightarrow d$ transitions and to less degree of $p\rightarrow d$ transitions (cf.~Table S4). The region immediately after \textbf{4} contains (contrary to 4c- and SR-CAM-B3LYP), transitions with significant intensity: the region is denoted \textbf{3} in Figures \ref{fig:trans_b3lyp_ll}(a) and \ref{fig:trans_b3lyp_sf}(a) and has for 4c- and SR-B3LYP transitions (at 4.05 and 4.02 eV or 306 and 308 nm) with $p\rightarrow d$ character. 

Overall, the  changes introduced when employing  SR-B3LYP are the same as for CAM-B3LYP, i.e., most of the intense transitions in the high-energy parts are predicted in quite reasonable correspondence to the 4c-B3LYP calculations. However, the low-energy parts differ as  the transitions in regions \textbf{2} and \textbf{1}  (of $\pi\rightarrow d$ character) also for  4c-B3LYP are considerably spin mixed. Thus, several  transitions in region \textbf{2} do not occur in the SR-calculation and region \textbf{1} is entirely missing (as seem for CAM-B3LYP).   

\textbf{Comparison with experiment:} The experimental spectrum has one strong absorption (without structure) at 285 nm (4.35 eV) assigned as a  $\pi\rightarrow d$ LMCT transition.\cite{mackay2006}  
 The charge-transfer character of this excitation would suggest that employing a range-separated functional is most appropriate.\cite{dreuw2004} Yet, judging only from most intense transitions at 4.51 eV (275 nm) with 4c-CAM-B3LYP and at 274 nm (4.52 eV) with 4c-B3LYP, little differences is seen for the two functionals, although  4c- and SR-B3LYP calculations obtain a range of transitions with high intensities around the experimental peak at 285 nm (regions \textbf{4} and \textbf{3} from 4.05--4.52 eV or 274--306 nm, cf.~Figures \ref{fig:trans_camb3lyp_sf} and \ref{fig:trans_b3lyp_sf}). Meanwhile, the range-separated functional only obtain one intense transition (cf.~Figures \ref{fig:trans_camb3lyp_ll}  and  \ref{fig:trans_camb3lyp_sf}). We therefore tentatively suggests that CAM-B3LYP best reproduces the experimental excitation profile, but the experimental resolution does not allow us to unequivocally conclude that CAM-B3LYP performs best for the \textit{trans}-Pt complex. Notably, the results for the \textit{cis}-Pt complex leads to the same conclusion, but are more clear (see below). 

 Direct comparison with experiment is not possible in the lower energy parts of the spectrum where the transitions have significantly lower intensity. We will in the following only consider the relativistic calculations (4c and SR) with CAM-B3LYP. As found in previous studies (see e.g.~Ref.~\citenum{sokolov2011}), the LMCT ($\pi\rightarrow d$ and $p\rightarrow d$) transitions extend into the regions above 300 nm.  Interstingly, light-induced activity has been observed for \textit{trans}-Pt around  365 nm (3.40 eV)\cite{mackay2006,ronconi2011} which corresponds well with regions \textbf{1}--\textbf{3}, having a number of transitions in the range 4.02--2.83 eV (308--438 nm).
 In particular, the  4c-CAM-B3LYP calculation have a $\pi\rightarrow d$ transition at 3.46 eV (359 nm; region \textbf{2}) which fits well with induced reactivity at 366 nm. While the SR calculation also displays a transition in this region (e.g.~in region \textbf{2} at 3.43 eV), the 4c calculation reveals that the density of states (with non-zero intensity) in this region is higher than the SR calculation due to inclusion of excitation with triplet character through spin-orbit coupling. Thus, several more states may participate in the photo-induced reactivity than expected from the SR (or NR) calculation. It is also noted that the  $\pi\rightarrow d$ transitions in \textbf{2} were found to be considerably spin-mixed (the same is true for \textbf{1} and \textbf{3}), which is not captured by the SR (or NR) calculation.
 
Finally, we note that no transitions were found at 647 nm, and the activity induced from light of this wavelength cannot be explained from our current calculations. 

\subsection{UV-vis spectra of \textit{cis}-Pt}

\begin{figure}[htb!]
\centering
\includegraphics[width=1\textwidth,trim={1cm 2.5cm 1cm 12cm},clip]{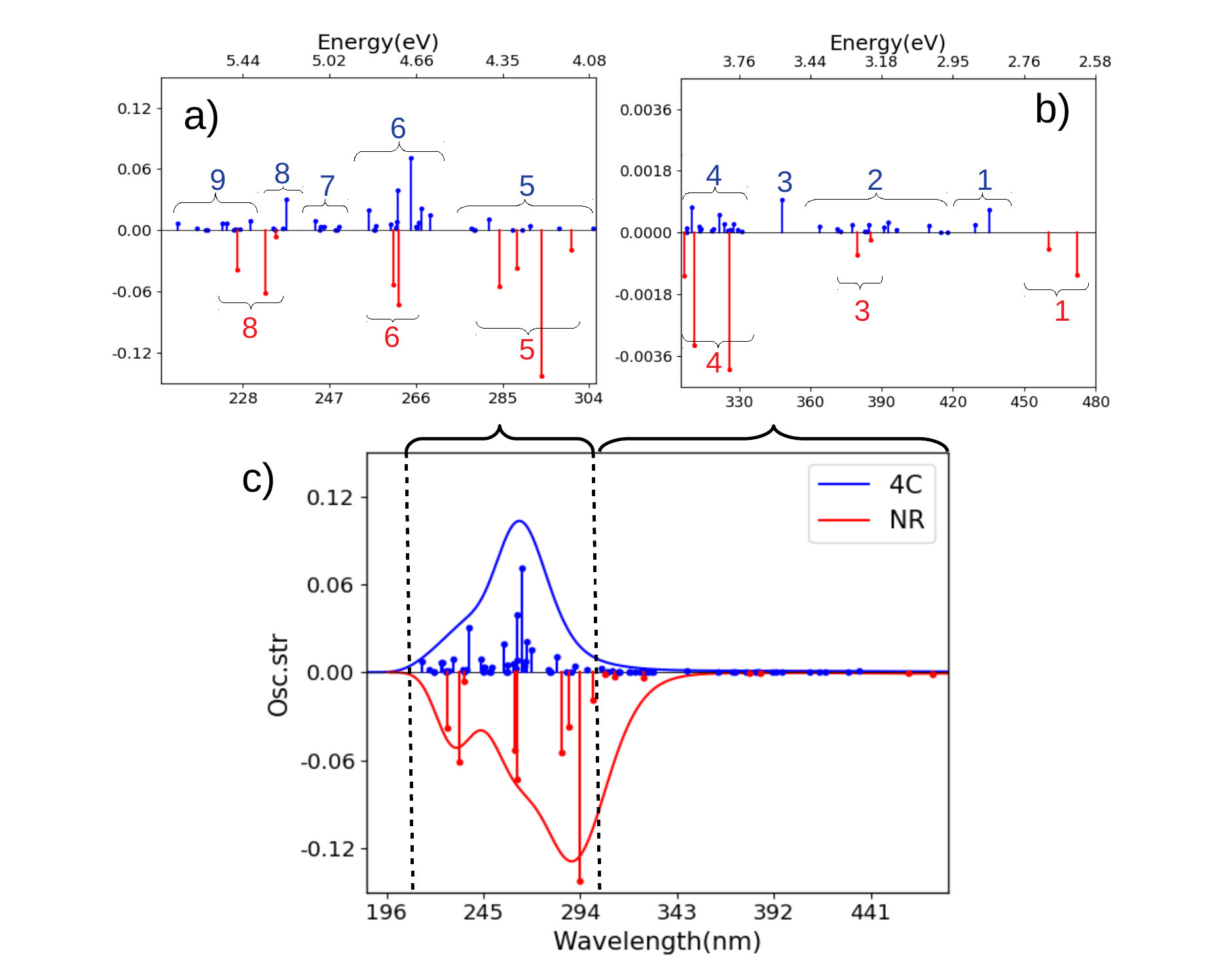}
\caption{Spectra for \textit{cis}-Pt calculated with CAM-B3LYP and NR or 4c  Hamiltonians. (a) and (b) are magnified for 210--306 nm and 306--480 nm. (c) shows the full spectrum. Assignments of main transitions within regions \textbf{1}--\textbf{9} are provided in Tables S7 and S8.  }
\label{fig:cis_camb3lyp_ll}
\end{figure}

\textbf{CAM-B3LYP results:} The spectra obtained with 4c and NR frameworks are shown in Figure \ref{fig:cis_camb3lyp_ll}. The spectra  contain  significantly more transitions with relatively high intensities, compared to \textit{trans}-Pt, and have therefore been divided into nine regions \textbf{1}--\textbf{9}, shown  in Figure \ref{fig:cis_camb3lyp_ll}(a) and \ref{fig:cis_camb3lyp_ll}(b). The full spectra are shown in Figure \ref{fig:cis_camb3lyp_ll}(c) and it can from this figure be seen that also for \textit{cis}-Pt, the NR calculation varies considerably from the 4c calculation. 
\begin{figure}[htb!]
\centering
\includegraphics[width=1\textwidth,trim={1cm 3cm 1cm 12cm},clip]{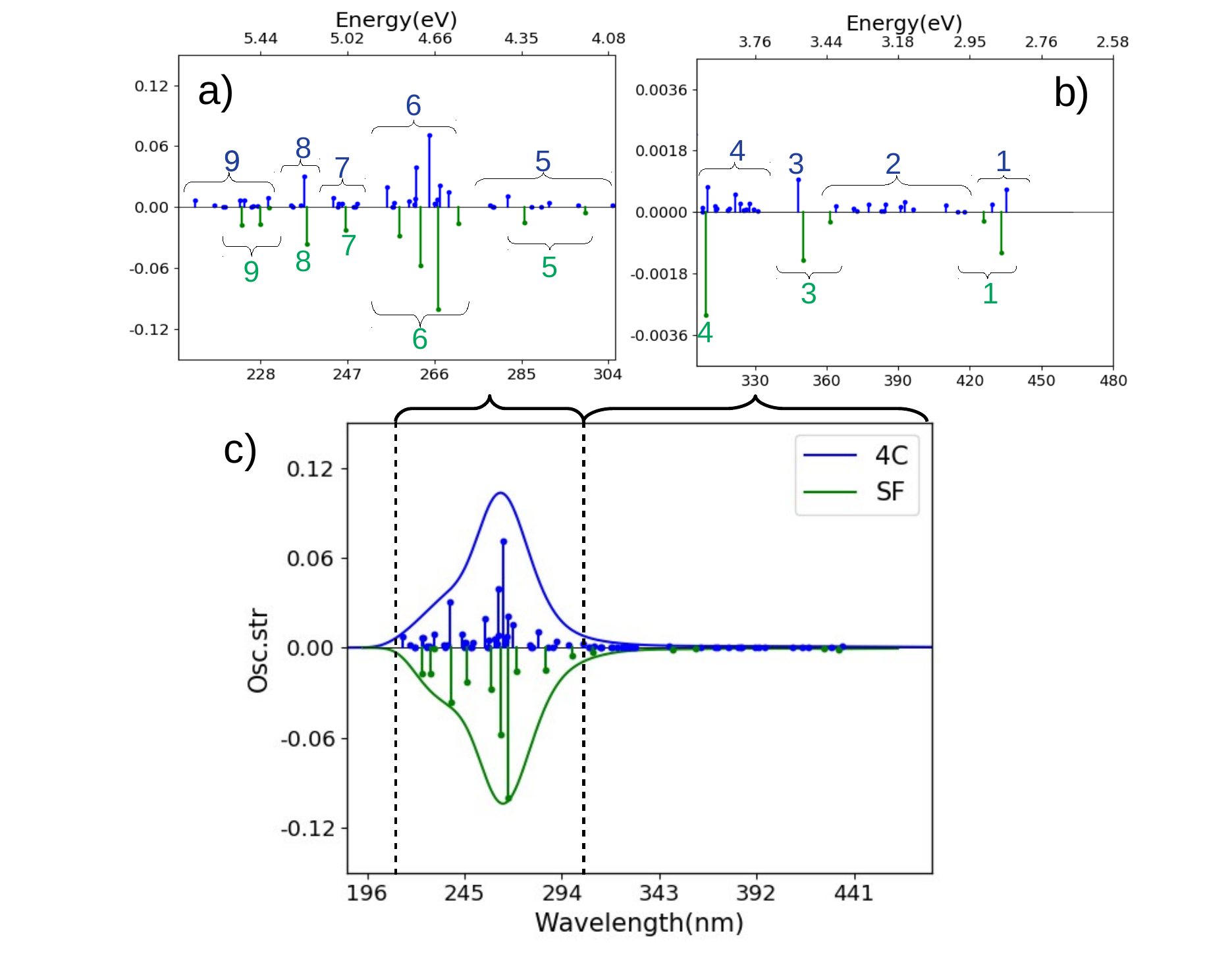}
\caption{ Comparison of spectra for \textit{cis}-Pt calculated with CAM-B3LYP and SR or 4c  Hamiltonians. a) and (b) are magnified for 210--306 nm and 306--480 nm. (c) shows the full spectrum. Assignments of main transitions are provided in Tables S7 and S9.   }
\label{fig:cis_camb3lyp_sf}
\end{figure}

The transitions at highest energy, \textbf{9}--\textbf{7} (5.80--5.40 eV or 214--230 nm) in the 4c calculation, are qualitatively different from their NR counterparts. For the latter we only see few transitions (of $d\rightarrow d$ character) at these high energies: they occur at 5.47--5.27  eV or 227--235 nm and are labeled \textbf{8} in Figure \ref{fig:cis_camb3lyp_ll}(a). The transitions labeled \textbf{9} in the 4c calculations  involve almost exclusively transitions of $d \rightarrow d$ character, consistent with their low intensity,  
while  \textbf{8} involves a mixture of transitions with $d \rightarrow d$ and $\pi\rightarrow d$ (LMCT) character.  

The next group of transitions (\textbf{7} at 5.08--4.98 eV or 244--249 nm) in Figure \ref{fig:cis_camb3lyp_ll}a are only seen in the 4c calculation 
and have relatively low intensity; the most intense transition has $\pi\rightarrow d$ character. 
Meanwhile, the most intense transitions in the spectrum are for the 4c calculation found in region \textbf{6} at  4.85--4.61 eV (256--269 nm) and are followed by several transitions of lower intensity (\textbf{5}) at 4.46--4.26 eV (278--291 nm).  Both regions \textbf{6} and \textbf{5} are of LMCT character with a mixtures of  $\pi\rightarrow d$ and $p\rightarrow d$ transitions, where region \textbf{5} has larger weight of the former and region \textbf{6} has larger weight of the latter. The NR  calculation also has intense transitions in regions \textbf{6} and  \textbf{5} with similar character (apart from the lower density of states): region \textbf{6} is at 4.75--4.73 eV (261--262 nm) and \textbf{5} is at 4.36--4.13 eV (284--300 nm); the transitions at \textbf{5} generally have considerably higher intensity compared to 4c counterpart.      
 
In the low-energy parts of the spectrum (Figure \ref{fig:cis_camb3lyp_ll}b), the first group of transitions (\textbf{4})  at 4.07--3.78 eV (305--328 nm) in the 4c calculation is a mixture of  transitions with $\pi\rightarrow d$ and $p\rightarrow d$ characters; the next region (\textbf{3} at 3.56 eV; 348 nm)  is of $p\rightarrow d$ character. In the NR case, regions \textbf{4} and \textbf{3} have similar characters as the 4c calculation, but \textbf{3} is significantly red-shifted (at 3.26--3.22 eV or 380--385 nm). Interestingly, region \textbf{2} in the 4c calculation consists of many low-intensity LMCT transitions (of mixed $\pi\rightarrow d$ and $p\rightarrow p$ character) at  3.28--3.02 eV eV (378--410 nm), but no match is found in the NR calculations. The last region (\textbf{1}) is in both 4c and NR calculations of LMCT character ($\pi\rightarrow d$), and is also red-shifted in the NR calculation (2.69--2.62 eV or 460--472 nm compared to 2.89--2.84 eV or 430--435 nm in the 4c calculation).  

As for the \textit{trans}-Pt complex, the NR calculation is both quantitatively and qualitatively different from the 4c calculation. We have therefore also investigated whether inclusion of only SR effects can remedy the errors seen in the NR calculation: as seen from Figure \ref{fig:cis_camb3lyp_sf}(c), the SR calculation overall reproduces the most dominant features of the spectrum in the high-energy parts, although there are important differences: as expected, spin-orbit coupling gives rise to significantly higher density of states in the 4c calculation and many  of these states have significant intensity, cf.~Figure \ref{fig:cis_camb3lyp_sf}(b). The higher density of states is evident for all groups of transitions in the high-energy regions (\textbf{9}--\textbf{6}), but perhaps most pronounced for region \textbf{6}: in the 4c calculation this region has five intense transitions (and a number of less intense ones), whereas the SR calculation only has four intense transitions in this region. Analyzing the underlying transitions shows that their character in 4c and SR calculations corresponds fairly well to each other ($\pi \rightarrow d $ and $p\rightarrow d$). However, the transitions in the 4c framework have both  large triplet character and contain spin-mixed orbitals (see Table S10 and Figure S4), causing the differences between 4c and SR (as well as NR) frameworks. 

In the lower-energy parts of the spectrum (regions \textbf{4}--\textbf{1} in Figure \ref{fig:cis_camb3lyp_sf}b), we also see differences between 4c and SR calculations: region \textbf{4} at 4.01 eV or 310 nm in the SR calculation  is almost exclusively of $\pi\rightarrow d$ character, compared to a mix between $\pi\rightarrow d$  and $p \rightarrow d$ in the 4c counterpart. Further, the transitions in  region \textbf{4} involve orbitals that are significantly spin-mixed (and some are also of dominant triplet character) in the 4c calculations, cf.~Table S7 and Figure S4. Accordingly, region \textbf{4} has a number of relatively intense transitions, not seen in the SR calculation. 
The character of transitions in regions \textbf{3} (mainly of $p\rightarrow d$) and \textbf{1} ($\pi\rightarrow d$) in the SR calculation corresponds reasonably well with the 4c calculation, while  no transitions occurs in region \textbf{2} in the SR (or NR) calculation. From the 4c calculation, the underlying transitions in region \textbf{2} are found to be considerably spin-mixed, explaining their absence in the NR and SR calculations.  

\begin{figure}[htb!]
\centering
\includegraphics[width=1\textwidth,trim={1cm 1.5cm 1cm 13.5cm},clip]{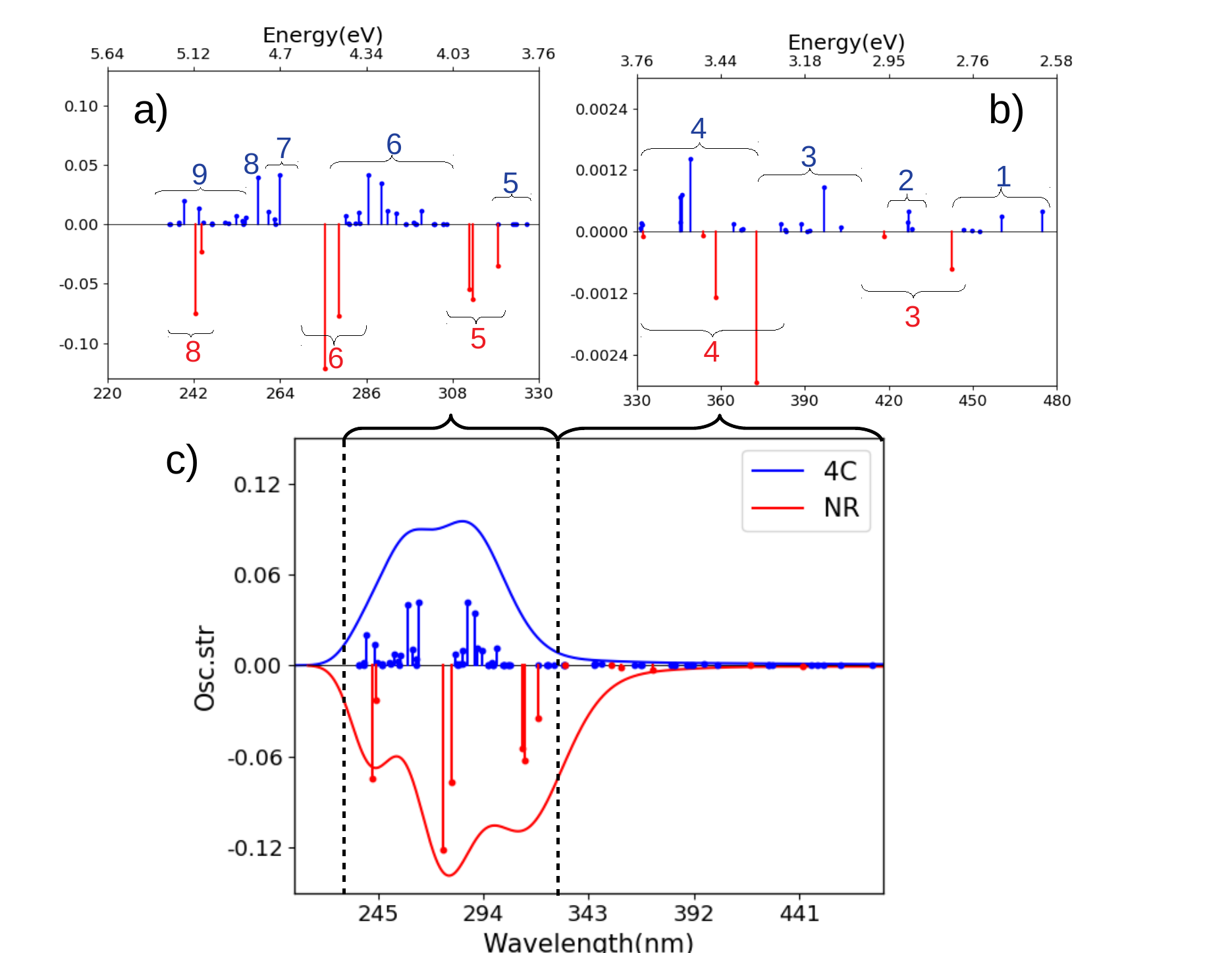}
\caption{Spectra for \textit{cis}-Pt calculated with B3LYP and NR or 4c  Hamiltonians. (a) and (b) are magnified for 220--330 nm and 330--480 nm. (c) shows the full spectrum. Assignments of main transitions within regions \textbf{1}--\textbf{9} are provided in Tables S10 and S11. }
\label{fig:cis_b3lyp_ll}
\end{figure}

\begin{figure}[htb!]
\centering
\includegraphics[width=1\textwidth,trim={1cm 1.5cm 1cm 13.5cm},clip]{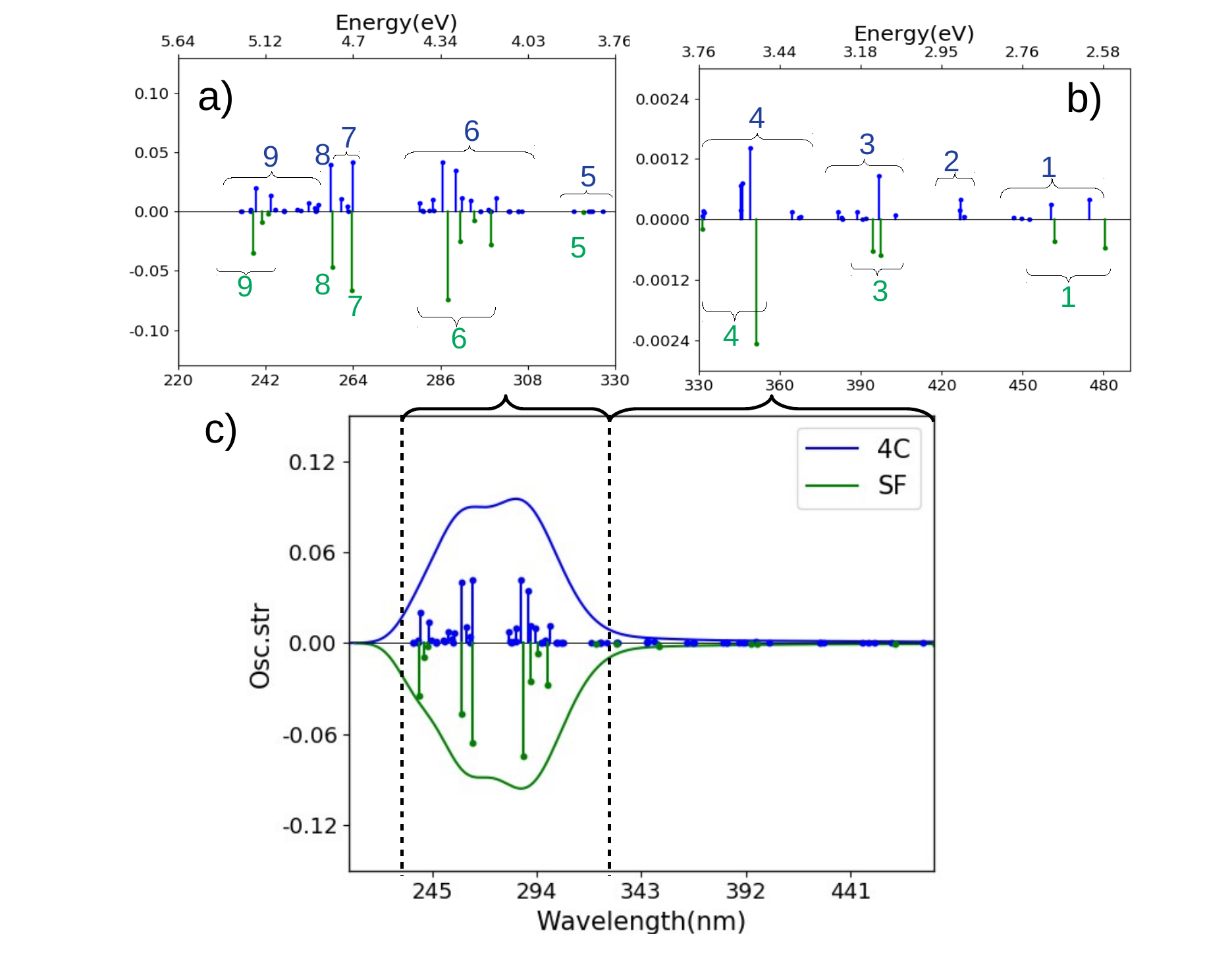}
\caption{Spectra for \textit{trans}-Pt calculated with B3LYP and SR or 4c  Hamiltonians.  (a) and (b) are magnified for 220--330 nm and 330--490 nm. (c) shows the full spectrum. Assignments of main transitions within regions \textbf{1}--\textbf{9} are provided in Tables S10 and S12.   }
\label{fig:cis_b3lyp_sf}
\end{figure}

\textbf{B3LYP results:} Results for 4c, NR and SR Hamiltonians and B3LYP are given in Figures \ref{fig:cis_b3lyp_ll} and \ref{fig:cis_b3lyp_sf}. We also here divide the spectrum into nine regions \textbf{1}--\textbf{9} (cf.~Figures \ref{fig:cis_b3lyp_ll}a and b as well as Figures \ref{fig:cis_b3lyp_sf}a and b).    
We start by comparing 4c-B3LYP with 4c-CAM-B3LYP in the high-energy regions \textbf{9}--\textbf{5} in Figure \ref{fig:cis_b3lyp_ll}(a). The underlying transitions  are roughly similar with the two functionals (cf.~Tables S7 and S11), although they on several occasions are predicted more intense with B3LYP, particular for regions \textbf{9}--\textbf{7}. To the extend comparison between the regions is possible, they are generally red-shifted (between 0.2--0.5 eV) in the 4c-B3LYP calculations, compared to 4c-CAM-B3LYP. Moving to the NR case, the two functionals yield  qualitatively similar spectra: as seen for NR-CAM-B3LYP, the NR-B3LYP calculation shows intense transitions in regions \textbf{8}, \textbf{6} and \textbf{5} and the character of the underlying transitions are similar, although the three regions  also for NR-B3LYP  are red-shifted with 0.2--0.5 eV, compared to NR-CAM-B3LYP.  The NR-B3LYP calculation also  overestimates the intensities in region \textbf{5} compared to the relativistic counterparts (as seen with CAM-B3LYP). 

For the low-energy part of the spectrum (regions \textbf{4}--\textbf{1}, Figure \ref{fig:cis_b3lyp_ll}),  the energy-shifts of 0.5--0.2 eV between regions in B3LYP and CAM-B3LYP seen for the high-energy parts remains for both 4c and NR calculations. Further, both 4c- and NR-B3LYP calculations obtain roughly the same character for the transitions as for CAM-B3LYP (i.e.~LMCT-transitions of $p\rightarrow d $ and $\pi\rightarrow d$ character). However, region \textbf{1} ($\pi\rightarrow d$ for 4c and NR-CAM-B3LYP) is not seen for NR-B3LYP.  

Introducing SR effects in the B3LYP calculations (Figure \ref{fig:cis_b3lyp_sf}) has the same effect as seen with CAM-B3LYP: the most intense features of the spectrum (regions \textbf{9}--\textbf{5}) from the 4c calculation are reproduced, and the transitions are of similar character. 
 Yet, a number of transitions with significant intensity are not seen in the SR calculation due to the lack of spin-orbit coupling. We highlight again  region \textbf{6} which (as for CAM-B3LYP) in the 4c framework contain both transition with dominant triplet character and orbitals that are considerably spin mixed (cf.~Table S10 and Figure S10). Ultimately, this leads to five intense transitions (and several more of lower intensity), compared to only four in the SR calculation. A similar scenario is found for the lower-energy regions (\textbf{4}--\textbf{1}); as for 4c-CAM-B3LYP the transitions in these regions are considerably spin-mixed  and region \textbf{2} is completely missing in SR (and NR) calculations.  

\textbf{Comparison with experiment:} The  B3LYP calculation predicts either three (NR-B3LYP) or two (4c- and SR-B3LYP) intense bands while  the experimental spectrum\cite{mackay2006} contains only one band (single peak at 256 nm or 4.84 eV): regions \textbf{7}--\textbf{8} at  4.80 eV--4.70 eV (259--264 nm) in the 4c-B3LYP calculation corresponds well with  experiment. However, 4c-B3LYP also predicts almost equally intense transitions at 4.36 eV or 284 nm (region \textbf{6}), which is not reflected in the experimental spectrum. A similar result was reported by Solokov and Schaefer\cite{sokolov2011} and Salassa et al.\cite{salassa2009}. The 4c- (or SR-) CAM-B3LYP calculation shows closer resemblance to experiment with one transition having significantly higher intensity than all other transitions (region \textbf{6} at 4.68 eV or 264 nm for the 4c calculation).  Thus, the 4c-CAM-B3LYP overall reproduces the experimental spectrum best. In the following we therefore focus on 4c- and SR-CAM-B3LYP
   
Analyzing the lower-energy parts of the spectrum with 4c-CAM-B3LYP, we note that the \textit{cis}-Pt complex reacts with DNA  after exposure to light at 366 nm, 457.9 nm and also 647.1 nm.\cite{kasparkova2003,muller2003,bednarski2006} From the spectra calculated with 4c-CAM-B3LYP (see Figure \ref{fig:cis_camb3lyp_ll} or \ref{fig:cis_camb3lyp_sf}) we see that regions \textbf{3} (3.56 eV or 348 nm) and \textbf{2} (3.28--3.02 eV or 378--411 nm) in the 4c calculations have transitions that corresponds well with 366 nm (3.39 eV). However, as seen for \textit{trans}-Pt, these regions are considerably spin mixed in the 4c calculation and hence has a denser manifold of states (especially for region \textbf{2}), compared to the SR calculation. Similar arguments holds for region \textbf{1} (2.89--2.85 eV or 429--435 nm) in the 4c-CAM-B3LYP calculation, where several transitions correspond well with transitions at 458 nm (2.7 eV). As for the \textit{trans}-Pt complex, we have not found any transitions at 647 nm. 
\section{Conclusion}

Using TD-DFT, we have carried out a systematic investigation of the effect of using NR, SR and 4c Hamiltonians in calculations of UV-vis spectra  on two Pt(IV) complexes. Both complexes are known to have light-activated activity against cancer cells. We have used both B3LYP and the long-range corrected CAM-B3LYP functionals. We find that 4c-CAM-B3LYP overall compares better with experimental spectra than 4c-B3LYP. However, the changes between NR, SR and 4c Hamiltonains are similar for the two functionals, ensuring that our conclusions are not merely artifacts from the functional choice. For both functionals, we find that (as might have been expected) a NR Hamiltonian does not reproduce the spectra obtained within a 4c framework. The SR calculations  perform  better. In fact, employing a SR Hamiltonian is sufficient to reproduce the main features of the experimental spectra, which are dominated by one strong absorption. However, the 4c-calculations shows that the underlying transitions are not always properly reproduced in the SR calculations, due to the lack of spin-orbit coupling: thus, intense bands in the higher energy parts (below 300 nm) as well as transitions of lower intensity (above 300 nm) can be considerably spin mixed and/or contain transitions of triplet character. This cannot be properly reproduced with SR and NR Hamiltonians. Notably, this can also be expected with ECPs if they are used together with a non-relativistic Hamiltonian.  

 Seeing that incident light with wavelength above 300 nm is usually employed to photo-activate \textit{trans}-Pt and  \textit{cis}-Pt, inclusion spin-orbit coupling seems pertinent to fully understand the activation mechanism. For instance, the 4c-CAM-B3LYP calculations predict a dense manifold of states with low, yet  non-zero intensities in regions around 366 nm (a wavelength use to activate both complexes). These transitions  either have triplet character and/or involve significantly spin-mixed orbitals (and a similar scenario occurs for \textit{cis}-Pt around 458 nm). 

While our present results show that it is advantageous to include spin-orbit coupling in investigations of Pt complexes used for PACT, it is probably not necessary to use a 4c framework;  we anticipate that various two-component methods will be sufficient. Further, we note that our present study did  not take any solvent effects into account. We have recently extended the so-called polarizable embedding model for this purpose\cite{hedegaard2017} and we plan to investigate the solvent effect in a forthcoming paper.

\section*{Acknowledgments}

E.D.H. acknowledges financial support from the European Commission (MetEmbed, Project No. 745967) and the Villum Foundation (Young Investigator Program, Grant No. 29412). 

\bibliography{article}

\begin{thebibliography}{60}%
\makeatletter
\providecommand \@ifxundefined [1]{%
 \@ifx{#1\undefined}
}%
\providecommand \@ifnum [1]{%
 \ifnum #1\expandafter \@firstoftwo
 \else \expandafter \@secondoftwo
 \fi
}%
\providecommand \@ifx [1]{%
 \ifx #1\expandafter \@firstoftwo
 \else \expandafter \@secondoftwo
 \fi
}%
\providecommand \natexlab [1]{#1}%
\providecommand \enquote  [1]{``#1''}%
\providecommand \bibnamefont  [1]{#1}%
\providecommand \bibfnamefont [1]{#1}%
\providecommand \citenamefont [1]{#1}%
\providecommand \href@noop [0]{\@secondoftwo}%
\providecommand \href [0]{\begingroup \@sanitize@url \@href}%
\providecommand \@href[1]{\@@startlink{#1}\@@href}%
\providecommand \@@href[1]{\endgroup#1\@@endlink}%
\providecommand \@sanitize@url [0]{\catcode `\\12\catcode `\$12\catcode
  `\&12\catcode `\#12\catcode `\^12\catcode `\_12\catcode `\%12\relax}%
\providecommand \@@startlink[1]{}%
\providecommand \@@endlink[0]{}%
\providecommand \url  [0]{\begingroup\@sanitize@url \@url }%
\providecommand \@url [1]{\endgroup\@href {#1}{\urlprefix }}%
\providecommand \urlprefix  [0]{URL }%
\providecommand \Eprint [0]{\href }%
\providecommand \doibase [0]{http://dx.doi.org/}%
\providecommand \selectlanguage [0]{\@gobble}%
\providecommand \bibinfo  [0]{\@secondoftwo}%
\providecommand \bibfield  [0]{\@secondoftwo}%
\providecommand \translation [1]{[#1]}%
\providecommand \BibitemOpen [0]{}%
\providecommand \bibitemStop [0]{}%
\providecommand \bibitemNoStop [0]{.\EOS\space}%
\providecommand \EOS [0]{\spacefactor3000\relax}%
\providecommand \BibitemShut  [1]{\csname bibitem#1\endcsname}%
\let\auto@bib@innerbib\@empty
\bibitem [{\citenamefont {Johnstone}, \citenamefont {Suntharalingam},\ and\
  \citenamefont {Lippard}(2016)}]{johnstone2016}%
  \BibitemOpen
  \bibfield  {author} {\bibinfo {author} {\bibfnamefont {T.~C.}\ \bibnamefont
  {Johnstone}}, \bibinfo {author} {\bibfnamefont {K.}~\bibnamefont
  {Suntharalingam}}, \ and\ \bibinfo {author} {\bibfnamefont {S.~J.}\
  \bibnamefont {Lippard}},\ }\href {\doibase 10.1021/acs.chemrev.5b00597}
  {\bibfield  {journal} {\bibinfo  {journal} {Chem. Rev.}\ }\textbf {\bibinfo
  {volume} {116}},\ \bibinfo {pages} {3436} (\bibinfo {year}
  {2016})}\BibitemShut {NoStop}%
\bibitem [{\citenamefont {Hall}\ and\ \citenamefont
  {Hambley}(2002)}]{hall2002}%
  \BibitemOpen
  \bibfield  {author} {\bibinfo {author} {\bibfnamefont {M.~D.}\ \bibnamefont
  {Hall}}\ and\ \bibinfo {author} {\bibfnamefont {T.~W.}\ \bibnamefont
  {Hambley}},\ }\href {\doibase 10.1016/S0010-8545(02)00026-7} {\bibfield
  {journal} {\bibinfo  {journal} {Coord. Chem. Rev.}\ }\textbf {\bibinfo
  {volume} {232}},\ \bibinfo {pages} {49} (\bibinfo {year} {2002})}\BibitemShut
  {NoStop}%
\bibitem [{\citenamefont {Graf}\ and\ \citenamefont
  {Lippard}(2012)}]{graf2012}%
  \BibitemOpen
  \bibfield  {author} {\bibinfo {author} {\bibfnamefont {N.}~\bibnamefont
  {Graf}}\ and\ \bibinfo {author} {\bibfnamefont {S.~J.}\ \bibnamefont
  {Lippard}},\ }\href {\doibase 10.1016/j.addr.2012.01.007} {\bibfield
  {journal} {\bibinfo  {journal} {Adv. Drug Deliv. Rev.}\ }\textbf {\bibinfo
  {volume} {64}},\ \bibinfo {pages} {993} (\bibinfo {year} {2012})}\BibitemShut
  {NoStop}%
\bibitem [{\citenamefont {Butler}\ and\ \citenamefont
  {Sadler}(2013)}]{butler2013}%
  \BibitemOpen
  \bibfield  {author} {\bibinfo {author} {\bibfnamefont {J.~S.}\ \bibnamefont
  {Butler}}\ and\ \bibinfo {author} {\bibfnamefont {P.~J.}\ \bibnamefont
  {Sadler}},\ }\href {\doibase 10.1016/j.cbpa.2013.01.004} {\bibfield
  {journal} {\bibinfo  {journal} {Curr. Opin. Chem. Biol.}\ }\textbf {\bibinfo
  {volume} {17}},\ \bibinfo {pages} {175} (\bibinfo {year} {2013})}\BibitemShut
  {NoStop}%
\bibitem [{\citenamefont {Farrer}, \citenamefont {Salassa},\ and\ \citenamefont
  {Sadler}(2009)}]{farrer2009}%
  \BibitemOpen
  \bibfield  {author} {\bibinfo {author} {\bibfnamefont {N.~J.}\ \bibnamefont
  {Farrer}}, \bibinfo {author} {\bibfnamefont {L.}~\bibnamefont {Salassa}}, \
  and\ \bibinfo {author} {\bibfnamefont {P.~J.}\ \bibnamefont {Sadler}},\
  }\href {\doibase 10.1039/b912898k} {\bibfield  {journal} {\bibinfo  {journal}
  {Dalton Trans.}\ }\textbf {\bibinfo {volume} {48}},\ \bibinfo {pages} {10690}
  (\bibinfo {year} {2009})}\BibitemShut {NoStop}%
\bibitem [{\citenamefont {Bonnet}(2018)}]{bonnet2018}%
  \BibitemOpen
  \bibfield  {author} {\bibinfo {author} {\bibfnamefont {S.}~\bibnamefont
  {Bonnet}},\ }\href {\doibase 10.1039/c8dt01585f} {\bibfield  {journal}
  {\bibinfo  {journal} {Dalton Trans.}\ }\textbf {\bibinfo {volume} {47}},\
  \bibinfo {pages} {10330} (\bibinfo {year} {2018})}\BibitemShut {NoStop}%
\bibitem [{\citenamefont {Brown}, \citenamefont {Brown},\ and\ \citenamefont
  {Walker}(2004)}]{brown2004}%
  \BibitemOpen
  \bibfield  {author} {\bibinfo {author} {\bibfnamefont {S.~B.}\ \bibnamefont
  {Brown}}, \bibinfo {author} {\bibfnamefont {E.~A.}\ \bibnamefont {Brown}}, \
  and\ \bibinfo {author} {\bibfnamefont {I.}~\bibnamefont {Walker}},\
  }\href@noop {} {\bibfield  {journal} {\bibinfo  {journal} {Lancet Oncol.}\
  }\textbf {\bibinfo {volume} {5}},\ \bibinfo {pages} {497} (\bibinfo {year}
  {2004})}\BibitemShut {NoStop}%
\bibitem [{\citenamefont {Chen}\ \emph {et~al.}(2015)\citenamefont {Chen},
  \citenamefont {Tian}, \citenamefont {He},\ and\ \citenamefont
  {Guo}}]{chen2015}%
  \BibitemOpen
  \bibfield  {author} {\bibinfo {author} {\bibfnamefont {H.}~\bibnamefont
  {Chen}}, \bibinfo {author} {\bibfnamefont {J.}~\bibnamefont {Tian}}, \bibinfo
  {author} {\bibfnamefont {W.}~\bibnamefont {He}}, \ and\ \bibinfo {author}
  {\bibfnamefont {Z.}~\bibnamefont {Guo}},\ }\href {\doibase 10.1021/ja511420n}
  {\bibfield  {journal} {\bibinfo  {journal} {J. Am. Chem. Soc.}\ }\textbf
  {\bibinfo {volume} {137}},\ \bibinfo {pages} {1539} (\bibinfo {year}
  {2015})}\BibitemShut {NoStop}%
\bibitem [{\citenamefont {Wilson}\ and\ \citenamefont
  {Hay}(2011)}]{wilson2011}%
  \BibitemOpen
  \bibfield  {author} {\bibinfo {author} {\bibfnamefont {W.~R.}\ \bibnamefont
  {Wilson}}\ and\ \bibinfo {author} {\bibfnamefont {M.~P.}\ \bibnamefont
  {Hay}},\ }\href {\doibase 10.1038/nrc3064} {\bibfield  {journal} {\bibinfo
  {journal} {Nat. Rev. Cancer}\ }\textbf {\bibinfo {volume} {11}},\ \bibinfo
  {pages} {393} (\bibinfo {year} {2011})}\BibitemShut {NoStop}%
\bibitem [{\citenamefont {Salassa}, \citenamefont {Phillips},\ and\
  \citenamefont {Sadler}(2009)}]{salassa2009}%
  \BibitemOpen
  \bibfield  {author} {\bibinfo {author} {\bibfnamefont {L.}~\bibnamefont
  {Salassa}}, \bibinfo {author} {\bibfnamefont {H.~I.~A.}\ \bibnamefont
  {Phillips}}, \ and\ \bibinfo {author} {\bibfnamefont {P.~J.}\ \bibnamefont
  {Sadler}},\ }\href {\doibase 10.1039/b912496a} {\bibfield  {journal}
  {\bibinfo  {journal} {Phys. Chem. Chem. Phys.}\ }\textbf {\bibinfo {volume}
  {11}},\ \bibinfo {pages} {10311} (\bibinfo {year} {2009})}\BibitemShut
  {NoStop}%
\bibitem [{\citenamefont {Mackay}\ \emph {et~al.}(2009)\citenamefont {Mackay},
  \citenamefont {Farrer}, \citenamefont {Salassa}, \citenamefont {Tai},
  \citenamefont {Deeth}, \citenamefont {Moggach}, \citenamefont {Wood},
  \citenamefont {Parsonsa},\ and\ \citenamefont {Sadler}}]{mackay2009}%
  \BibitemOpen
  \bibfield  {author} {\bibinfo {author} {\bibfnamefont {F.~S.}\ \bibnamefont
  {Mackay}}, \bibinfo {author} {\bibfnamefont {N.~J.}\ \bibnamefont {Farrer}},
  \bibinfo {author} {\bibfnamefont {L.}~\bibnamefont {Salassa}}, \bibinfo
  {author} {\bibfnamefont {H.-C.}\ \bibnamefont {Tai}}, \bibinfo {author}
  {\bibfnamefont {R.~J.}\ \bibnamefont {Deeth}}, \bibinfo {author}
  {\bibfnamefont {S.~A.}\ \bibnamefont {Moggach}}, \bibinfo {author}
  {\bibfnamefont {P.~A.}\ \bibnamefont {Wood}}, \bibinfo {author}
  {\bibfnamefont {S.}~\bibnamefont {Parsonsa}}, \ and\ \bibinfo {author}
  {\bibfnamefont {P.~J.}\ \bibnamefont {Sadler}},\ }\href {\doibase
  10.1039/b820550g} {\bibfield  {journal} {\bibinfo  {journal} {Dalton Trans.}\
  }\textbf {\bibinfo {volume} {13}},\ \bibinfo {pages} {2315} (\bibinfo {year}
  {2009})}\BibitemShut {NoStop}%
\bibitem [{\citenamefont {Farrer}\ \emph {et~al.}(2010)\citenamefont {Farrer},
  \citenamefont {Woods}, \citenamefont {Salassa}, \citenamefont {Zhao},
  \citenamefont {Robinson}, \citenamefont {Clarkson}, \citenamefont {Mackay},\
  and\ \citenamefont {Sadler}}]{farrer2010}%
  \BibitemOpen
  \bibfield  {author} {\bibinfo {author} {\bibfnamefont {N.~J.}\ \bibnamefont
  {Farrer}}, \bibinfo {author} {\bibfnamefont {J.~A.}\ \bibnamefont {Woods}},
  \bibinfo {author} {\bibfnamefont {L.}~\bibnamefont {Salassa}}, \bibinfo
  {author} {\bibfnamefont {Y.}~\bibnamefont {Zhao}}, \bibinfo {author}
  {\bibfnamefont {K.~S.}\ \bibnamefont {Robinson}}, \bibinfo {author}
  {\bibfnamefont {G.}~\bibnamefont {Clarkson}}, \bibinfo {author}
  {\bibfnamefont {F.~S.}\ \bibnamefont {Mackay}}, \ and\ \bibinfo {author}
  {\bibfnamefont {P.~J.}\ \bibnamefont {Sadler}},\ }\href {\doibase
  10.1002/anie.201003399} {\bibfield  {journal} {\bibinfo  {journal} {Angew.
  Chem. Int. Ed.}\ }\textbf {\bibinfo {volume} {49}},\ \bibinfo {pages} {8905}
  (\bibinfo {year} {2010})}\BibitemShut {NoStop}%
\bibitem [{\citenamefont {Westendorf}\ \emph {et~al.}(2011)\citenamefont
  {Westendorf}, \citenamefont {Zerzankova}, \citenamefont {Salassa},
  \citenamefont {Sadler}, \citenamefont {Brabec},\ and\ \citenamefont
  {Bednarski}}]{westendorf2011}%
  \BibitemOpen
  \bibfield  {author} {\bibinfo {author} {\bibfnamefont {A.~F.}\ \bibnamefont
  {Westendorf}}, \bibinfo {author} {\bibfnamefont {L.}~\bibnamefont
  {Zerzankova}}, \bibinfo {author} {\bibfnamefont {L.}~\bibnamefont {Salassa}},
  \bibinfo {author} {\bibfnamefont {P.~J.}\ \bibnamefont {Sadler}}, \bibinfo
  {author} {\bibfnamefont {V.}~\bibnamefont {Brabec}}, \ and\ \bibinfo {author}
  {\bibfnamefont {P.~J.}\ \bibnamefont {Bednarski}},\ }\href {\doibase
  10.1016/j.jinorgbio.2011.01.003} {\bibfield  {journal} {\bibinfo  {journal}
  {J. Inorg. Biochem.}\ }\textbf {\bibinfo {volume} {105}},\ \bibinfo {pages}
  {652} (\bibinfo {year} {2011})}\BibitemShut {NoStop}%
\bibitem [{\citenamefont {Westendorf}\ \emph {et~al.}(2012)\citenamefont
  {Westendorf}, \citenamefont {Woods}, \citenamefont {Korpis}, \citenamefont
  {Farrer}, \citenamefont {Salassa}, \citenamefont {Robinson}, \citenamefont
  {Appleyard}, \citenamefont {Murray}, \citenamefont {Gr{\"{u}}nert},
  \citenamefont {Thompson}, \citenamefont {Sadler},\ and\ \citenamefont
  {Bednarski}}]{westendorf2012}%
  \BibitemOpen
  \bibfield  {author} {\bibinfo {author} {\bibfnamefont {A.~F.}\ \bibnamefont
  {Westendorf}}, \bibinfo {author} {\bibfnamefont {J.~A.}\ \bibnamefont
  {Woods}}, \bibinfo {author} {\bibfnamefont {K.}~\bibnamefont {Korpis}},
  \bibinfo {author} {\bibfnamefont {N.~J.}\ \bibnamefont {Farrer}}, \bibinfo
  {author} {\bibfnamefont {L.}~\bibnamefont {Salassa}}, \bibinfo {author}
  {\bibfnamefont {K.}~\bibnamefont {Robinson}}, \bibinfo {author}
  {\bibfnamefont {V.}~\bibnamefont {Appleyard}}, \bibinfo {author}
  {\bibfnamefont {K.}~\bibnamefont {Murray}}, \bibinfo {author} {\bibfnamefont
  {R.}~\bibnamefont {Gr{\"{u}}nert}}, \bibinfo {author} {\bibfnamefont {A.~M.}\
  \bibnamefont {Thompson}}, \bibinfo {author} {\bibfnamefont {P.~J.}\
  \bibnamefont {Sadler}}, \ and\ \bibinfo {author} {\bibfnamefont {P.~J.}\
  \bibnamefont {Bednarski}},\ }\href {\doibase 10.1158/1535-7163.MCT-11-0959}
  {\bibfield  {journal} {\bibinfo  {journal} {Mol. Cancer Ther.}\ }\textbf
  {\bibinfo {volume} {11}},\ \bibinfo {pages} {1894} (\bibinfo {year}
  {2012})}\BibitemShut {NoStop}%
\bibitem [{\citenamefont {Zhao}\ \emph
  {et~al.}(2013{\natexlab{a}})\citenamefont {Zhao}, \citenamefont {Woods},
  \citenamefont {Farrer}, \citenamefont {Robinson}, \citenamefont {Pracharova},
  \citenamefont {Ka{\v{s}}p{\'{a}}rkov{\'{a}}}, \citenamefont {Novakova},
  \citenamefont {Li}, \citenamefont {Salassa}, \citenamefont {Pizarro},
  \citenamefont {Clarkson}, \citenamefont {Song}, \citenamefont {Brabec},\ and\
  \citenamefont {Sadler}}]{zhao2013a}%
  \BibitemOpen
  \bibfield  {author} {\bibinfo {author} {\bibfnamefont {Y.}~\bibnamefont
  {Zhao}}, \bibinfo {author} {\bibfnamefont {J.~A.}\ \bibnamefont {Woods}},
  \bibinfo {author} {\bibfnamefont {N.~J.}\ \bibnamefont {Farrer}}, \bibinfo
  {author} {\bibfnamefont {K.~S.}\ \bibnamefont {Robinson}}, \bibinfo {author}
  {\bibfnamefont {J.}~\bibnamefont {Pracharova}}, \bibinfo {author}
  {\bibfnamefont {J.}~\bibnamefont {Ka{\v{s}}p{\'{a}}rkov{\'{a}}}}, \bibinfo
  {author} {\bibfnamefont {O.}~\bibnamefont {Novakova}}, \bibinfo {author}
  {\bibfnamefont {H.}~\bibnamefont {Li}}, \bibinfo {author} {\bibfnamefont
  {L.}~\bibnamefont {Salassa}}, \bibinfo {author} {\bibfnamefont {A.~M.}\
  \bibnamefont {Pizarro}}, \bibinfo {author} {\bibfnamefont {G.~J.}\
  \bibnamefont {Clarkson}}, \bibinfo {author} {\bibfnamefont {L.}~\bibnamefont
  {Song}}, \bibinfo {author} {\bibfnamefont {V.}~\bibnamefont {Brabec}}, \ and\
  \bibinfo {author} {\bibfnamefont {P.~J.}\ \bibnamefont {Sadler}},\ }\href
  {\doibase 10.1002/chem.201300374} {\bibfield  {journal} {\bibinfo  {journal}
  {Chem. Eur. J.}\ }\textbf {\bibinfo {volume} {19}},\ \bibinfo {pages} {9578}
  (\bibinfo {year} {2013}{\natexlab{a}})}\BibitemShut {NoStop}%
\bibitem [{\citenamefont {Zhao}\ \emph
  {et~al.}(2013{\natexlab{b}})\citenamefont {Zhao}, \citenamefont {Farrer},
  \citenamefont {Li}, \citenamefont {Butler}, \citenamefont {Mcquitty},
  \citenamefont {Habtemariam}, \citenamefont {Wang},\ and\ \citenamefont
  {Sadler}}]{zhao2013b}%
  \BibitemOpen
  \bibfield  {author} {\bibinfo {author} {\bibfnamefont {Y.}~\bibnamefont
  {Zhao}}, \bibinfo {author} {\bibfnamefont {N.~J.}\ \bibnamefont {Farrer}},
  \bibinfo {author} {\bibfnamefont {H.}~\bibnamefont {Li}}, \bibinfo {author}
  {\bibfnamefont {J.~S.}\ \bibnamefont {Butler}}, \bibinfo {author}
  {\bibfnamefont {R.~J.}\ \bibnamefont {Mcquitty}}, \bibinfo {author}
  {\bibfnamefont {A.}~\bibnamefont {Habtemariam}}, \bibinfo {author}
  {\bibfnamefont {F.}~\bibnamefont {Wang}}, \ and\ \bibinfo {author}
  {\bibfnamefont {P.~J.}\ \bibnamefont {Sadler}},\ }\href {\doibase
  10.1002/ange.201307505} {\bibfield  {journal} {\bibinfo  {journal} {Angew.
  Chem. Int. Ed.}\ }\textbf {\bibinfo {volume} {125}},\ \bibinfo {pages}
  {13878} (\bibinfo {year} {2013}{\natexlab{b}})}\BibitemShut {NoStop}%
\bibitem [{\citenamefont {Shaili}\ \emph {et~al.}(2019)\citenamefont {Shaili},
  \citenamefont {Salassa}, \citenamefont {Woods}, \citenamefont {Clarkson},
  \citenamefont {Sadler},\ and\ \citenamefont {Farrer}}]{shaili2019}%
  \BibitemOpen
  \bibfield  {author} {\bibinfo {author} {\bibfnamefont {E.}~\bibnamefont
  {Shaili}}, \bibinfo {author} {\bibfnamefont {L.}~\bibnamefont {Salassa}},
  \bibinfo {author} {\bibfnamefont {J.~A.}\ \bibnamefont {Woods}}, \bibinfo
  {author} {\bibfnamefont {G.}~\bibnamefont {Clarkson}}, \bibinfo {author}
  {\bibfnamefont {P.~J.}\ \bibnamefont {Sadler}}, \ and\ \bibinfo {author}
  {\bibfnamefont {N.~J.}\ \bibnamefont {Farrer}},\ }\href {\doibase
  10.1039/c9sc02644d} {\bibfield  {journal} {\bibinfo  {journal} {Chem. Sci.}\
  }\textbf {\bibinfo {volume} {10}},\ \bibinfo {pages} {8610} (\bibinfo {year}
  {2019})}\BibitemShut {NoStop}%
\bibitem [{\citenamefont {Sokolov}\ and\ \citenamefont
  {Schaefer~III}(2011)}]{sokolov2011}%
  \BibitemOpen
  \bibfield  {author} {\bibinfo {author} {\bibfnamefont {A.~Y.}\ \bibnamefont
  {Sokolov}}\ and\ \bibinfo {author} {\bibfnamefont {H.~F.}\ \bibnamefont
  {Schaefer~III}},\ }\href {\doibase 10.1039/c1dt10493d} {\bibfield  {journal}
  {\bibinfo  {journal} {Dalton Trans.}\ }\textbf {\bibinfo {volume} {40}},\
  \bibinfo {pages} {7571} (\bibinfo {year} {2011})}\BibitemShut {NoStop}%
\bibitem [{\citenamefont {Liu}\ and\ \citenamefont {Xiao}(2018)}]{liu2018}%
  \BibitemOpen
  \bibfield  {author} {\bibinfo {author} {\bibfnamefont {W.}~\bibnamefont
  {Liu}}\ and\ \bibinfo {author} {\bibfnamefont {Y.}~\bibnamefont {Xiao}},\
  }\href@noop {} {\bibfield  {journal} {\bibinfo  {journal} {Chem. Soc. Rev.}\
  }\textbf {\bibinfo {volume} {47}},\ \bibinfo {pages} {4481} (\bibinfo {year}
  {2018})}\BibitemShut {NoStop}%
\bibitem [{\citenamefont {Saue}\ \emph {et~al.}(2020)\citenamefont {Saue},
  \citenamefont {Bast}, \citenamefont {Gomes}, \citenamefont {Jensen},
  \citenamefont {Visscher}, \citenamefont {Aucar}, \citenamefont {Di~Remigio},
  \citenamefont {Dyall}, \citenamefont {Eliav}, \citenamefont {Fasshauer},
  \citenamefont {Fleig}, \citenamefont {Halbert}, \citenamefont {Hedeg{\aa}rd},
  \citenamefont {Helmich-Paris}, \citenamefont {Iliaš}, \citenamefont {Jacob},
  \citenamefont {Knecht}, \citenamefont {L{\ae}dahl}, \citenamefont {Vidal},
  \citenamefont {Nayak}, \citenamefont {Olejniczak}, \citenamefont {Olsen},
  \citenamefont {Pernpointner}, \citenamefont {Senjean}, \citenamefont {Shee},
  \citenamefont {Sunaga},\ and\ \citenamefont {van Stralen}}]{saue2020}%
  \BibitemOpen
  \bibfield  {author} {\bibinfo {author} {\bibfnamefont {T.}~\bibnamefont
  {Saue}}, \bibinfo {author} {\bibfnamefont {R.}~\bibnamefont {Bast}}, \bibinfo
  {author} {\bibfnamefont {A.~S.~P.}\ \bibnamefont {Gomes}}, \bibinfo {author}
  {\bibfnamefont {H.~J.~{\relax Aa}.}\ \bibnamefont {Jensen}}, \bibinfo
  {author} {\bibfnamefont {L.}~\bibnamefont {Visscher}}, \bibinfo {author}
  {\bibfnamefont {I.~A.}\ \bibnamefont {Aucar}}, \bibinfo {author}
  {\bibfnamefont {R.}~\bibnamefont {Di~Remigio}}, \bibinfo {author}
  {\bibfnamefont {K.~G.}\ \bibnamefont {Dyall}}, \bibinfo {author}
  {\bibfnamefont {E.}~\bibnamefont {Eliav}}, \bibinfo {author} {\bibfnamefont
  {E.}~\bibnamefont {Fasshauer}}, \bibinfo {author} {\bibfnamefont
  {T.}~\bibnamefont {Fleig}}, \bibinfo {author} {\bibfnamefont
  {L.}~\bibnamefont {Halbert}}, \bibinfo {author} {\bibfnamefont {E.~D.}\
  \bibnamefont {Hedeg{\aa}rd}}, \bibinfo {author} {\bibfnamefont
  {B.}~\bibnamefont {Helmich-Paris}}, \bibinfo {author} {\bibfnamefont
  {M.}~\bibnamefont {Iliaš}}, \bibinfo {author} {\bibfnamefont {C.~R.}\
  \bibnamefont {Jacob}}, \bibinfo {author} {\bibfnamefont {S.}~\bibnamefont
  {Knecht}}, \bibinfo {author} {\bibfnamefont {J.~K.}\ \bibnamefont
  {L{\ae}dahl}}, \bibinfo {author} {\bibfnamefont {M.~L.}\ \bibnamefont
  {Vidal}}, \bibinfo {author} {\bibfnamefont {M.~K.}\ \bibnamefont {Nayak}},
  \bibinfo {author} {\bibfnamefont {M.}~\bibnamefont {Olejniczak}}, \bibinfo
  {author} {\bibfnamefont {J.~M.~H.}\ \bibnamefont {Olsen}}, \bibinfo {author}
  {\bibfnamefont {M.}~\bibnamefont {Pernpointner}}, \bibinfo {author}
  {\bibfnamefont {B.}~\bibnamefont {Senjean}}, \bibinfo {author} {\bibfnamefont
  {A.}~\bibnamefont {Shee}}, \bibinfo {author} {\bibfnamefont {A.}~\bibnamefont
  {Sunaga}}, \ and\ \bibinfo {author} {\bibfnamefont {J.~N.~P.}\ \bibnamefont
  {van Stralen}},\ }\href {\doibase 10.1063/5.0004844} {\bibfield  {journal}
  {\bibinfo  {journal} {J. Chem. Phys.}\ }\textbf {\bibinfo {volume} {152}},\
  \bibinfo {pages} {204104} (\bibinfo {year} {2020})}\BibitemShut {NoStop}%
\bibitem [{\citenamefont {Repisk{\'y}}\ \emph {et~al.}(2020)\citenamefont
  {Repisk{\'y}}, \citenamefont {Komorovsk{\'y}}, \citenamefont {Kadek},
  \citenamefont {Konecny}, \citenamefont {Ekstr{\"o}m}, \citenamefont {Malkin},
  \citenamefont {Kaupp}, \citenamefont {Ruud}, \citenamefont {Malkina},\ and\
  \citenamefont {Malkin}}]{repisky2020}%
  \BibitemOpen
  \bibfield  {author} {\bibinfo {author} {\bibfnamefont {M.}~\bibnamefont
  {Repisk{\'y}}}, \bibinfo {author} {\bibfnamefont {S.}~\bibnamefont
  {Komorovsk{\'y}}}, \bibinfo {author} {\bibfnamefont {M.}~\bibnamefont
  {Kadek}}, \bibinfo {author} {\bibfnamefont {L.}~\bibnamefont {Konecny}},
  \bibinfo {author} {\bibfnamefont {U.}~\bibnamefont {Ekstr{\"o}m}}, \bibinfo
  {author} {\bibfnamefont {E.}~\bibnamefont {Malkin}}, \bibinfo {author}
  {\bibfnamefont {M.}~\bibnamefont {Kaupp}}, \bibinfo {author} {\bibfnamefont
  {K.}~\bibnamefont {Ruud}}, \bibinfo {author} {\bibfnamefont {O.~L.}\
  \bibnamefont {Malkina}}, \ and\ \bibinfo {author} {\bibfnamefont {V.~G.}\
  \bibnamefont {Malkin}},\ }\href {\doibase 10.1063/5.0005094} {\bibfield
  {journal} {\bibinfo  {journal} {J. Chem. Phys.}\ }\textbf {\bibinfo {volume}
  {152}},\ \bibinfo {pages} {184101} (\bibinfo {year} {2020})}\BibitemShut
  {NoStop}%
\bibitem [{\citenamefont {Gao}\ \emph {et~al.}(2004)\citenamefont {Gao},
  \citenamefont {Liu}, \citenamefont {Song},\ and\ \citenamefont
  {Liu}}]{gao2004}%
  \BibitemOpen
  \bibfield  {author} {\bibinfo {author} {\bibfnamefont {J.}~\bibnamefont
  {Gao}}, \bibinfo {author} {\bibfnamefont {W.}~\bibnamefont {Liu}}, \bibinfo
  {author} {\bibfnamefont {B.}~\bibnamefont {Song}}, \ and\ \bibinfo {author}
  {\bibfnamefont {C.}~\bibnamefont {Liu}},\ }\href@noop {} {\bibfield
  {journal} {\bibinfo  {journal} {J. Chem. Phys.}\ }\textbf {\bibinfo {volume}
  {121}},\ \bibinfo {pages} {6658} (\bibinfo {year} {2004})}\BibitemShut
  {NoStop}%
\bibitem [{\citenamefont {Gao}\ \emph {et~al.}(2005)\citenamefont {Gao},
  \citenamefont {Zou}, \citenamefont {Liu}, \citenamefont {Xiao}, \citenamefont
  {Peng}, \citenamefont {Song},\ and\ \citenamefont {Liu}}]{gao2005}%
  \BibitemOpen
  \bibfield  {author} {\bibinfo {author} {\bibfnamefont {J.}~\bibnamefont
  {Gao}}, \bibinfo {author} {\bibfnamefont {W.}~\bibnamefont {Zou}}, \bibinfo
  {author} {\bibfnamefont {W.}~\bibnamefont {Liu}}, \bibinfo {author}
  {\bibfnamefont {Y.}~\bibnamefont {Xiao}}, \bibinfo {author} {\bibfnamefont
  {D.}~\bibnamefont {Peng}}, \bibinfo {author} {\bibfnamefont {B.}~\bibnamefont
  {Song}}, \ and\ \bibinfo {author} {\bibfnamefont {C.}~\bibnamefont {Liu}},\
  }\href@noop {} {\bibfield  {journal} {\bibinfo  {journal} {J. Chem. Phys.}\
  }\textbf {\bibinfo {volume} {123}},\ \bibinfo {pages} {054102} (\bibinfo
  {year} {2005})}\BibitemShut {NoStop}%
\bibitem [{\citenamefont {Bast}, \citenamefont {Jensen},\ and\ \citenamefont
  {Saue}(2009)}]{bast2009}%
  \BibitemOpen
  \bibfield  {author} {\bibinfo {author} {\bibfnamefont {R.}~\bibnamefont
  {Bast}}, \bibinfo {author} {\bibfnamefont {H.~J.~{\relax Aa}.}\ \bibnamefont
  {Jensen}}, \ and\ \bibinfo {author} {\bibfnamefont {T.}~\bibnamefont
  {Saue}},\ }\href@noop {} {\bibfield  {journal} {\bibinfo  {journal} {Int. J.
  Quantum Chem.}\ }\textbf {\bibinfo {volume} {109}},\ \bibinfo {pages} {2091}
  (\bibinfo {year} {2009})}\BibitemShut {NoStop}%
\bibitem [{\citenamefont {Wang}\ \emph {et~al.}(2005)\citenamefont {Wang},
  \citenamefont {Ziegler}, \citenamefont {{Van Lenthe}}, \citenamefont {{Van
  Gisbergen}},\ and\ \citenamefont {Baerends}}]{wang2005}%
  \BibitemOpen
  \bibfield  {author} {\bibinfo {author} {\bibfnamefont {F.}~\bibnamefont
  {Wang}}, \bibinfo {author} {\bibfnamefont {T.}~\bibnamefont {Ziegler}},
  \bibinfo {author} {\bibfnamefont {E.}~\bibnamefont {{Van Lenthe}}}, \bibinfo
  {author} {\bibfnamefont {S.}~\bibnamefont {{Van Gisbergen}}}, \ and\ \bibinfo
  {author} {\bibfnamefont {E.~J.}\ \bibnamefont {Baerends}},\ }\href {\doibase
  10.1063/1.1899143} {\bibfield  {journal} {\bibinfo  {journal} {J. Chem.
  Phys.}\ }\textbf {\bibinfo {volume} {122}},\ \bibinfo {pages} {204103}
  (\bibinfo {year} {2005})}\BibitemShut {NoStop}%
\bibitem [{\citenamefont {Peng}, \citenamefont {Zou},\ and\ \citenamefont
  {Liu}(2005)}]{peng2005}%
  \BibitemOpen
  \bibfield  {author} {\bibinfo {author} {\bibfnamefont {D.}~\bibnamefont
  {Peng}}, \bibinfo {author} {\bibfnamefont {W.}~\bibnamefont {Zou}}, \ and\
  \bibinfo {author} {\bibfnamefont {W.}~\bibnamefont {Liu}},\ }\href {\doibase
  10.1063/1.2047554} {\bibfield  {journal} {\bibinfo  {journal} {J. Chem.
  Phys.}\ }\textbf {\bibinfo {volume} {123}},\ \bibinfo {pages} {144101}
  (\bibinfo {year} {2005})}\BibitemShut {NoStop}%
\bibitem [{\citenamefont {K{\"u}hn}\ and\ \citenamefont
  {Weigend}(2013)}]{kuhn2013}%
  \BibitemOpen
  \bibfield  {author} {\bibinfo {author} {\bibfnamefont {M.}~\bibnamefont
  {K{\"u}hn}}\ and\ \bibinfo {author} {\bibfnamefont {F.}~\bibnamefont
  {Weigend}},\ }\href@noop {} {\bibfield  {journal} {\bibinfo  {journal} {J.
  Chem. Theory Comput.}\ }\textbf {\bibinfo {volume} {19}},\ \bibinfo {pages}
  {5341} (\bibinfo {year} {2013})}\BibitemShut {NoStop}%
\bibitem [{\citenamefont {Egidi}\ \emph {et~al.}(2016)\citenamefont {Egidi},
  \citenamefont {Goings}, \citenamefont {Frisch},\ and\ \citenamefont
  {Li}}]{egidi2016}%
  \BibitemOpen
  \bibfield  {author} {\bibinfo {author} {\bibfnamefont {F.}~\bibnamefont
  {Egidi}}, \bibinfo {author} {\bibfnamefont {J.~J.}\ \bibnamefont {Goings}},
  \bibinfo {author} {\bibfnamefont {M.~J.}\ \bibnamefont {Frisch}}, \ and\
  \bibinfo {author} {\bibfnamefont {X.}~\bibnamefont {Li}},\ }\href {\doibase
  10.1021/acs.jctc.6b00474} {\bibfield  {journal} {\bibinfo  {journal} {J.
  Chem. Theory Comput.}\ }\textbf {\bibinfo {volume} {12}},\ \bibinfo {pages}
  {3711} (\bibinfo {year} {2016})}\BibitemShut {NoStop}%
\bibitem [{\citenamefont {M{\"{u}}ller}\ \emph {et~al.}(2003)\citenamefont
  {M{\"{u}}ller}, \citenamefont {Schr{\"{o}}der}, \citenamefont {Parkinson},
  \citenamefont {Kratochwil}, \citenamefont {Coxall}, \citenamefont {Parkin},
  \citenamefont {Parsons},\ and\ \citenamefont {Sadler}}]{muller2003}%
  \BibitemOpen
  \bibfield  {author} {\bibinfo {author} {\bibfnamefont {P.}~\bibnamefont
  {M{\"{u}}ller}}, \bibinfo {author} {\bibfnamefont {B.}~\bibnamefont
  {Schr{\"{o}}der}}, \bibinfo {author} {\bibfnamefont {J.~A.}\ \bibnamefont
  {Parkinson}}, \bibinfo {author} {\bibfnamefont {N.~A.}\ \bibnamefont
  {Kratochwil}}, \bibinfo {author} {\bibfnamefont {R.~A.}\ \bibnamefont
  {Coxall}}, \bibinfo {author} {\bibfnamefont {A.}~\bibnamefont {Parkin}},
  \bibinfo {author} {\bibfnamefont {S.}~\bibnamefont {Parsons}}, \ and\
  \bibinfo {author} {\bibfnamefont {P.~J.}\ \bibnamefont {Sadler}},\
  }\href@noop {} {\bibfield  {journal} {\bibinfo  {journal} {Angew. Chem. Int.
  Ed.}\ }\textbf {\bibinfo {volume} {42}},\ \bibinfo {pages} {335} (\bibinfo
  {year} {2003})}\BibitemShut {NoStop}%
\bibitem [{\citenamefont {Mackay}\ \emph {et~al.}(2006)\citenamefont {Mackay},
  \citenamefont {Woods}, \citenamefont {Moseley}, \citenamefont {Ferguson},
  \citenamefont {Dawson}, \citenamefont {Parsons},\ and\ \citenamefont
  {Sadler}}]{mackay2006}%
  \BibitemOpen
  \bibfield  {author} {\bibinfo {author} {\bibfnamefont {F.~S.}\ \bibnamefont
  {Mackay}}, \bibinfo {author} {\bibfnamefont {J.~A.}\ \bibnamefont {Woods}},
  \bibinfo {author} {\bibfnamefont {H.}~\bibnamefont {Moseley}}, \bibinfo
  {author} {\bibfnamefont {J.}~\bibnamefont {Ferguson}}, \bibinfo {author}
  {\bibfnamefont {A.}~\bibnamefont {Dawson}}, \bibinfo {author} {\bibfnamefont
  {S.}~\bibnamefont {Parsons}}, \ and\ \bibinfo {author} {\bibfnamefont
  {P.~J.}\ \bibnamefont {Sadler}},\ }\href {\doibase 10.1002/chem.200501601}
  {\bibfield  {journal} {\bibinfo  {journal} {Chem. Eur. J.}\ }\textbf
  {\bibinfo {volume} {12}},\ \bibinfo {pages} {3155} (\bibinfo {year}
  {2006})}\BibitemShut {NoStop}%
\bibitem [{\citenamefont {Bednarski}\ \emph {et~al.}(2006)\citenamefont
  {Bednarski}, \citenamefont {Gr{\"{u}}nert}, \citenamefont {Zielzki},
  \citenamefont {Wellner}, \citenamefont {Mackay},\ and\ \citenamefont
  {Sadler}}]{bednarski2006}%
  \BibitemOpen
  \bibfield  {author} {\bibinfo {author} {\bibfnamefont {P.~J.}\ \bibnamefont
  {Bednarski}}, \bibinfo {author} {\bibfnamefont {R.}~\bibnamefont
  {Gr{\"{u}}nert}}, \bibinfo {author} {\bibfnamefont {M.}~\bibnamefont
  {Zielzki}}, \bibinfo {author} {\bibfnamefont {A.}~\bibnamefont {Wellner}},
  \bibinfo {author} {\bibfnamefont {F.~S.}\ \bibnamefont {Mackay}}, \ and\
  \bibinfo {author} {\bibfnamefont {P.~J.}\ \bibnamefont {Sadler}},\ }\href
  {\doibase 10.1016/j.chembiol.2005.10.011} {\bibfield  {journal} {\bibinfo
  {journal} {Chem. Biol.}\ }\textbf {\bibinfo {volume} {13}},\ \bibinfo {pages}
  {61} (\bibinfo {year} {2006})}\BibitemShut {NoStop}%
\bibitem [{\citenamefont {Imran}\ \emph {et~al.}(2018)\citenamefont {Imran},
  \citenamefont {Ayub}, \citenamefont {Butler},\ and\ \citenamefont
  {Rehman}}]{imran2018}%
  \BibitemOpen
  \bibfield  {author} {\bibinfo {author} {\bibfnamefont {M.}~\bibnamefont
  {Imran}}, \bibinfo {author} {\bibfnamefont {W.}~\bibnamefont {Ayub}},
  \bibinfo {author} {\bibfnamefont {I.~S.}\ \bibnamefont {Butler}}, \ and\
  \bibinfo {author} {\bibfnamefont {Z.-u.}\ \bibnamefont {Rehman}},\ }\href
  {\doibase 10.1016/j.ccr.2018.08.009} {\bibfield  {journal} {\bibinfo
  {journal} {Coord. Chem. Rev.}\ }\textbf {\bibinfo {volume} {376}},\ \bibinfo
  {pages} {405} (\bibinfo {year} {2018})}\BibitemShut {NoStop}%
\bibitem [{\citenamefont {Ka{\v{s}}p{\'{a}}rkov{\'{a}}}\ \emph
  {et~al.}(2003)\citenamefont {Ka{\v{s}}p{\'{a}}rkov{\'{a}}}, \citenamefont
  {Mackay}, \citenamefont {Brabec},\ and\ \citenamefont
  {Sadler}}]{kasparkova2003}%
  \BibitemOpen
  \bibfield  {author} {\bibinfo {author} {\bibfnamefont {J.}~\bibnamefont
  {Ka{\v{s}}p{\'{a}}rkov{\'{a}}}}, \bibinfo {author} {\bibfnamefont {F.~S.}\
  \bibnamefont {Mackay}}, \bibinfo {author} {\bibfnamefont {V.}~\bibnamefont
  {Brabec}}, \ and\ \bibinfo {author} {\bibfnamefont {P.~J.}\ \bibnamefont
  {Sadler}},\ }\href {\doibase 10.1007/s00775-003-0474-3} {\bibfield  {journal}
  {\bibinfo  {journal} {J Biol. Inorg. Chem.}\ }\textbf {\bibinfo {volume}
  {8}},\ \bibinfo {pages} {741} (\bibinfo {year} {2003})}\BibitemShut {NoStop}%
\bibitem [{\citenamefont {Mackay}\ \emph {et~al.}(2007)\citenamefont {Mackay},
  \citenamefont {Woods}, \citenamefont {Heringov{\'{a}}}, \citenamefont
  {Ka{\v{s}}p{\'{a}}rkov{\'{a}}}, \citenamefont {Pizarro}, \citenamefont
  {Moggach}, \citenamefont {Parsons}, \citenamefont {Brabec},\ and\
  \citenamefont {Sadler}}]{mackay2007}%
  \BibitemOpen
  \bibfield  {author} {\bibinfo {author} {\bibfnamefont {F.~S.}\ \bibnamefont
  {Mackay}}, \bibinfo {author} {\bibfnamefont {J.~A.}\ \bibnamefont {Woods}},
  \bibinfo {author} {\bibfnamefont {P.}~\bibnamefont {Heringov{\'{a}}}},
  \bibinfo {author} {\bibfnamefont {J.}~\bibnamefont
  {Ka{\v{s}}p{\'{a}}rkov{\'{a}}}}, \bibinfo {author} {\bibfnamefont {A.~M.}\
  \bibnamefont {Pizarro}}, \bibinfo {author} {\bibfnamefont {S.~A.}\
  \bibnamefont {Moggach}}, \bibinfo {author} {\bibfnamefont {S.}~\bibnamefont
  {Parsons}}, \bibinfo {author} {\bibfnamefont {V.}~\bibnamefont {Brabec}}, \
  and\ \bibinfo {author} {\bibfnamefont {P.~J.}\ \bibnamefont {Sadler}},\
  }\href@noop {} {\bibfield  {journal} {\bibinfo  {journal} {Proc. Nat. Acad.
  Sci. U.S.A.}\ }\textbf {\bibinfo {volume} {104}},\ \bibinfo {pages} {20743}
  (\bibinfo {year} {2007})}\BibitemShut {NoStop}%
\bibitem [{\citenamefont {Pracharova}\ \emph {et~al.}(2012)\citenamefont
  {Pracharova}, \citenamefont {Zerzankova}, \citenamefont {Stepankova},
  \citenamefont {Novakova}, \citenamefont {Farrer}, \citenamefont {Sadler},
  \citenamefont {Brabec},\ and\ \citenamefont
  {Ka{\v{s}}p{\'{a}}rkov{\'{a}}}}]{pracharova2012}%
  \BibitemOpen
  \bibfield  {author} {\bibinfo {author} {\bibfnamefont {J.}~\bibnamefont
  {Pracharova}}, \bibinfo {author} {\bibfnamefont {L.}~\bibnamefont
  {Zerzankova}}, \bibinfo {author} {\bibfnamefont {J.}~\bibnamefont
  {Stepankova}}, \bibinfo {author} {\bibfnamefont {O.}~\bibnamefont
  {Novakova}}, \bibinfo {author} {\bibfnamefont {N.~J.}\ \bibnamefont
  {Farrer}}, \bibinfo {author} {\bibfnamefont {P.~J.}\ \bibnamefont {Sadler}},
  \bibinfo {author} {\bibfnamefont {V.}~\bibnamefont {Brabec}}, \ and\ \bibinfo
  {author} {\bibfnamefont {J.}~\bibnamefont {Ka{\v{s}}p{\'{a}}rkov{\'{a}}}},\
  }\href {\doibase 10.1021/tx300057y} {\bibfield  {journal} {\bibinfo
  {journal} {Chem. Res. Toxicol.}\ }\textbf {\bibinfo {volume} {25}},\ \bibinfo
  {pages} {1099} (\bibinfo {year} {2012})}\BibitemShut {NoStop}%
\bibitem [{\citenamefont {Min}\ \emph {et~al.}(2014)\citenamefont {Min},
  \citenamefont {Li}, \citenamefont {Liu}, \citenamefont {Yeow},\ and\
  \citenamefont {Xing}}]{min2014}%
  \BibitemOpen
  \bibfield  {author} {\bibinfo {author} {\bibfnamefont {Y.}~\bibnamefont
  {Min}}, \bibinfo {author} {\bibfnamefont {J.}~\bibnamefont {Li}}, \bibinfo
  {author} {\bibfnamefont {F.}~\bibnamefont {Liu}}, \bibinfo {author}
  {\bibfnamefont {E.~K.~L.}\ \bibnamefont {Yeow}}, \ and\ \bibinfo {author}
  {\bibfnamefont {B.}~\bibnamefont {Xing}},\ }\href {\doibase
  10.1002/anie.201308834} {\bibfield  {journal} {\bibinfo  {journal} {Angew.
  Chem. Int. Ed.}\ }\textbf {\bibinfo {volume} {53}},\ \bibinfo {pages} {1012}
  (\bibinfo {year} {2014})}\BibitemShut {NoStop}%
\bibitem [{\citenamefont {Gandioso}\ \emph {et~al.}(2015)\citenamefont
  {Gandioso}, \citenamefont {Shaili}, \citenamefont {Massaguer}, \citenamefont
  {Artigas}, \citenamefont {Gonz{\'{a}}lez-Canto}, \citenamefont {Woods},
  \citenamefont {Sadler},\ and\ \citenamefont {{Vicente
  March{\'{a}}n}}}]{gandioso2015}%
  \BibitemOpen
  \bibfield  {author} {\bibinfo {author} {\bibfnamefont {A.}~\bibnamefont
  {Gandioso}}, \bibinfo {author} {\bibfnamefont {E.}~\bibnamefont {Shaili}},
  \bibinfo {author} {\bibfnamefont {A.}~\bibnamefont {Massaguer}}, \bibinfo
  {author} {\bibfnamefont {G.}~\bibnamefont {Artigas}}, \bibinfo {author}
  {\bibfnamefont {A.}~\bibnamefont {Gonz{\'{a}}lez-Canto}}, \bibinfo {author}
  {\bibfnamefont {J.~A.}\ \bibnamefont {Woods}}, \bibinfo {author}
  {\bibfnamefont {P.~J.}\ \bibnamefont {Sadler}}, \ and\ \bibinfo {author}
  {\bibnamefont {{Vicente March{\'{a}}n}}},\ }\href {\doibase
  10.1039/c5cc03180j} {\bibfield  {journal} {\bibinfo  {journal} {Chem.
  Commun.}\ }\textbf {\bibinfo {volume} {51}},\ \bibinfo {pages} {9169}
  (\bibinfo {year} {2015})}\BibitemShut {NoStop}%
\bibitem [{\citenamefont {Imberti}\ \emph {et~al.}(2020)\citenamefont
  {Imberti}, \citenamefont {Zhang}, \citenamefont {Huang},\ and\ \citenamefont
  {Sadler}}]{imberti2020}%
  \BibitemOpen
  \bibfield  {author} {\bibinfo {author} {\bibfnamefont {C.}~\bibnamefont
  {Imberti}}, \bibinfo {author} {\bibfnamefont {P.}~\bibnamefont {Zhang}},
  \bibinfo {author} {\bibfnamefont {H.}~\bibnamefont {Huang}}, \ and\ \bibinfo
  {author} {\bibfnamefont {P.~J.}\ \bibnamefont {Sadler}},\ }\href {\doibase
  10.1002/anie.201905171} {\bibfield  {journal} {\bibinfo  {journal} {Angew.
  Chem. Int. Ed.}\ }\textbf {\bibinfo {volume} {59}},\ \bibinfo {pages} {61}
  (\bibinfo {year} {2020})}\BibitemShut {NoStop}%
\bibitem [{\citenamefont {Ronconi}\ and\ \citenamefont
  {Sadler}(2008)}]{ronconi2008}%
  \BibitemOpen
  \bibfield  {author} {\bibinfo {author} {\bibfnamefont {L.}~\bibnamefont
  {Ronconi}}\ and\ \bibinfo {author} {\bibfnamefont {P.~J.}\ \bibnamefont
  {Sadler}},\ }\href {\doibase 10.1039/B714216A} {\bibfield  {journal}
  {\bibinfo  {journal} {Chem. Commun.}\ }\textbf {\bibinfo {volume} {2}},\
  \bibinfo {pages} {235} (\bibinfo {year} {2008})}\BibitemShut {NoStop}%
\bibitem [{\citenamefont {Ronconi}\ and\ \citenamefont
  {Sadler}(2011)}]{ronconi2011}%
  \BibitemOpen
  \bibfield  {author} {\bibinfo {author} {\bibfnamefont {L.}~\bibnamefont
  {Ronconi}}\ and\ \bibinfo {author} {\bibfnamefont {P.~J.}\ \bibnamefont
  {Sadler}},\ }\href {\doibase 10.1039/C0DT00546K} {\bibfield  {journal}
  {\bibinfo  {journal} {Dalton Trans.}\ }\textbf {\bibinfo {volume} {40}},\
  \bibinfo {pages} {262} (\bibinfo {year} {2011})}\BibitemShut {NoStop}%
\bibitem [{\citenamefont {Ahlrichs}\ \emph {et~al.}(1989)\citenamefont
  {Ahlrichs}, \citenamefont {B{\"a}r}, \citenamefont {H{\"a}ser}, \citenamefont
  {Horn},\ and\ \citenamefont {K{\"o}lmel}}]{Ahlrichs1989}%
  \BibitemOpen
  \bibfield  {author} {\bibinfo {author} {\bibfnamefont {R.}~\bibnamefont
  {Ahlrichs}}, \bibinfo {author} {\bibfnamefont {M.}~\bibnamefont {B{\"a}r}},
  \bibinfo {author} {\bibfnamefont {M.}~\bibnamefont {H{\"a}ser}}, \bibinfo
  {author} {\bibfnamefont {H.}~\bibnamefont {Horn}}, \ and\ \bibinfo {author}
  {\bibfnamefont {C.}~\bibnamefont {K{\"o}lmel}},\ }\href@noop {} {\bibfield
  {journal} {\bibinfo  {journal} {Chem. Phys. Lett.}\ }\textbf {\bibinfo
  {volume} {162}},\ \bibinfo {pages} {165} (\bibinfo {year}
  {1989})}\BibitemShut {NoStop}%
\bibitem [{\citenamefont {Tao}\ \emph {et~al.}(2003)\citenamefont {Tao},
  \citenamefont {Perdew}, \citenamefont {Staroverov},\ and\ \citenamefont
  {Scuseria}}]{Tao2003}%
  \BibitemOpen
  \bibfield  {author} {\bibinfo {author} {\bibfnamefont {J.}~\bibnamefont
  {Tao}}, \bibinfo {author} {\bibfnamefont {J.~P.}\ \bibnamefont {Perdew}},
  \bibinfo {author} {\bibnamefont {Staroverov}}, \ and\ \bibinfo {author}
  {\bibfnamefont {G.~E.}\ \bibnamefont {Scuseria}},\ }\href@noop {} {\bibfield
  {journal} {\bibinfo  {journal} {Phys. Rev. Lett.}\ }\textbf {\bibinfo
  {volume} {91}},\ \bibinfo {pages} {146401} (\bibinfo {year}
  {2003})}\BibitemShut {NoStop}%
\bibitem [{\citenamefont {Sch\"{a}fer}, \citenamefont {Horn},\ and\
  \citenamefont {Ahlrichs}(1992)}]{Schafer1992}%
  \BibitemOpen
  \bibfield  {author} {\bibinfo {author} {\bibfnamefont {A.}~\bibnamefont
  {Sch\"{a}fer}}, \bibinfo {author} {\bibfnamefont {H.}~\bibnamefont {Horn}}, \
  and\ \bibinfo {author} {\bibfnamefont {R.}~\bibnamefont {Ahlrichs}},\ }\href
  {\doibase 10.1063/1.463096} {\bibfield  {journal} {\bibinfo  {journal} {J.
  Chem. Phys.}\ }\textbf {\bibinfo {volume} {97}},\ \bibinfo {pages} {2571}
  (\bibinfo {year} {1992})}\BibitemShut {NoStop}%
\bibitem [{\citenamefont {Eichkorn}\ \emph {et~al.}(1995)\citenamefont
  {Eichkorn}, \citenamefont {Oliver}, \citenamefont {{\"O}hm}, \citenamefont
  {H{\"a}ser},\ and\ \citenamefont {Ahlrichs}}]{eichkorn1995}%
  \BibitemOpen
  \bibfield  {author} {\bibinfo {author} {\bibfnamefont {K.}~\bibnamefont
  {Eichkorn}}, \bibinfo {author} {\bibfnamefont {T.}~\bibnamefont {Oliver}},
  \bibinfo {author} {\bibfnamefont {H.}~\bibnamefont {{\"O}hm}}, \bibinfo
  {author} {\bibfnamefont {M.}~\bibnamefont {H{\"a}ser}}, \ and\ \bibinfo
  {author} {\bibfnamefont {R.}~\bibnamefont {Ahlrichs}},\ }\href@noop {}
  {\bibfield  {journal} {\bibinfo  {journal} {Chem. Phys. Lett.}\ }\textbf
  {\bibinfo {volume} {240}},\ \bibinfo {pages} {283} (\bibinfo {year}
  {1995})}\BibitemShut {NoStop}%
\bibitem [{\citenamefont {Eichkorn}\ \emph {et~al.}(1997)\citenamefont
  {Eichkorn}, \citenamefont {Weigend}, \citenamefont {Treutler},\ and\
  \citenamefont {Ahlrichs}}]{eichkorn1997}%
  \BibitemOpen
  \bibfield  {author} {\bibinfo {author} {\bibfnamefont {K.}~\bibnamefont
  {Eichkorn}}, \bibinfo {author} {\bibfnamefont {F.}~\bibnamefont {Weigend}},
  \bibinfo {author} {\bibfnamefont {O.}~\bibnamefont {Treutler}}, \ and\
  \bibinfo {author} {\bibfnamefont {R.}~\bibnamefont {Ahlrichs}},\ }\href@noop
  {} {\bibfield  {journal} {\bibinfo  {journal} {Theor. Chem. Acc.}\ }\textbf
  {\bibinfo {volume} {97}},\ \bibinfo {pages} {119} (\bibinfo {year}
  {1997})}\BibitemShut {NoStop}%
\bibitem [{DIR()}]{DIRAC18}%
  \BibitemOpen
  \href@noop {} {}\bibinfo {note} {{DIRAC}, a relativistic ab initio electronic
  structure program, Release {DIRAC18} (2018), written by T.~Saue, L.~Visscher,
  H.~J.~{\relax Aa}.~Jensen, and R.~Bast, with contributions from V.~Bakken,
  K.~G.~Dyall, S.~Dubillard, U.~Ekstr{\"o}m, E.~Eliav, T.~Enevoldsen,
  E.~Fa{\ss}hauer, T.~Fleig, O.~Fossgaard, A.~S.~P.~Gomes, E.~D.~Hedeg{\aa}rd,
  T.~Helgaker, J.~Henriksson, M.~Ilia{\v{s}}, C.~R.~Jacob, S.~Knecht,
  S.~Komorovsk{\'y}, O.~Kullie, J.~K.~L{\ae}rdahl, C.~V.~Larsen, Y.~S.~Lee,
  H.~S.~Nataraj, M.~K.~Nayak, P.~Norman, G.~Olejniczak, J.~Olsen,
  J.~M.~H.~Olsen, Y.~C.~Park, J.~K.~Pedersen, M.~Pernpointner, R.~Di~Remigio,
  K.~Ruud, P.~Sa{\l}ek, B.~Schimmelpfennig, A.~Shee, J.~Sikkema,
  A.~J.~Thorvaldsen, J.~Thyssen, J.~van~Stralen, S.~Villaume, O.~Visser,
  T.~Winther, and S.~Yamamoto (available at
  \url{https://doi.org/10.5281/zenodo.2253986}, see also
  \url{http://www.diracprogram.org})}\BibitemShut {NoStop}%
\bibitem [{\citenamefont {Andrae}\ \emph {et~al.}(1990)\citenamefont {Andrae},
  \citenamefont {H{\"a}u{\ss}ermann}, \citenamefont {Dolg}, \citenamefont
  {Stoll},\ and\ \citenamefont {Preu{\ss}}}]{Andrae1990a}%
  \BibitemOpen
  \bibfield  {author} {\bibinfo {author} {\bibfnamefont {D.}~\bibnamefont
  {Andrae}}, \bibinfo {author} {\bibfnamefont {U.}~\bibnamefont
  {H{\"a}u{\ss}ermann}}, \bibinfo {author} {\bibfnamefont {M.}~\bibnamefont
  {Dolg}}, \bibinfo {author} {\bibfnamefont {H.}~\bibnamefont {Stoll}}, \ and\
  \bibinfo {author} {\bibfnamefont {H.}~\bibnamefont {Preu{\ss}}},\ }\href@noop
  {} {\bibfield  {journal} {\bibinfo  {journal} {Theor. Chim. Acta}\ }\textbf
  {\bibinfo {volume} {77}},\ \bibinfo {pages} {123} (\bibinfo {year}
  {1990})}\BibitemShut {NoStop}%
\bibitem [{\citenamefont {Weigend}\ and\ \citenamefont
  {Ahlrichs}(2005)}]{Weigend2005a}%
  \BibitemOpen
  \bibfield  {author} {\bibinfo {author} {\bibfnamefont {F.}~\bibnamefont
  {Weigend}}\ and\ \bibinfo {author} {\bibfnamefont {R.}~\bibnamefont
  {Ahlrichs}},\ }\href@noop {} {\bibfield  {journal} {\bibinfo  {journal}
  {Phys. Chem. Chem. Phys.}\ }\textbf {\bibinfo {volume} {7}},\ \bibinfo
  {pages} {3297} (\bibinfo {year} {2005})}\BibitemShut {NoStop}%
\bibitem [{\citenamefont {Dyall}\ and\ \citenamefont
  {Gomes}(2009)}]{Dyall2009}%
  \BibitemOpen
  \bibfield  {author} {\bibinfo {author} {\bibfnamefont {K.~G.}\ \bibnamefont
  {Dyall}}\ and\ \bibinfo {author} {\bibfnamefont {A.~S.~P.}\ \bibnamefont
  {Gomes}},\ }\href@noop {} {\bibfield  {journal} {\bibinfo  {journal} {Theor.
  Chem. Acc.}\ }\textbf {\bibinfo {volume} {125}},\ \bibinfo {pages} {97}
  (\bibinfo {year} {2009})}\BibitemShut {NoStop}%
\bibitem [{\citenamefont {Visscher}(1997)}]{Visscher1997}%
  \BibitemOpen
  \bibfield  {author} {\bibinfo {author} {\bibfnamefont {L.}~\bibnamefont
  {Visscher}},\ }\href@noop {} {\bibfield  {journal} {\bibinfo  {journal}
  {Theor. Chim. Acta}\ }\textbf {\bibinfo {volume} {98}},\ \bibinfo {pages}
  {68} (\bibinfo {year} {1997})}\BibitemShut {NoStop}%
\bibitem [{\citenamefont {Yanai}, \citenamefont {Tew},\ and\ \citenamefont
  {Handy}(2004)}]{Takeshi2004}%
  \BibitemOpen
  \bibfield  {author} {\bibinfo {author} {\bibfnamefont {T.}~\bibnamefont
  {Yanai}}, \bibinfo {author} {\bibfnamefont {D.~P.}\ \bibnamefont {Tew}}, \
  and\ \bibinfo {author} {\bibfnamefont {N.~C.}\ \bibnamefont {Handy}},\ }\href
  {\doibase https://doi.org/10.1016/j.cplett.2004.06.011} {\bibfield  {journal}
  {\bibinfo  {journal} {Chem. Phys. Lett.}\ }\textbf {\bibinfo {volume}
  {393}},\ \bibinfo {pages} {51} (\bibinfo {year} {2004})}\BibitemShut
  {NoStop}%
\bibitem [{\citenamefont {Becke}(1988)}]{Becke1988}%
  \BibitemOpen
  \bibfield  {author} {\bibinfo {author} {\bibfnamefont {A.~D.}\ \bibnamefont
  {Becke}},\ }\href {\doibase 10.1103/PhysRevA.38.3098} {\bibfield  {journal}
  {\bibinfo  {journal} {Phys. Rev. A}\ }\textbf {\bibinfo {volume} {38}},\
  \bibinfo {pages} {3098} (\bibinfo {year} {1988})}\BibitemShut {NoStop}%
\bibitem [{\citenamefont {Lee}, \citenamefont {Yang},\ and\ \citenamefont
  {Parr}(1988)}]{Lee1988}%
  \BibitemOpen
  \bibfield  {author} {\bibinfo {author} {\bibfnamefont {C.}~\bibnamefont
  {Lee}}, \bibinfo {author} {\bibfnamefont {W.}~\bibnamefont {Yang}}, \ and\
  \bibinfo {author} {\bibfnamefont {R.~G.}\ \bibnamefont {Parr}},\ }\href
  {\doibase 10.1103/PhysRevB.37.785} {\bibfield  {journal} {\bibinfo  {journal}
  {Phys. Rev. B}\ }\textbf {\bibinfo {volume} {37}},\ \bibinfo {pages} {785}
  (\bibinfo {year} {1988})}\BibitemShut {NoStop}%
\bibitem [{\citenamefont {Becke}(1993)}]{Becke1993}%
  \BibitemOpen
  \bibfield  {author} {\bibinfo {author} {\bibfnamefont {A.~D.}\ \bibnamefont
  {Becke}},\ }\href {\doibase 10.1063/1.464913} {\bibfield  {journal} {\bibinfo
   {journal} {J. Chem. Phys.}\ }\textbf {\bibinfo {volume} {98}},\ \bibinfo
  {pages} {5648} (\bibinfo {year} {1993})}\BibitemShut {NoStop}%
\bibitem [{\citenamefont {L{\'e}vy-Leblond}(1967)}]{Levy-leblond1967}%
  \BibitemOpen
  \bibfield  {author} {\bibinfo {author} {\bibfnamefont {J.-M.}\ \bibnamefont
  {L{\'e}vy-Leblond}},\ }\href@noop {} {\bibfield  {journal} {\bibinfo
  {journal} {Commun. Math. Phys.}\ }\textbf {\bibinfo {volume} {6}},\ \bibinfo
  {pages} {286} (\bibinfo {year} {1967})}\BibitemShut {NoStop}%
\bibitem [{\citenamefont {Dyall}(1994)}]{Dyall1994}%
  \BibitemOpen
  \bibfield  {author} {\bibinfo {author} {\bibfnamefont {K.~G.}\ \bibnamefont
  {Dyall}},\ }\href {\doibase 10.1063/1.466508} {\bibfield  {journal} {\bibinfo
   {journal} {J. Chem. Phys.}\ }\textbf {\bibinfo {volume} {100}},\ \bibinfo
  {pages} {2118} (\bibinfo {year} {1994})}\BibitemShut {NoStop}%
\bibitem [{\citenamefont {Visscher}\ and\ \citenamefont
  {Saue}(2000)}]{Visscher2000}%
  \BibitemOpen
  \bibfield  {author} {\bibinfo {author} {\bibfnamefont {L.}~\bibnamefont
  {Visscher}}\ and\ \bibinfo {author} {\bibfnamefont {T.}~\bibnamefont
  {Saue}},\ }\href {\doibase 10.1063/1.1288371} {\bibfield  {journal} {\bibinfo
   {journal} {J. Chem. Phys.}\ }\textbf {\bibinfo {volume} {113}},\ \bibinfo
  {pages} {3996} (\bibinfo {year} {2000})}\BibitemShut {NoStop}%
\bibitem [{\citenamefont {Garino}\ and\ \citenamefont
  {Salassa}(2013)}]{garino2013}%
  \BibitemOpen
  \bibfield  {author} {\bibinfo {author} {\bibfnamefont {C.}~\bibnamefont
  {Garino}}\ and\ \bibinfo {author} {\bibfnamefont {L.}~\bibnamefont
  {Salassa}},\ }\href@noop {} {\bibfield  {journal} {\bibinfo  {journal} {Phil.
  Trans. R. Soc. A}\ }\textbf {\bibinfo {volume} {371}},\ \bibinfo {pages}
  {20120134} (\bibinfo {year} {2013})}\BibitemShut {NoStop}%
\bibitem [{\citenamefont {Dreuw}\ and\ \citenamefont
  {Head-Gordon}(2004)}]{dreuw2004}%
  \BibitemOpen
  \bibfield  {author} {\bibinfo {author} {\bibfnamefont {A.}~\bibnamefont
  {Dreuw}}\ and\ \bibinfo {author} {\bibfnamefont {M.}~\bibnamefont
  {Head-Gordon}},\ }\href {\doibase 10.1021/ja039556n} {\bibfield  {journal}
  {\bibinfo  {journal} {J. Am. Chem. Soc.}\ }\textbf {\bibinfo {volume}
  {126}},\ \bibinfo {pages} {4007} (\bibinfo {year} {2004})}\BibitemShut
  {NoStop}%
\bibitem [{\citenamefont {Hedeg{\aa}rd}\ \emph {et~al.}(2017)\citenamefont
  {Hedeg{\aa}rd}, \citenamefont {Bast}, \citenamefont {Kongsted}, \citenamefont
  {Olsen},\ and\ \citenamefont {Jensen}}]{hedegaard2017}%
  \BibitemOpen
  \bibfield  {author} {\bibinfo {author} {\bibfnamefont {E.~D.}\ \bibnamefont
  {Hedeg{\aa}rd}}, \bibinfo {author} {\bibfnamefont {R.}~\bibnamefont {Bast}},
  \bibinfo {author} {\bibfnamefont {J.}~\bibnamefont {Kongsted}}, \bibinfo
  {author} {\bibfnamefont {J.~M.~H.}\ \bibnamefont {Olsen}}, \ and\ \bibinfo
  {author} {\bibfnamefont {H.~J.~{\relax Aa}.}\ \bibnamefont {Jensen}},\ }\href
  {\doibase 10.1021/acs.jctc.7b00162} {\bibfield  {journal} {\bibinfo
  {journal} {J. Chem. Theory Comput.}\ }\textbf {\bibinfo {volume} {13}},\
  \bibinfo {pages} {2870} (\bibinfo {year} {2017})}\BibitemShut {NoStop}%
\end{thebibliography}%

\end{document}



\title[]{Supporting information: The importance of relativistic effects for platinum complexes with light-activated activity against cancer cells.}

\author{Joel Creutzberg}
\affiliation{Division of Theoretical Chemistry, Lund University, Lund, Sweden}
\author{Erik Donovan Hedeg{\aa}rd}
\email{erik.hedegard@teokem.lu.se}
\affiliation{Division of Theoretical Chemistry, Lund University, Lund, Sweden}

\date{\today}

{\centering \textbf{Supporting information: The importance of relativistic effects for platinum complexes with light-activated activity against cancer cells.}
} 

\section*{Assignment of electronic transitions}

\begin{table}[H]
\begin{center}
\hspace*{-2.0cm}
\centering
\scalebox{0.7}{
\begin{tabular}{cc}
\begin{tabular}[t]{ |c c c c |}
 \hline
 \hline
  \multicolumn{4}{r}{4c}      \\  
 \hline
 \hline
 Energy nm (eV)  & Osc.~str- ($x10^{-4}$) & Orbitals (weight) & Interval  \\ 
 \hline
 217.2 (5.71) & 92.0 & $ \pi(4) \rightarrow 82$  0.31  & \textbf{6} \\
 & &  $ \pi(3) \rightarrow 87$ 0.28   & \\
& &  $ \pi(4) \rightarrow 84 $  0.27   &  \\
 \hline
270.3  (4.59) & 135 & p(2) $\rightarrow$ d(4) 0.42 & \textbf{5} \\
& & p(2) $\rightarrow$ d(3)  0.39 & \\
& & p(1) $\rightarrow$ d(4) 0.24    &  \\
260.2 (4.77) & 119 & p(2) $\rightarrow$ d(3)  0.52   & \\
& & $\pi$(3) $\rightarrow$ d(4) 0.32  & \\
& & p(1) $\rightarrow$ d(3) 0.25    &  \\
250.3 (4.95) & 240 & $\pi$(2) $\rightarrow$ d(4)  0.54   & \\
& & p(1) $\rightarrow$ d(3) 0.29  & \\
& & p(2) $\rightarrow$ d(4) 0.20    &  \\
249.0 (4.98) & 102 & $\pi$(2)  $\rightarrow$ d(4)  0.53   & \\
& & p(2) $\rightarrow$ d(4)  0.30 & \\
& & $\pi$(3) $\rightarrow$ d(4) 0.20    &  \\
245.0 (5.06) & 138 & p(1) $\rightarrow$ d(3)  0.46  & \\
& & $\pi$(3) $\rightarrow$ d(4)  0.40 & \\
& & $\pi$(2) $\rightarrow$ d(4) 0.26    &  \\
\hline
274.8 (4.51) & 3145 & $\pi$(2) $\rightarrow$ d(3)  0.62   & \textbf{4} \\
& & p(2) $\rightarrow$ d(3)  0.14 & \\
& & $\pi$(2) $\rightarrow$ d(4) 0.12    &  \\
\hline
292.7  (4.24) & 0.00 & $d/p \rightarrow d(3)$  0.45  & \textbf{3} \\
& &  $ \pi(1) \rightarrow d(3) $  0.34 & \\
& &  $ \pi(4) \rightarrow 87$   0.13   &  \\
286.5  (4.33) & 0.00 & $d/p \rightarrow d(3)$  0.53  & \\
& &  $ \pi(4) \rightarrow 87 $  0.16 & \\
& &  $ \pi(4) \rightarrow 86$  0.14    &  \\
275.8  (4.50) & 9.00 & $d/p \rightarrow d(3)$  0.38  & \\
& &  $\pi(1) \rightarrow d(3)$  0.30 & \\
& &  $ \pi(1) \rightarrow d(4) $ 0.23     &  \\
\hline
\end{tabular}
\begin{tabular}[t]{ c c c c |}
 \hline
 \hline
  \multicolumn{4}{c}{}         \\ 
\hline
\hline
 \hline
 Energy nm (eV)  & Osc.~str. ($x10^{-4}$) & Orbitals (weight) & Interval  \\ 
 \hline 
\hline
358.9  (3.46) & 17.0 & $\pi(3)\rightarrow d(3)$  0.65  & \textbf{2}\\
& &  $ \pi(2) \rightarrow d(3)$ 0.13 & \\
& &  $ \pi(4) \rightarrow 85$   0.12  &  \\
324.2  (3.82) & 0.00 & $\pi(3) \rightarrow 86$  0.42  & \\
& &  $ \pi(4) \rightarrow 84 $ 0.32 & \\
& &  $ \pi(4) \rightarrow 85 $ 0.25    &  \\
308.1  (4.02) & 5.00 & $p(1) \rightarrow d(4)$  0.45  & \\
& &  $ \pi(3) \rightarrow d(4)$ 0.38 & \\
& &  $p(2) \rightarrow d(4)$   0.23  &  \\
308.0  (4.03) & 2.00 & $p(1) \rightarrow d(4)$  0.45  & \\
& &  $ \pi(3) \rightarrow d(4)$ 0.37 & \\
& &  $p(2) \rightarrow d(4)$ 0.23    &  \\
307.8  (4.03) & 0.00 & $p(1) \rightarrow d(4)$  0.46  & \\
& &  $ \pi(3) \rightarrow d(4) $ 0.37 & \\
& &  $p(2) \rightarrow d(4)$   0.23  &  \\
\hline
450.9  (2.75) & 0.00 & $\pi(4) \rightarrow d(3)$  0.65  & \textbf{1}\\
& &  $ p(3) \rightarrow d(3) $ 0.19 & \\
& &  $ d/p \rightarrow d(3) $ 0.09     &  \\
425.4  (2.91) & 0.00 & $\pi(2) \rightarrow d(3)$  0.66  & \\
& &  $ \pi(3) \rightarrow d(3) $ 0.14 & \\
& &  $\pi(1) \rightarrow 85 $    0.09 &  \\
411.4  (3.01) & 0.00 & $p(3) \rightarrow d(4)$  0.54 & \\
& &  $ p(3)  \rightarrow  d(3) $ 0.28 & \\
& &  $ \pi(4) \rightarrow d(4) $0.27     &  \\
406.1  (3.05) & 2.00 & $\pi(3) \rightarrow d(3)$  0.63 & \\
& &  $ \pi(4) \rightarrow 85 $ 0.15 & \\
& &  $ \pi(2) \rightarrow d(3) $  0.13   &  \\
402.0  (3.08) & 0.00 & $p(3) \rightarrow d(3)$  0.61  & \\
& &  $ p(3) \rightarrow d(4) $ 0.24 & \\
& &  $ p(3) \rightarrow 87 $   0.12  &  \\
\hline
\end{tabular}
\end{tabular}
}
\caption*{TABLE S1:\footnotesize Transitions and assignments for the \textit{trans}-Pt complex, calculated with 4c-CAM-B3LYP. The labels \textbf{1}--\textbf{6} are shown in Figures 3 and 4. In each region, the five most intense transitions were picked.
}
\end{center}
\end{table}

\begin{table}[H]
\begin{center}
\hspace*{-2.0cm}
\centering
\scalebox{0.8}{
\begin{tabular}{ |c c c c |}
 \hline
 \hline
  \multicolumn{4}{c}{LL}        \\ 
 \hline
 \hline
 Energy nm (eV)  & Osc.~str.($x10^{-4}$) & Orbitals (weight) & Interval  \\ 
 \hline
296.6 (4.18) & 343 & p(2) $\rightarrow$ d(2)  0.53  & \textbf{5} \\
& & $\pi$(2) $\rightarrow$ d(3) 0.28  & \\
& & $\pi$(2) $\rightarrow$ d(3) 0.25    &  \\
289.4 (4.28) & 253 & $\pi$(3) $\rightarrow$ d(3)  0.41  & \\
& & p(1) $\rightarrow$ d(2) 0.39   & \\
& & p(2) $\rightarrow$ d(2) 0.27    &  \\
280.7 (4.42) & 1423 & p(1) $\rightarrow$ d(2)  0.50  & \\
& & $\pi$(3) $\rightarrow$ d(3)  0.35  & \\
& & $\pi$(3) $\rightarrow$ d(2) 0.22    &  \\
277.7 (4.46) & 32.0 & $\pi$(1) $\rightarrow$ d(2)  0.51  & \\
& & d/p $\rightarrow$ d(2) 0.33   & \\
& & d(1) $\rightarrow$ d(2) 0.25    &  \\
277.3 (4.47) & 737 & $\pi$(2) $\rightarrow$ d(3)  0.58  & \\
& & p(1) $\rightarrow$ d(3)  0.21  & \\
& & p(2) $\rightarrow$ d(2) 0.20    &  \\
\hline
303.1 (4.09) & 2383 & $\pi$(3) $\rightarrow$ d(2)  0.43  & \textbf{4} \\
& & $\pi$(2) $\rightarrow$ d(2) 0.33   & \\
& & $\pi$(3) $\rightarrow$ d(3) 0.30    &  \\
\hline
403.8 (3.07) & 13.0 & $\pi$(2) $\rightarrow$ d(2)  0.54  & \textbf{2} \\
& & $\pi$(3) $\rightarrow$ d(2) 0.43   & \\
& & $\pi$(4) $\rightarrow$ $85$ 0.08    &  \\
\hline
\end{tabular}
}
\caption*{TABLE S2:\footnotesize{  Transitions and assignments for the \textit{trans}-Pt complex, calculated with NR-CAM-B3LYP. The labels \textbf{2}--\textbf{5} are shown in Figure 3. In each region, the five most intense transitions were picked.}
}
\end{center}
\end{table}

\begin{table}[H]
\begin{center}
\hspace*{-2.0cm}
\centering
\scalebox{0.7}{
\begin{tabular}{ |c c c c |}
 \hline
 \hline
  \multicolumn{4}{c}{SF}       \\ 
 \hline
 \hline
 Energy nm (eV)  & Osc.~str.($x10^{-4}$) & Orbitals (weight) & Interval  \\ 
 \hline
217.2  (5.71) & 0.0 & $\pi(3)$ $\rightarrow$ $\pi(5)$  0.33  & \textbf{6} \\
& & d/p(1) $\rightarrow$ d(4)  0.27  & \\
& & $\pi(2)$ $\rightarrow$ $84$ 0.27    &  \\
217.0  (5.71) & 94.0 & $\pi(3)$ $\rightarrow$ 82  0.33  & \\
& & $\pi(2)$ $\rightarrow$ $\pi(5)$ 0.32   & \\
& & $\pi(3)$ $\rightarrow$ $84$  0.29  &  \\
\hline
261.8 (4.74) & 135 & 75 $\rightarrow$ 80  0.48  & \textbf{5} \\
& & $\pi(2)$ $\rightarrow$ d(4) 0.36   & \\
& & p(1) $\rightarrow$ d(3) 0.31    &  \\
258.0 (4.81) & 66.0 & p(2) $\rightarrow$ d(3)  0.44  & \\
& & $\pi(2)$ $\rightarrow$ d(4) 0.30   & \\
& & p(1) $\rightarrow$ d(3) 0.29    &  \\
251.8 (4.92) & 371 & $\pi$(1) $\rightarrow$ d(4)  0.54  & \\
& & p(1) $\rightarrow$ d(3) 0.31   & \\
& & p(2) $\rightarrow$ d(3) 0.18   &  \\
247.0 (5.02) & 164 & $\pi(3)$ $\rightarrow$ d(4)  0.46  & \\
& & p(1) $\rightarrow$ d(3) 0.42   & \\
& & $\pi(1)$ $\rightarrow$ d(4) 0.26  &  \\
239.6 (5.17) & 42.0 & p(2) $\rightarrow$ d(4)  0.61  & \\
& & $\pi(1)$ $\rightarrow$ d(4) 0.21  & \\
& & $\pi(2)$ $\rightarrow$ d(4) 0.20  &  \\
\hline
275.4 (4.50) & 3383 & $\pi(1)$ $\rightarrow$ d(3)  0.64  & \textbf{4}\\
& & p(2) $\rightarrow$ d(3) 0.14  & \\
& & $\pi(1)$ $\rightarrow$ d(4) 0.12  &  \\
\hline
280.7  (4.42) & 1.00 & $73 \rightarrow 80$  0.51  & \textbf{3} \\
& &  $ d/p(1) \rightarrow d(3) $ 0.33  & \\
& &  $ d(2) \rightarrow d(3) $ 0.26  &  \\
\hline
368.0  (3.37) & 0.00 & $p(3) \rightarrow d(3)$  0.64  & \textbf{2}\\
& &  $ p(3) \rightarrow d(4) $ 0.23  & \\
& &  $ d(1) \rightarrow d(3) $ 0.11  &  \\
361.5 (3.43) & 17.0 & $\pi(2) \rightarrow d(3)$  0.67  & \\
& &  $\pi(3) \rightarrow 85 $  0.12 & \\
& &  $ \pi(3) \rightarrow 84$ 0.11  &  \\
350.7  (3.54) & 0.00 & $p(3) \rightarrow d(4)$  0.61  & \\
& &  $p(3) \rightarrow d(3)$ 0.24  & \\
& &  $ \pi(3) \rightarrow d(4)$ 0.20  &  \\
311.8 (3.98) & 0.00 & $\pi(3) \rightarrow d(4)$  0.61  & \\
& &  $ d/p(1) \rightarrow d(4)$  0.23 & \\
& &  $ p(3) \rightarrow d(4)$ 0.20  &  \\
\hline
\end{tabular}
}
\caption*{TABLE S3: \footnotesize  Transitions and assignments for the \textit{trans}-Pt complex, calculated with SR-CAM-B3LYP. The labels \textbf{2}--\textbf{5} are shown in Figure 4. In each region, the five most intense transitions were picked.}
\end{center}
\end{table}

\begin{table}[H]
\begin{center}
\hspace*{-2.0cm}
\scalebox{0.5}{
\begin{tabular}{cc}
\begin{tabular}[t]{ |c c c c |}
 \hline
 \hline
  \multicolumn{4}{c}{4c}    \\ 
 \hline
 \hline
 Energy nm (eV)  & Osc.~str.($x10^{-4}$) & Orbitals (weight) & Interval  \\ 
 \hline
 234.8  (5.28) & 78.0 & $\pi(3) \rightarrow 82$  0.53  & \textbf{6} \\
  &  & $\pi(3) \rightarrow 84$ 0.34  &     \\
 &  & $\pi(2) \rightarrow 87$ 0.19  &     \\
 \hline
269.3 (4.60) & 21.0 & $p(1) \rightarrow d(4)$  0.45  &  \textbf{5} \\
 &  & $p(1) \rightarrow d(3)$ 0.35  &     \\
 &  & $p(2) \rightarrow d(4)$  0.28 &     \\
268.1 (4.62) & 13.0 & $p(1) \rightarrow d(4)$  0.49  & \\
&  & $p(3) \rightarrow d(3)$  0.34 &     \\
 &  & $\pi(2) \rightarrow d(4)$  0.25 &     \\
267.9 (4.63) & 119 & $p(1) \rightarrow d(4)$  0.49   & \\
&  & $p(1) \rightarrow d(3)$ 0.32  &     \\
 &  & $\pi(2) \rightarrow d(4)$ 0.24  &     \\
262.5 (4.72) & 74.0 & $p(2) \rightarrow d(4)$  0.62  & \\
&  & $p(1) \rightarrow d(4)$  0.23  &     \\
 &  & $\pi(1) \rightarrow d(4)$  0.20 &     \\
253.4 (4.89) & 16.0 & $\pi(3) \rightarrow 82$  0.38  & \\
&  & $\pi(3) \rightarrow 84$ 0.34  &     \\
&  & $\pi(2) \rightarrow 87$ 0.25  &     \\
\hline
294.0 (4.22) & 168 & $\pi(2) \rightarrow d(4)$  0.50  &  \textbf{4}\\
&  & $p(2) \rightarrow d(4)$  0.28 &     \\
 &  & $p(1) \rightarrow d(3)$  0.24 &     \\
293.0 (4.23) & 83.0 & $p(2) \rightarrow d(4)$  0.45  & \\
&  & $p(1) \rightarrow d(3)$  0.29 &     \\
 &  & $\pi(1) \rightarrow d(4)$  0.27 &     \\
290.1 (4.27) & 1232 & $\pi(1) \rightarrow d(3)$  0.40  & \\
&  & $\pi(1) \rightarrow d(4)$  0.38 &     \\
 &  & $p(2) \rightarrow d(3)$  0.27 &     \\
287.5 (4.31) & 86.0 & $p(1) \rightarrow d(3)$  0.43  & \\
&  & $\pi(2) \rightarrow d(4)$  0.31 &     \\
 &  & $\pi(1) \rightarrow d(4)$  0.30 &     \\
274.1  (4.52) & 1565 & $\pi(1) \rightarrow d(4)$  0.51  & \\
&  & $\pi(1) \rightarrow d(3)$  0.32 &     \\
 &  & $p(2) \rightarrow d(4)$  0.19 &     \\
\hline
\end{tabular}
\begin{tabular}[t]{ |c c c c |}
 \hline
 \hline
  \multicolumn{4}{c}{4c}    \\ 
 \hline
 \hline
 Energy nm (eV)  & Osc.~str.($x10^{-4}$) & Orbitals (weight) & Interval  \\ 
 \hline
306.1  (4.05) & 361 & $p(2) \rightarrow d(3)$  0.62  & \textbf{3}\\
&  & $\pi(1) \rightarrow d(3)$  0.31 &     \\
 &  & $p(2) \rightarrow d(4)$  0.07 &     \\
\hline
393.8 (3.15) & 11.0 & $\pi(2) \rightarrow d(3)$  0.70  & \textbf{2}\\
 &  & $ \pi(1) \rightarrow d(3) $  0.06 &     \\
 &  & $ \pi(3) \rightarrow 82 $ 0.05  &     \\
387.2 (3.20) & 0.00 & $\pi(3) \rightarrow d(4)$  0.5552 & \\
 &  & $ p(3) \rightarrow d(4) $  0.38 &     \\
 &  & $ d/p(1) \rightarrow d(4) $ 0.13  &     \\
346.0 (3.58) & 0.00 & $\pi/d(1) \rightarrow d(3)$  0.64  & \\
 &  & $ \pi(3) \rightarrow d(4) $  0.15  &     \\
 &  & $ d/p(1) \rightarrow d(3) $ 0.11  &     \\
344.6 (3.60) & 0.00 & $\pi(3) \rightarrow d(4)$  0.50 & \\
 &  & $ p(3) \rightarrow d(4) $  0.40 &     \\
 &  & $ \pi/d(1) \rightarrow d(3) $ 0.24  &     \\
328.7 (3.77) & 2.00 & $\pi(2) \rightarrow d(4)$  0.50  & \\
 &  & $ p(1) \rightarrow d(4) $  0.38 &     \\
 &  & $ p(1) \rightarrow d(3) $ 0.18  &     \\
328.6 (3.77) & 2.00 & $\pi(2) \rightarrow d(4)$  0.51  & \\
 &  & $ p(1) \rightarrow d(4) $  0.38 &     \\
 &  & $ p(1) \rightarrow d(3) $ 0.18  &     \\
328.4 (3.77) & 0.00 & $\pi(2) \rightarrow d(4)$ 0.51  & \\
 &  & $ p(1) \rightarrow d(4) $  0.38 &     \\
 &  & $ p(1) \rightarrow d(3) $ 0.18  &     \\
 \hline
 437.6  (2.83) & 0.00 & $\pi(2) \rightarrow d(3)$  0.54  &  \textbf{1}\\
 &  & $ \pi(1) \rightarrow d(3) $  0.43 &     \\
 &  & $ \pi(2) \rightarrow d(4) $ 0.06  &     \\
436.8 (2.84) & 1.00 & $\pi(2)  \rightarrow d(3)$  0.51  & \\
&  & $ \pi(1) \rightarrow d(3) $  0.46 &     \\
 &  & $ p(2) \rightarrow d(3) $ 0.06  &     \\
435.9 (2.84) & 0.00 & $\pi(1) \rightarrow d(3)$  0.53  & \\
&  & $ \pi(2) \rightarrow d(3) $  0.44 &     \\
 &  & $ \pi(3) \rightarrow 84 $ 0.07  &     \\
435.1 (2.85) & 1.00 & $\pi(1) \rightarrow d(3)$  0.52  & \\
&  & $ \pi(2) \rightarrow d(3) $ 0.44  &     \\
 &  & $ p(3) \rightarrow d(4) $  0.10  &     \\
435.0 (2.85) & 0.00 & $p(3) \rightarrow d(4)$  0.52  & \\
&  & $ \pi(3) \rightarrow d(4) $ 0.36  &     \\
 &  & $ p(3) \rightarrow d(3) $ 0.24  &     \\
 \hline
\end{tabular}
\end{tabular}
}
\caption*{TABLE S4: \footnotesize  Transitions and assignments for the \textit{trans}-Pt complex, calculated with 4c-B3LYP. The labels \textbf{1}--\textbf{5} are shown in Figures 5 and 6. In each region, the five most intense transitions were picked. }
\end{center}
\end{table}

\begin{table}[H]
\begin{center}
\hspace*{-2.0cm}
\centering
\scalebox{0.5}{
\begin{tabular}{ |c c c c |}
 \hline
 \hline
  \multicolumn{4}{c}{LL}    \\ 
 \hline
 \hline
 Energy nm (eV)  & Osc.~str.($x10^{-4}$) & Orbitals (weight) & Interval  \\ 
 \hline
 299.0 (4.15) & 2012 & $p(2) \rightarrow d/p(2)$  0.50  & \textbf{5} \\
 &  & $ \pi(2) \rightarrow d(3) $  0.39 &     \\
 &  & $ \pi(2) \rightarrow d/p(2) $ 0.22  &     \\
294.7  (4.21) & 202 & $p(2) \rightarrow d/p(2)$  0.47  & \\
&  & $ \pi(2) \rightarrow d(3) $  0.37 &     \\
 &  & $ \pi(2) \rightarrow d/p(2) $ 0.34  &     \\
 \hline
 340.0 (3.65) & 256 & $p(2) \rightarrow d(3)$  0.64  & \textbf{4} \\
 &  & $ \pi(2) \rightarrow d(3) $ 0.27  &     \\
 &  & $ \pi(2) \rightarrow d/p(2) $ 0.12  &     \\
329.8 (3.76) & 8.00 & $\pi(3) \rightarrow d/p(2)$  0.62  & \\
 &  & $ p(1) \rightarrow d(3) $   0.31&     \\
 &  & $ \pi(2) \rightarrow d(3) $  0.09 &     \\
326.4 (3.80) & 536 & $\pi(2) \rightarrow d/p(2)$  0.57  & \\
 &  & $ \pi(2) \rightarrow d(3) $  0.32 &     \\
 &  & $ p(2) \rightarrow d(3) $  0.25 &     \\
318.1 (3.90) & 32.0 & $p(1) \rightarrow d(3)$  0.62  & \\
 &  & $ \pi(3) \rightarrow d/p(2) $  0.29 &     \\
 &  & $ p(1) \rightarrow d/p(2) $  0.15 &     \\
\hline
447.8  (2.77) & 8.00 & $\pi(3) \rightarrow d(3)$  0.70  & \textbf{2} \\
&  & $ \pi(2) \rightarrow d(3) $  0.04 &     \\
 &  & $ \pi(4) \rightarrow 84 $  0.04 &     \\
\hline
\end{tabular}
}
\caption*{TABLE S5: \footnotesize  Transitions and assignments for the \textit{trans}-Pt complex, calculated with NR-B3LYP. The labels \textbf{1}--\textbf{5} are shown in Figure 5. In each region, the five most intense transitions were picked.  }
\end{center}
\end{table}

\begin{table}[H]
\begin{center}
\hspace*{-2.0cm}
\centering
\scalebox{0.8}{
\begin{tabular}{ |c c c c |}
 \hline
 \hline
  \multicolumn{4}{c}{SF}       \\ 
 \hline
 \hline
 Energy nm (eV)  & Osc.~str.($x10^{-4}$) & Orbitals (weight) & Interval  \\ 
 \hline
234.3517 (5.29) & 77.0 & $\pi(3) \rightarrow 82 $ 0.53 & \textbf{6}\\
&  & $ \pi(3) \rightarrow \pi(3) $  0.34 &     \\
 &  & $ \pi(2) \rightarrow 87 $ 0.19  &     \\
\hline
265.1  (4.68) & 103 & $p(2) \rightarrow d(4)$  0.67  & \textbf{5} \\
&  & $ \pi(1) \rightarrow d(4) $  0.20 &     \\
 &  & $  \pi(1) \rightarrow d(3) $ 0.06  &     \\
\hline
295.5 (4.20) & 199 & $\pi(2) \rightarrow d(4)$  0.65  & \textbf{4} \\
&  & $  p(1) \rightarrow d(3) $  0.20 &     \\
 &  & $  \pi(1) \rightarrow d(3) $ 0.15  &     \\
293.5 (4.22) & 1.00 & $d/\pi(1) \rightarrow d(3)$  0.58  & \\
&  & $  d/p(1) \rightarrow d(3) $  0.31 &     \\
 &  & $  d(2) \rightarrow d(3) $ 0.19  &     \\
290.9 (4.26) & 1546 & $\pi(1) \rightarrow d(3)$  0.45  & \\
&  & $  \pi(1) \rightarrow d(4) $ 0.42  &     \\
 &  & $  p(2) \rightarrow d(3) $  0.28 &     \\
280.8 (4.42) & 37.0 & $p(1) \rightarrow d(3)$  0.66 & \\
&  & $ \pi(2) \rightarrow d(4) $  0.19 &     \\
 &  & $ p(1)  \rightarrow d(4) $ 0.13  &     \\
275.5 (4.50) & 1875 & $\pi(1) \rightarrow d(4)$  0.52  & \\
&  & $ \pi(1) \rightarrow  d(3) $  0.36 &     \\
 &  & $ p(2)  \rightarrow d(4) $ 0.19  &     \\
\hline
308.2 (4.02) & 35.0 & $p(2) \rightarrow d(3)$  0.63  &  \textbf{3}\\
&  & $ \pi(1) \rightarrow 89  $  0.31 &     \\
 &  & $ \pi(1)  \rightarrow d(4) $ 0.06  &     \\
\hline
397.2 (3.12) & 11.0 & $\pi(2) \rightarrow d(3)$  0.70  & \textbf{2} \\
&  & $\pi(3) \rightarrow 82  $  0.05 &     \\
 &  & $  \pi(3) \rightarrow 84 $ 0.05  &     \\
\hline
\end{tabular}
}
\caption*{TABLE S6: \footnotesize Transitions and assignments for the \textit{trans}-Pt complex, calculated with SR-B3LYP. The labels \textbf{1}--\textbf{5} are shown in Figure 6. In each region, the five most intense transitions were picked.   }
\end{center}
\end{table}

\begin{table}[H]
\begin{center}
\hspace*{-2.0cm}
\centering
\scalebox{0.4}{
\begin{tabular}{cc}
\begin{tabular}[t]{ |c c c c |}
 \hline
 \hline
  \multicolumn{4}{r}{4c}      \\ 
 \hline
 \hline
 Energy nm (eV)  & Osc.~str.($x10^{-4}$) & Orbitals (weight) & Interval  \\ 
 \hline
 229.6  (5.40) & 89.0 & $d(2) \rightarrow d/p(2)$ 0.28  & \textbf{9} \\
& &  $ d(2) \rightarrow d(3) $ 0.26  & \\
& &  $ p(1) \rightarrow d/p(2) $ 0.23  & \\
224.4 (5.53) & 69.0 & $d(2) \rightarrow d/p(2)$  0.25  & \\
& &  $ d(1) \rightarrow  d(3) $ 0.20 & \\
& &  $ d(2) \rightarrow  d(3) $ 0.20  & \\
223.5 (5.55) & 69.0 & $\pi(3) \rightarrow 85$  0.24  & \\
& &  $ d(2) \rightarrow d/p(2)  $ 0.21 & \\
& &  $ d(1) \rightarrow d(3) $ 0.20  & \\
217.9 (5.69) & 20.0 & $d(1) \rightarrow d(3)$  0.38  & \\
& &  $ d/p(1) \rightarrow  d(3) $ 0.30 & \\
& &  $ d(2) \rightarrow d(3) $  0.20 & \\
213.7 (5.80) & 70.0 & $ d(2) \rightarrow d/p(2)$  0.25 & \\
& &  $ \pi(2) \rightarrow 85 $ 0.19 & \\
& &  $ d/p(1) \rightarrow d/p(2) $ 0.17  & \\
\hline
237.5 (5.22) & 307 & $\pi(2) \rightarrow d/p(2)$  0.5803  & \textbf{8} \\
& &  $ d(2) \rightarrow d(3) $ 0.12  & \\
& &  $ \pi(4) \rightarrow \pi(5) $ 0.11   & \\
236.7  (5.24) & 18.0 & $\pi(2) \rightarrow d/p(2)$  0.39  & \\
& &  $ d(2) \rightarrow d(3) $  0.34 & \\
& &  $ p(2) \rightarrow d(3) $ 0.18   & \\
235.0  (5.27) & 5.00 & $\pi(2) \rightarrow d/p(2)$  0.40  & \\
& &  $ d(2) \rightarrow d(3) $ 0.31 & \\
& &  $ d(1) \rightarrow d(3) $  0.21  & \\
234.6 (5.28) & 20.0 & $\pi(2) \rightarrow d/p(2)$  0.39  & \\
& &  $ d(2) \rightarrow d(3) $ 0.30 & \\
& &  $ d(1) \rightarrow d(3) $  0.20  & \\
\hline
249.1 (4.98) & 36.0 & $\pi(1) \rightarrow d/p(2)$ 0.25 & \textbf{7} \\
& &  $ \pi(3) \rightarrow 85 $ 0.23 & \\
& &  $ \pi(4) \rightarrow  85 $ 0.22  & \\
248.6 (4.99) & 4.00 & $\pi(4) \rightarrow 85$ 0.25  & \\
& &  $ \pi(3) \rightarrow 85 $ 0.22 & \\
& &  $ d(2) \rightarrow d/p(2) $  0.19 & \\
245.9  (5.04) & 38.0 & $\pi(1) \rightarrow d/p(2)$  0.29  & \\
& &  $ \pi(3) \rightarrow 85 $ 0.20 & \\
& &  $ d(2) \rightarrow d(3) $ 0.19   & \\
245.1 (5.06) & 34.0 & $d(2) \rightarrow d(3)$  0.26 & \\
& &  $ \pi(3) \rightarrow 85  $ 0.24 & \\
& &  $ p(3) \rightarrow d/p(2)$ 0.21  & \\
243.8 (5.08) & 94.0 & $\pi(1) \rightarrow d/p(2)$ 0.43  & \\
& &  $ d(2) \rightarrow d(3) $ 0.16  & \\
& &  $ p(1) \rightarrow d/p(2) $ 0.16  & \\
\hline
269.2  (4.61) & 152 & $\pi(1) \rightarrow d(3)$  0.39  & \textbf{6} \\
& &  $p(1) \rightarrow d(3) $ 0.31 & \\
& &  $p(2) \rightarrow d(3)$ 0.30  & \\
267.2  (4.64) & 211 & $p(1) \rightarrow d(3)$  0.43  & \\
& &  $\pi(2) \rightarrow d(3) $ 0.24 & \\
& &  $\pi(1) \rightarrow d/p(2)$ 0.17  & \\
264.8 (4.68) & 711 & $\pi(2) \rightarrow d(3)$  0.44 & \\
& &  $p(1) \rightarrow d(3) $ 0.28 & \\
& &  $p(2) \rightarrow d/p(2)$ 0.21  & \\
262.0  (4.73) & 389 & $\pi(1) \rightarrow d/p(2)$  0.40 & \\
& &  $\pi(2) \rightarrow d(3) $ 0.33 & \\
& &  $p(1) \rightarrow d$ 0.23  & \\
255.6 (4.85) & 194 & $p(1) \rightarrow d(3)$  0.51  & \\
& &  $\pi(1) \rightarrow d(3) $ 0.25 & \\
& &  $\pi(2) \rightarrow d/p(2)$ 0.21  & \\
\hline
\end{tabular}
\begin{tabular}[t]{|c c c c|}
 \hline
 \hline
  \multicolumn{4}{c}{}          \\ 
 \hline
 \hline
 Energy nm (eV) & Osc.~str.($x10^{-4}$) & Orbitals (weight) & Interval  \\ 
 \hline
 291.1 (4.26) & 40.0 & $p(3) \rightarrow d/p(2)$  0.30  & \textbf{5}\\
& &  $\pi(3) \rightarrow d/p(2) $ 0.25 & \\
& &  $\pi(1) \rightarrow d(3)$ 0.23  & \\
287.2 (4.32) & 6.00 & $p(2) \rightarrow d(3) $ 0.34  & \\
& &  $p(3) \rightarrow d/p(2) $ 0.27 & \\
& &  $\pi(4) \rightarrow d/p(2)$ 0.23  & \\
282.1 (4.40) & 105 & $p(3) \rightarrow d/p(2)$  0.33 & \\
& &  $\pi(4) \rightarrow d/p(2) $ 0.26 & \\
& &  $p(1) \rightarrow d/p(2)$ 0.21  & \\
278.7 (4.45) & 5.00 & $p(1) \rightarrow d/p(2)$  0.37  & \\
& &  $p(2) \rightarrow d/p(2) $ 0.35 & \\
& &  $\pi(3) \rightarrow d/p(2)$ 0.19  & \\
278.2 (4.46) & 17.0 & $p(1) \rightarrow d/p(2)$  0.31 & \\
& &  $p(2) \rightarrow d/p(2) $ 0.30 & \\
& &  $\pi(3) \rightarrow d/p(2)$ 0.27  & \\
\hline
327.8 (3.78) & 3.00 & $\pi(3) \rightarrow d/p(2)$  0.25 & \textbf{4} \\
& &  $\pi(3) \rightarrow 86 $ 0.24 & \\
& &  $\pi(3) \rightarrow \pi(5)$ 0.24  & \\
323.9  (3.83) & 2.00 & $\pi(3) \rightarrow \pi(5)$  0.33 & \\
& &  $\pi(3) \rightarrow 86 $ 0.28 & \\
& &  $p(2) \rightarrow d(3)$ 0.19  & \\
321.6 (3.86) & 5.00 & $p(2) \rightarrow d(3)$  0.28 & \\
& &  $d(2) \rightarrow d(3) $ 0.24 & \\
& &  $p(3) \rightarrow d/p(2)$ 0.21  & \\
313.4 (3.96) & 2.00 & $\pi(4) \rightarrow 86$  0.26  & \\
& &  $p(2) \rightarrow d(3) $ 0.26 & \\
& &  $\pi(4) \rightarrow \pi(5)$ 0.24  & \\
310.1 (4.00) & 7.00 & $p(2) \rightarrow d/p(2)$  0.36 & \\
& &  $p(1) \rightarrow d/p(2) $ 0.34 & \\
& &  $\pi(3) \rightarrow d/p(2)$ 0.27  & \\
305.0 (4.07) & 23.0 & $p(1) \rightarrow d/p(2)$  0.34  & \\
& &  $\pi(3) \rightarrow d/p(2) $ 0.30 & \\
& &  $p(2) \rightarrow d/p(2)$ 0.29  & \\
\hline
348.0 (3.56) & 9.00 & $p(3) \rightarrow d(3)$  0.54  & \textbf{3} \\
& &  $\pi(3) \rightarrow d/p(2) $ 0.23 & \\
& &  $\pi(2) \rightarrow d(3)$ 0.17  & \\
\hline
409.9 (3.02) & 2.00 & $p(3) \rightarrow d/p(2)$  0.40  & \textbf{2}\\
& &  $\pi(3) \rightarrow d(3) $ 0.37 & \\
& &  $\pi(4) \rightarrow d/p(2)$ 0.23  & \\
392.9 (3.16) & 3.00 & $p(3) \rightarrow d(3)$  0.46 & \\
& &  $\pi(1) \rightarrow d(3) $ 0.25 & \\
& &  $\pi(2) \rightarrow d(3)$ 0.21  & \\
391.0 (3.17) & 2.00 & $p(3) \rightarrow d(3)$  0.49  & \\
& &  $\pi(2) \rightarrow d(3) $ 0.29 & \\
& &  $\pi(3) \rightarrow d/p(2)$ 0.21  & \\
384.6 (3.22) & 2.00 & $\pi(1) \rightarrow d(3)$  0.46  & \\
& &  $\pi(3) \rightarrow d/p(2) $ 0.23 & \\
& &  $p(3) \rightarrow d/p(2)$ 0.18  & \\
377.6 (3.28) & 2.00 & $\pi(2) \rightarrow d(3)$  0.45  & \\
& &  $p(3) \rightarrow d(3) $ 0.29 & \\
& &  $\pi(1) \rightarrow d/p(2)$ 0.21  & \\
\hline
435.3 (2.85) & 7.00 & $\pi(4) \rightarrow d(3)$  0.45  & \textbf{1}\\
& &  $\pi(4) \rightarrow d/p(2) $ 0.32 & \\
& &  $\pi(3) \rightarrow d(3)$ 0.29  & \\
429.5  (2.89) & 2.00 & $\pi(3) \rightarrow d(3)$  0.50  & \\
& &  $\pi(4) \rightarrow d/p(2) $ 0.30 & \\
& &  $\pi(4) \rightarrow d(3)$ 0.23  & \\
\hline
\end{tabular}
\end{tabular}
}
\caption*{TABLE S7:\footnotesize   Transitions and assignments for the \textit{cis}-Pt complex, calculated with 4c CAM-B3LYP. The labels \textbf{1}--\textbf{9} are shown in Figures 7 and 8. In each region, the five most intense transitions were picked.   }
\end{center}
\end{table}

\begin{table}[H]
\begin{center}
\hspace*{-2.0cm}
\centering
\scalebox{0.7}{
\begin{tabular}{cc}
\begin{tabular}[t]{ |c c c c |}
 \hline
 \hline
  \multicolumn{4}{r}{LL}    \\ 
 \hline
 \hline
 Energy nm (eV)  & Osc.~str.($x10^{-4}$) & Orbitals (weight) & Interval  \\ 
 \hline
 235.2 (5.27) & 64.0 & $d/p (2) \rightarrow d(2)$  0.42 & \textbf{8} \\
& &  $d/p (1) \rightarrow d(2) $ 0.35 & \\
& &  $\pi(2) \rightarrow 85$ 0.16  & \\
232.9  (5.32) & 614 & $p(1) \rightarrow d/p(3)$  0.37  & \\
& &  $d/p(2) \rightarrow d/p(3) $ 0.32 & \\
& &  $d/p (1) \rightarrow d/p(3)$ 0.25  & \\
226.8 (5.47) & 384 & $d(1) \rightarrow d(2)$  0.46  & \\
& &  $d/p (2) \rightarrow d(2) $ 0.26 & \\
& &  $\pi (2) \rightarrow 85$ 0.20  & \\
\hline
262.1 (4.73) & 729 & $75 \rightarrow d/p(3)$  0.53  &  \textbf{6}\\
& &  $\pi(1) \rightarrow d/p(3) $ 0.29 & \\
& &  $p(2) \rightarrow d/p(3)$ 0.23  & \\
261.0 (4.75) & 532 & $\pi(1) \rightarrow d/p(3)$ 0.57 & \\
& &  $75 \rightarrow d/p(3) $ 0.24 & \\
& &  $p(2) \rightarrow d/p(3)$ 0.17  & \\
\hline
300.2 (4.13) & 192 & $p(1) \rightarrow d(2)$  0.56  & \textbf{5} \\
& &  $75 \rightarrow d(2) $ 0.26 & \\
& &  $d(1) \rightarrow d(2)$ 0.15  & \\
293.5 (4.22) & 1423 & $\pi(1) \rightarrow d(2)$  0.58 & \\
& &  $75 \rightarrow d/p(3) $ 0.17 & \\
& &  $p(1) \rightarrow d(2)$ 0.16  & \\
288.1 (4.30) & 372 & $75 \rightarrow d/p(3)$  0.42 & \\
& &  $p(2) \rightarrow d/p(3) $ 0.30 & \\
& &  $p(1) \rightarrow d/p(3)$ 0.28  & \\
284.4 (4.36) & 545 & $75 \rightarrow d(2)$  0.37  & \\
& &  $p(1) \rightarrow d/p(3) $ 0.27 & \\
& &  $p(2) \rightarrow d/p(3)$ 0.25  & \\
\hline
\end{tabular}
\begin{tabular}[t]{ c c c c |}
 \hline
 \hline
  \multicolumn{4}{c}{}        \\ 
 \hline
 \hline
 Energy nm (eV)  & Osc.~str.($x10^{-4}$) & Orbitals (weight) & Interval  \\ 
 \hline
325.8 (3.81) & 40.0 & $\pi(3) \rightarrow d/p(3)$  0.46 & \textbf{4} \\
& &  $\pi(2) \rightarrow d/p(3) $ 0.37 & \\
& &  $p(2) \rightarrow d/p(3)$ 0.23  & \\
311.3 (3.98) & 33.0 & $p(2) \rightarrow d(2)$  0.38  & \\
& &  $\pi(2) \rightarrow d/p(3) $ 0.32 & \\
& &  $\pi(3) \rightarrow d/p(3)$ 0.26  & \\
306.9 (4.04) & 13.0 & $p(2) \rightarrow d(2)$  0.51 & \\
& &  $p(3) \rightarrow d/p(3)$ 0.26 & \\
& &  $\pi(2) \rightarrow d/p(3)$ 0.22  & \\
 \hline
 385.2 (3.22) & 2.00 & $p(3) \rightarrow d/p(3)$  0.48 & \textbf{3} \\
& &  $p(3) \rightarrow d(2) $ 0.32 & \\
& &  $\pi(3) \rightarrow d/p(3)$ 0.27  & \\
379.8 (3.26) & 6.00 & $p(3) \rightarrow d(2)$  0.57  & \\
& &  $p(3) \rightarrow d/p(3) $ 0.29 & \\
& &  $\pi(2) \rightarrow d/p(3)$ 0.19  & \\
\hline
472.2 (2.63) & 12.0 & $\pi(3) \rightarrow d(2)$  0.65  & \textbf{1} \\
& &  $\pi(3) \rightarrow d/p(3) $ 0.21 & \\
& &  $p(3) \rightarrow d(2)$ 0.07  & \\
460.4 (2.69) & 5.00 & $\pi(2) \rightarrow d(2)$  0.65  & \\
& &  $\pi(2) \rightarrow d/p(3) $ 0.19 & \\
& &  $p(3) \rightarrow d(2)$ 0.10  & \\
\hline
\end{tabular}
\end{tabular}
}
\caption*{TABLE S8: \footnotesize  Transitions and assignments for the \textit{cis}-Pt complex, calculated with NR-CAM-B3LYP. The labels \textbf{1}--\textbf{9} are shown in Figure 7. In each region, the five most intense transitions were picked. }
\end{center}
\end{table}

\begin{table}[H]
\begin{center}
\hspace*{-2.0cm}
\centering
\scalebox{0.6}{
\begin{tabular}{cc}
\begin{tabular}[t]{ |c c c c |}
 \hline
 \hline
  \multicolumn{4}{r}{SF}    \\ 
 \hline
 \hline
 Energy nm (eV)  & Osc.~str.($x10^{-4}$) & Orbitals (weight) & Interval  \\ 
 \hline
229.9 (5.39) & 7.00 & $d(2) \rightarrow d(3)$  0.41 & \textbf{9}\\
& &  $ d(1) \rightarrow d(3) $ 0.22 & \\
& &  $ \pi(4) \rightarrow 83 $ 0.17  & \\
227.9 (5.44) & 170 & $p(1) \rightarrow d/p(2)$  0.30  & \\
& &  $ d(2) \rightarrow d/p(2) $ 0.26 & \\
& &  $ \pi(3) \rightarrow 85 $ 0.17  & \\
223.9 (5.54) & 174 & $d(1) \rightarrow d(3)$  0.34  & \\
& &  $ d(2) \rightarrow d(3) $ 0.30 & \\
& &  $ \pi(1) \rightarrow d(3) $ 0.20  & \\
\hline
238.1 (5.21) & 363 & $\pi(2) \rightarrow d/p(2)$  0.64 & \textbf{8}\\
& &  $ \pi(4) \rightarrow 84 $ 0.12 & \\
& &  $ \pi(4) \rightarrow 86 $ 0.11  & \\
\hline
246.5 (5.03) & 225 & $\pi(1) \rightarrow d/p(2)$  0.61 & \textbf{7} \\
& &  $ d(2) \rightarrow d/p(2) $ 0.19 & \\
& &  $ p(1) \rightarrow d/p(2) $  0.18 & \\
\hline
271.1 (4.57) & 157 & $\pi(1) \rightarrow d(3)$  0.43  & \textbf{6} \\
& &  $ p(1) \rightarrow d(3) $ 0.31  & \\
& &  $ p(2) \rightarrow d(3) $ 0.27  & \\
266.8 (4.65) & 1001 & $\pi(2) \rightarrow d(3)$  0.49 & \\
& &  $ p(2) \rightarrow d/p(2) $ 0.35 & \\
& &  $ \pi(3) \rightarrow d/p(2) $ 0.18  & \\
263.0 (4.71) & 576 & $\pi(2) \rightarrow d(3)$  0.40  & \\
& &  $ p(2) \rightarrow d/p(2) $ 0.38 & \\
& &  $ p(1) \rightarrow d(3) $ 0.29  & \\
258.4 (4.80) & 279 & $p(1) \rightarrow d(3)$  0.53  & \\
& &  $ \pi(1) \rightarrow d(3) $ 0.27 & \\
& &  $ p(2) \rightarrow d(3)$  0.20 & \\
\hline
\end{tabular}
\begin{tabular}[t]{ c c c c |}
 \hline
 \hline
 \multicolumn{4}{c}{}        \\ 
 \hline
 \hline
 Energy nm (eV)  & Osc.~str.($x10^{-4}$) & Orbitals (weight) & Interval  \\ 
 \hline
299.0  (4.15) & 53.0 & $p(2) \rightarrow d(3)$  0.56  & \textbf{5}\\
& &  $ d(2) \rightarrow d(3) $ 0.27 & \\
& &  $ \pi(1) \rightarrow d(3) $  0.20 & \\
285.6 (4.34) & 152 & $p(3) \rightarrow d/p(2) $  0.43  & \\
& &  $ \pi(4) \rightarrow d/p(2) $ 0.27 & \\
& &  $ \pi(1) \rightarrow d(3) $  0.27 & \\
 \hline
309.5  (4.01) & 30.0 & $\pi(3) \rightarrow d/p(2)$  0.40  & \textbf{4}\\
& &  $ \pi(4)  \rightarrow d/p(2) $ 0.35 & \\
& &  $ p(2) \rightarrow d/p(2) $ 0.26  & \\
 \hline
 361.4 (3.43) & 3.00 & $p(3) \rightarrow d/p(2)$  0.45  & \textbf{3} \\
& &  $\pi(4) \rightarrow d/p(2) $ 0.37 & \\
& &  $\pi(3) \rightarrow d/p(2)$ 0.26  & \\
350.1 (3.54) & 14.0 & $p(3) \rightarrow d(3)$  0.62 & \\
& &  $\pi(3) \rightarrow d/p(2) $ 0.23 & \\
& &  $\pi(4) \rightarrow d/p(2)$ 0.11  & \\
\hline
 433.3 (2.86) & 12.0 & $\pi(4) \rightarrow d(3)$  0.64 & \textbf{1} \\
& &  $\pi(4) \rightarrow d/p(2) $ 0.23 & \\
& &  $p(3) \rightarrow d(3)$ 0.09  & \\
425.7 (2.91) & 3.00 & $\pi(3) \rightarrow d(3)$  0.64 & \\
& &  $\pi(3) \rightarrow d/p(2) $ 0.22 & \\
& &  $p(3) \rightarrow d(3)$ 0.08  & \\
\hline
\end{tabular}
\end{tabular}
}
\caption*{TABLE S9: \footnotesize   Transitions and assignments for the \textit{cis}-Pt complex, calculated with SR-CAM-B3LYP. The labels \textbf{1}--\textbf{9} are shown in Figure 8. In each region, the five most intense transitions were picked.  }
\end{center}
\end{table}

\begin{table}[H]
\begin{center}
\hspace*{-2.0cm}
\centering
\scalebox{0.4}{
\begin{tabular}{cc}
\begin{tabular}[t]{ |c c c c }
 \hline
 \hline
  \multicolumn{4}{c}{4c}    \\ 
 \hline
 \hline
 Energy nm (eV)  & Osc.~str.($x10^{-4}$) & Orbitals (weight) & Interval  \\ 
 \hline
  255.3  (4.86) & 62.0 & $d(2) \rightarrow  d(3)$  0.24  & \textbf{9} \\
  & &  $ d/\pi(1) \rightarrow 87 $ 0.21 & \\
& &  $ d/\pi(1) \rightarrow 83 $ 0.21   & \\
254.3 (4.88) & 28.0 & $d(2) \rightarrow d/p(2)$  0.27 0. & \\
  & &  $ d/\pi(1) \rightarrow  83 $ 0.24 & \\
& &  $ d/\pi(1) \rightarrow 85 $ 0.22  & \\
252.7 (4.91) & 74.0 & $\pi(3) \rightarrow 85$  0.27  & \\
  & &  $ \pi(3) \rightarrow 83 $ 0.22 & \\
& &  $ \pi(2) \rightarrow d/p(2) $  0.19 & \\
243.2 (5.10) & 136 & $d(2) \rightarrow d(3)$  0.33  & \\
  & &  $ d(2) \rightarrow d/p(2) $ 0.26 & \\
& &  $ p(1) \rightarrow d/p(2) $  0.22 & \\
239.5 (5.18) & 200 & $d(2) \rightarrow d/p(2)$  0.32  & \\
  & &  $ d(2) \rightarrow d(3) $ 0.29 & \\
& &  $ p(1) \rightarrow d/p(2) $ 0.29  & \\
 \hline
 258.5 (4.80) & 397 & $\pi(2) \rightarrow d/p(2)$  0.49  &  \textbf{8}\\
   & &  $d(2) \rightarrow  d(3)$ 0.29 & \\
& &  $ p(2) \rightarrow d/p(2) $ 0.22  & \\
 \hline
  263.9  (4.70) & 415 & $\pi(1)  \rightarrow d/p(2)$  0.45  &  \textbf{7}\\
    & &  $ d(2) \rightarrow d(3) $ 0.27 & \\
& &  $ \pi(2) \rightarrow d/p(2)$ 0.26  & \\
262.8 (4.72) & 2.00 & $d(2) \rightarrow d(3)$  0.52  & \\
  & &  $ \pi(2) \rightarrow d/p(2) $ 0.23 & \\
& &  $ p(2) \rightarrow d/p(2) $ 0.14  & \\
262.7 (4.72) & 44.0 & $d(2) \rightarrow d(3)$  0.48  & \\
  & &  $ \pi(1) \rightarrow d/p(2) $ 0.28 & \\
& &  $  p(2) \rightarrow d/p(2) $  0.17 & \\
261.0 (4.75) & 106 & $d(2) \rightarrow d(3)$  0.51  & \\
  & &  $ \pi(2) \rightarrow  d/p(2) $ 0.24 & \\
& &  $ p(2) \rightarrow d(3) $ 0.16   & \\
\hline 
300.1 (4.13) & 114 &  $p(1) \rightarrow d(3)$  0.38  & \textbf{6} \\
  & &  $ \pi(1) \rightarrow  d(3) $ 0.38 & \\
& &  $ \pi(2) \rightarrow d(3) $ 0.22  & \\
291.4 (4.26) & 113 & $p(1) \rightarrow d(3)$  0.59  & \\
  & &  $ \pi(1) \rightarrow  d(3)$ 0.22 & \\
& &  $ p(2) \rightarrow d(3) $  0.16 & \\
289.9 (4.28) & 347 & $p(2) \rightarrow d/p(2)$  0.42  & \\
  & &  $ \pi(2) \rightarrow d(2) $ 0.36 & \\
& &  $ \pi(2) \rightarrow d/p(2) $  0.27 & \\
286.4 (4.33) & 415 & $\pi(2) \rightarrow d(3)$ 0.37  & \\
  & &  $ p(2) \rightarrow d/p(2) $ 0.35 & \\
& &  $ \pi(1) \rightarrow d/p(2) $ 0.26  & \\
284.1 (4.36) & 100 & $\pi(1) \rightarrow d/p(2)$  0.53  & \\
  & &  $ \pi(2) \rightarrow d(3) $ 0.22 & \\
& &  $ \pi(2) \rightarrow d/p(2) $ 0.20  & \\
 \hline
 \end{tabular}
 \begin{tabular}[t]{ |c c c c |}
 \hline
 \hline
  \multicolumn{4}{c}{4c}    \\ 
 \hline
 \hline
 Energy nm (eV)  & Osc.~str.($x10^{-4}$) & Orbitals (weight) & Interval  \\ 
 \hline
326.9  (3.79) & 1.00 & $p(2) \rightarrow d(3)$  0.45  & \textbf{5}\\
  & &  $p(3) \rightarrow d/p(2) $ 0.31 & \\
& &  $ \pi(1) \rightarrow d(3) $ 0.27  & \\
324.3 (3.82) & 3.00 & $p(1) \rightarrow d/p(2)$  0.36  & \\
  & &  $ p(2) \rightarrow d/p(2) $ 0.36 & \\
& &  $ p(3) \rightarrow d/p(2) $ 0.28  & \\
323.8 (3.83) & 1.00 & $p(1) \rightarrow d/p(2)$  0.36  & \\
  & &  $ p(2) \rightarrow d/p(2) $ 0.36 & \\
& &  $ p(3) \rightarrow d/p(2) $ 0.27  & \\
323.3 (3.83) & 1.00 & $p(1) \rightarrow d/p(2)$  0.39  & \\
  & &  $ p(2) \rightarrow d/p(2) $ 0.39 & \\
& &  $ p(3) \rightarrow d/p(2) $ 0.21  & \\
319.5 (3.88) & 3.00 & $p(3) \rightarrow d/p(2)$  0.55  & \\
  & &  $ p(2) \rightarrow d(3) $ 0.29 & \\
& &  $ \pi(1) \rightarrow d(3) $ 0.15  & \\
\hline
349.1  (3.55) & 14.0 & $\pi(3) \rightarrow d/p(2)$  0.48  & \textbf{4}\\
  & &  $ d/\pi(1) \rightarrow d/p(2) $ 0.26 & \\
& &  $ p(3) \rightarrow d/p(2) $ 0.24  & \\
346.1  (3.58) & 7.0 & $p(2) \rightarrow d(3)$  0.37  & \\
  & &  $ \pi(3) \rightarrow  d/p(2) $ 0.31  & \\
& &  $ \pi(2) \rightarrow d(3) $ 0.28  & \\
345.4  (3.59) & 2.0 & $p(2) \rightarrow d(3)$  0.38  & \\
  & &  $ p(3) \rightarrow d/p(2) $ 0.35 & \\
& &  $ \pi(2) \rightarrow d(3) $  0.24 & \\
345.3 (3.59) & 7.00 & $p(2) \rightarrow d(3)$  0.35  & \\
  & &  $ p(3) \rightarrow d/p(2) $ 0.35 & \\
& &  $\pi(2) \rightarrow d(3) $ 0.25  & \\
331.6 (3.74) & 2.00 & $p(3) \rightarrow d/p(2)$  0.38  & \\
  & &  $ p(2) \rightarrow d(3) $ 0.32 & \\
& &  $ p(2) \rightarrow d/p(2) $ 0.28   & \\
\hline
403.0  (3.08) & 1.00 & $d/\pi(1) \rightarrow d/p(2)$  0.54  &  \textbf{3}\\
  & &  $p(3) \rightarrow d(3) $ 0.18 & \\
& &  $ p(3) \rightarrow  d/p(2) $ 0.18  & \\
396.9 (3.12) & 9.00 & $p(3) \rightarrow d(3)$  0.56  & \\
  & &  $ \pi(1) \rightarrow d(3) $ 0.27 & \\
& &  $\pi(2) \rightarrow d(3) $ 0.18  & \\
388.6 (3.19) & 2.00 & $\pi(1) \rightarrow d(3)$  0.46  & \\
  & &  $ p(3) \rightarrow d(3) $ 0.32 & \\
& &  $ p(2) \rightarrow d(3) $ 0.24  & \\
381.5 (3.25) & 1.00 & $\pi(2) \rightarrow d(3)$  0.44 & \\
  & &  $ \pi(1) \rightarrow d(3) $ 0.28 & \\
& &  $ p(2) \rightarrow d(3) $ 0.25   & \\
\hline
428.4  (2.89) & 1.00 & $p(3) \rightarrow d(3)$  0.62  & \textbf{2} \\
  & &  $ \pi(3) \rightarrow d/p(2)  $ 0.26 & \\
& &  $ \pi(1) \rightarrow d(3) $ 0.10   & \\
427.0 (2.90) & 4.00 & $p(3) \rightarrow d(3)$  0.60  & \\
  & &  $ \pi(3) \rightarrow d/p(2) $ 0.25 & \\
& &  $ \pi(3) \rightarrow d(3) $ 0.17  & \\
426.7 (2.91) & 2.00 &  $ p(3) \rightarrow d(3)$ 0.61  & \\
  & &  $ \pi(3) \rightarrow d/p(2) $ 0.27 & \\
& &  $ \pi(3) \rightarrow d(3) $ 0.14  & \\
\hline
563.1 (2.20) & 1.00 & $ d/\pi(1) \rightarrow d(3)$   0.67  & \textbf{1} \\
  & &  $ \pi(3) \rightarrow d/p(2) $ 0.12 & \\
& &  $ \pi(3) \rightarrow d(3)$  0.09  & \\
552.0 (2.25) & 1.00 & $d/\pi(1) \rightarrow d(3)$  0.66  & \\
  & &  $ \pi(3) \rightarrow d(3) $ 0.15 & \\
& &  $ d/\pi(1) \rightarrow d/p(2) $ 0.10  & \\
532.5 (2.33) & 1.00 & $\pi(3) \rightarrow d(3)$  0.48  & \\
  & &  $ d/\pi(1) \rightarrow d(3) $ 0.47 & \\
& &  $ d/\pi(1) \rightarrow d/p(2) $ 0.12  & \\
474.9 (2.61) & 4.00 & $d/\pi(1) \rightarrow d(3)$  0.48  & \\
  & &  $\pi(3) \rightarrow d(3) $ 0.41 & \\
& &  $ d/\pi(1) \rightarrow d/p(2)$ 0.28  & \\
460.5 (2.69) & 3.00 & $\pi(3) \rightarrow d(3)$  0.59  & \\
  & &  $ d/\pi(1) \rightarrow d/p(2) $ 0.27 & \\
& &  $ d/\pi(1) \rightarrow d(3)$  0.19  & \\
\hline
\end{tabular}
\end{tabular}
}
\caption*{TABLE S10: \footnotesize  Transitions and assignments for the \textit{cis}-Pt complex, calculated with 4c-B3LYP. The labels \textbf{1}--\textbf{9} are shown in Figures 9 and 10. In each region, the five most intense transitions were picked.  }
\end{center}
\end{table}

\begin{table}[H]
\begin{center}
\hspace*{-2.0cm}
\centering
\scalebox{0.5}{
\begin{tabular}{ |c c c c |}
 \hline
 \hline
  \multicolumn{4}{c}{LL}    \\ 
 \hline
 \hline
 Energy nm (eV)  & Osc.~str.($x10^{-4}$) & Orbitals (weight) & Interval  \\ 

\hline
 &  &   & \textbf{8} \\
 244.0  (5.08) & 226 & $d/p(2) \rightarrow d(2)$ 0.60 & \\
   & &  $ d/p(1) \rightarrow d(2) $ 0.20 & \\
& &  $ \pi(1) \rightarrow d(2) $ 0.13   & \\
242.4 (5.11) & 749 & $p(1) \rightarrow d/p(3)$  0.45 & \\
 & &  $ d/p(2) \rightarrow d/p(3) $ 0.41 & \\
& &  $ \pi(1) \rightarrow d/p(3) $  0.15  & \\
\hline
 &  &   &  \textbf{6} \\
279.0  (4.44) & 768 & $\pi(2) \rightarrow d/p(3)$  0.55  & \\
 & &  $ \pi(1) \rightarrow d(2) $ 0.23 & \\
& &  $ d/p(2) \rightarrow d(2) $ 0.19   & \\
275.5 (4.50) & 1217 & $\pi(1) \rightarrow d/p(3)$  0.56  & \\
 & &  $ \pi(2) \rightarrow d(2) $ 0.23 & \\
& &  $ p(2) \rightarrow d/p(3)$ 0.20    & \\
\hline
&  &  & \textbf{5} \\
319.5  (3.88) & 348 & $p(2) \rightarrow d/p(3)$  0.58  & \\
 & &  $ \pi(2) \rightarrow d(2) $ 0.37 & \\
& &  $ p(1) \rightarrow d/p(3)$  0.075  & \\
313.2 (3.96) & 630 & $\pi(2)  \rightarrow d(2)$  0.41  & \\
 & &  $ \pi(2) \rightarrow d/p(3) $ 0.31 & \\
& &  $ \pi(1) \rightarrow d(2)$ 0.27   & \\
312.2 (3.97) & 549 & $\pi(1) \rightarrow d(2)$  0.46  & \\
 & &  $ \pi(2) \rightarrow d(2) $ 0.31 & \\
& &  $ \pi(1) \rightarrow d/p(3)$  0.22  & \\
\hline
&  &   & \textbf{4} \\
372.8  (3.33) & 29.0 & $\pi(2) \rightarrow d/p(3)$  0.55  & \\
 & &  $ \pi(3) \rightarrow d/p(3) $ 0.38 & \\
& &  $ p(1) \rightarrow d/p(3) $ 0.10   & \\
358.0 (3.46) & 13.0 & $p(2) \rightarrow d(2)$  0.50  & \\
 & &  $ p(3) \rightarrow d/p(3) $ 0.35 & \\
& &  $ \pi(3) \rightarrow d/p(3)$ 0.20   & \\
353.7  (3.51) & 1.00 & $p(3) \rightarrow d/p(3)$  0.46 & \\
& &  $ p(2) \rightarrow d(2) $ 0.42 & \\
& &  $ \pi(3) \rightarrow d/p(3)$ 0.19   & \\
332.1 (3.73) & 1.00 & $p(1) \rightarrow d(2)$  0.68  & \\
 & &  $ \pi(1) \rightarrow d(2) $ 0.18 & \\
& &  $ p(2) \rightarrow d/p(3)$ 0.04   & \\
\hline
&  &   & \textbf{3} \\
519.4  (2.39) & 8.00 & $\pi(3) \rightarrow d(2)$  0.69  &
\\
 & &  $ \pi(3) \rightarrow d/p(3) $ 0.11 & \\
& &  $ \pi(2) \rightarrow d/p(3) $ 0.04   & \\
504.0  (2.46) & 3.00 & $\pi(2) \rightarrow d(2)$  0.69  & \\
 & &  $ \pi(2) \rightarrow d/p(3) $ 0.10 & \\
& &  $ \pi(3) \rightarrow d/p(3) $ 0.07   & \\
442.4  (2.80) & 7.00 & $p(3) \rightarrow d(2)$ 0.70  & \\
 & &  $ \pi(3) \rightarrow d/p(3) $ 0.07 & \\
& &  $ \pi(2) \rightarrow d(2) $  0.04  & \\
418.2  (2.96) & 1.00 & $\pi(3) \rightarrow d/p(3)$  0.49  & \\
 & &  $ p(3) \rightarrow d/p(3) $ 0.38 & \\
& &  $ \pi(2) \rightarrow d/p(3)$  0.29  & \\
\hline
\end{tabular}
}
\caption*{TABLE S11: \footnotesize Transitions and assignments for the \textit{cis}-Pt complex, calculated with NR-B3LYP. The labels \textbf{1}--\textbf{9} are shown in Figure 9. In each region, the five most intense transitions were picked. }
\end{center}
\end{table}

\begin{table}[H]
\begin{center}
\hspace*{-2.0cm}
\centering
\scalebox{0.6}{
\begin{tabular}{cc}
\begin{tabular}[t]{ |c c c c |}
 \hline
 \hline
  \multicolumn{4}{c}{SF}       \\ 
 \hline
 \hline
 Energy nm (eV)  & Osc.~str.($x10^{-4}$) & Orbitals (weight) & Interval  \\ 
 \hline
  &  &   &  \textbf{9}\\
 242.7  (5.11) & 19.0 & $d/\pi(2) \rightarrow 82$  0.62  & \\
 & &  $ d/\pi() \rightarrow 82 $ 0.23 & \\
& &  $ \pi/p(1) \rightarrow 82$  0.14  & \\
241.1 (5.14) & 91.0 & $d(2) \rightarrow d(3)$  0.54 & \\
 & &  $ d/\pi(1) \rightarrow 82 $ 0.20 & \\
& &  $ d/\pi(2) \rightarrow 83$ 0.16   & \\
238.9 (5.19) & 352 & $p(1) \rightarrow d/p(2)$ 0.44  & \\
 & &  $ d(2) \rightarrow d/p(2) $ 0.31 & \\
& &  $ d/\pi(2) \rightarrow 82 $ 0.18   & \\
 \hline
  &  &   & \textbf{8} \\
 258.8  (4.79) & 468 & $\pi(2) \rightarrow d/p(2)$  0.60  & \\
  & &  $ d(2) \rightarrow d(3) $ 0.20 & \\
& &  $ p(2) \rightarrow d/p(2)$  0.18  & \\
 \hline
  &  &   & \textbf{7} \\
 263.7  (4.70) & 662 & $\pi(1) \rightarrow d/p(2)$  0.59  & \\
  & &  $ p(2) \rightarrow d/p(2) $ 0.19 & \\
& &  $ d(2) \rightarrow d/p(2) $ 0.17   & \\
 \hline
 \end{tabular}
\begin{tabular}[t]{ |c c c c |}
 \hline
 \hline
  \multicolumn{4}{c}{SF}       \\ 
 \hline
 \hline
 Energy nm (eV)  & Osc.~str.($x10^{-4}$) & Orbitals (weight) & Interval  \\ 
 \hline
& &   & \textbf{6} \\
298.8 (4.15) & 280 & $\pi(1) \rightarrow d(3)$  0.53  & \\
 & &  $p(1)  \rightarrow d(3) $ 0.26 & \\
& &  $ p(2) \rightarrow d(3)$  0.19  & \\
294.5 (4.21) & 72.0 & $p(1) \rightarrow d(3)$  0.65  & \\
 & &  $ \pi(1) \rightarrow d(3) $ 0.19 & \\
& &  $p(2)  \rightarrow d(3)$   0.12 & \\
291.0 (4.26) & 254 & $p(2) \rightarrow d/p(2)$  0.55 & \\
 & &  $ \pi(2) \rightarrow d(3) $ 0.28 & \\
& &  $ \pi(2) \rightarrow d/p(2)$  0.21  & \\
287.8 (4.31) & 743 & $\pi(2) \rightarrow d(3)$  0.52  & \\
 & &  $ p(2) \rightarrow d/p(2) $ 0.29 & \\
& &  $ \pi(1) \rightarrow d/p(2) $  0.27  & \\
\hline
 &  &  &  \textbf{5}\\
322.0  (3.85) & 6.00 & $\pi/p(1) \rightarrow d/p(2)$  0.61  & \\
 & &  $p(2)  \rightarrow d(3) $ 0.22 & \\
& &  $ d/\pi(2) \rightarrow d/p(2)$  0.19  & \\
\hline
 &  &   & \textbf{4} \\
 351.2  (3.53) & 25.0 & $d/\pi(1) \rightarrow d/p(2)$  0.62  & \\
  & &  $ d/\pi(2) \rightarrow d/p(2) $ 0.23 & \\
& &  $ \pi/p(1) \rightarrow d(3)$ 0.14   & \\
331.3  (3.74) & 2.00 & $p(2) \rightarrow d(3)$  0.58  & \\
 & &  $ \pi(1) \rightarrow d(3) $ 0.23 & \\
& &  $ \pi/p(1) \rightarrow d/p(2)$ 0.22   & \\
\hline
 &  &  & \textbf{3}\\
397.4  (3.12) & 7.00 & $\pi/p(19 \rightarrow 80$  0.59  & \\
 & &  $ d/\pi(2) \rightarrow d/p(2) $ 0.36 & \\
& &  $ \pi/p(1)  \rightarrow d/p(2) $  0.12  & \\
394.6 (3.14) & 6.00 & $d/\pi(2) \rightarrow d/p(2)$  0.48  & \\
 & &  $ \pi/p(1) \rightarrow d(3)  $ 0.36 & \\
& &  $ d/\pi(1) \rightarrow d/p(2)$ 0.25   & \\
\hline
 &  &  & \textbf{1}\\
480.5  (2.58) & 6.00 & $d/\pi(2) \rightarrow d(3)$  0.69  & \\
 & &  $ d/\pi(2) \rightarrow  d/p(2) $ 0.08 & \\
& &  $ d/\pi(1) \rightarrow d/p(2)$  0.08  & \\
461.9  (2.68) & 4.00 & $d/\pi(29 \rightarrow d(3)$  0.68  & \\
 & &  $ d/\pi(2) \rightarrow d/p(2)  $ 0.14 & \\
& &  $ d/\pi(1) \rightarrow d/p(2)$ 0.08   & \\
\hline
\end{tabular}
\end{tabular}
}
\caption*{TABLE S12: \footnotesize  Transitions and assignments for the \textit{cis}-Pt complex, calculated with SR-B3LYP. The labels \textbf{1}--\textbf{9} are shown in Figure 10.  In each region, the five most intense transitions were picked. }
\end{center}
\end{table}

\newpage

\section*{Orbital for CAM-B3LYP calculations}

\begin{figure}[H]
\includegraphics[width=0.8\textwidth]{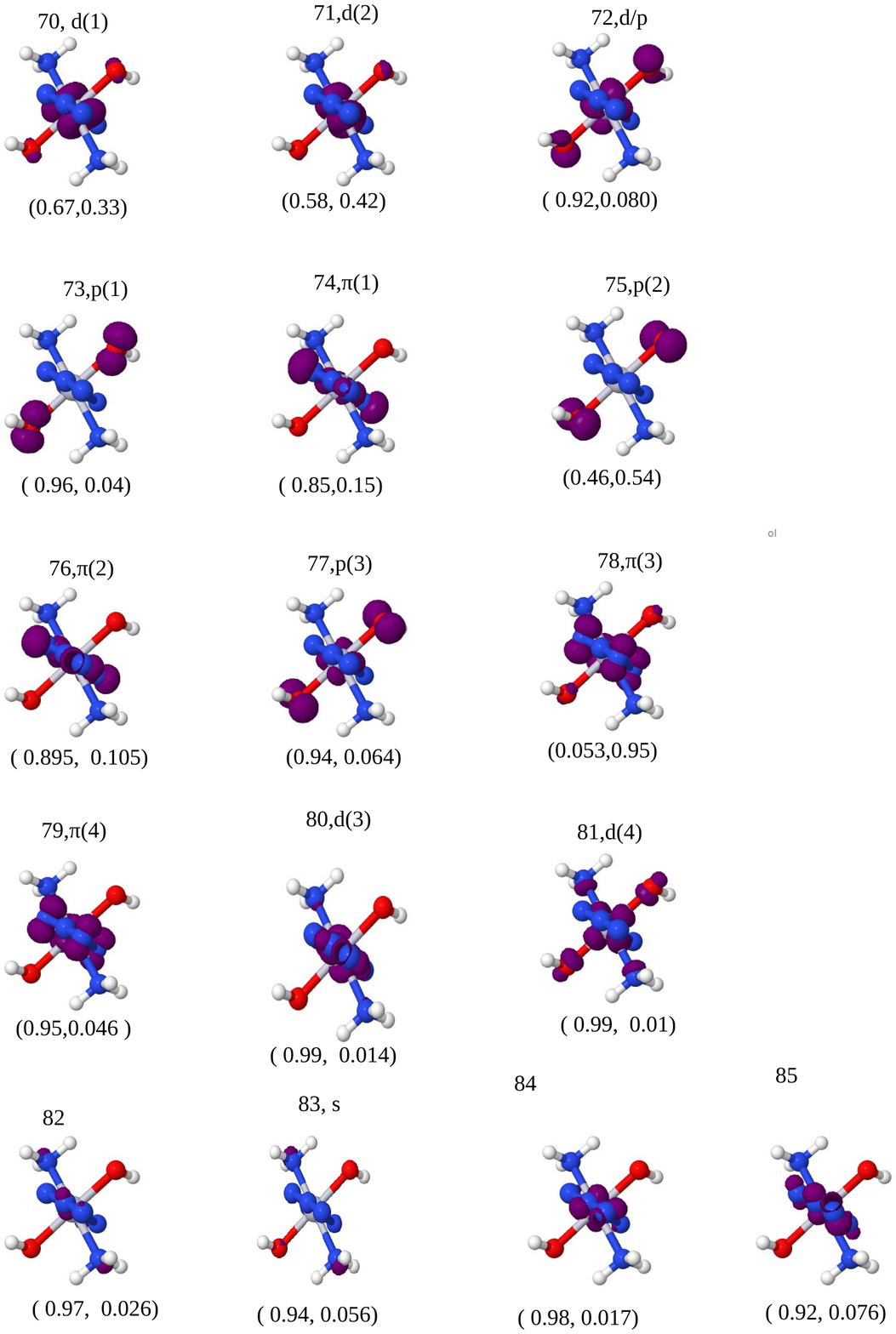}
\caption*{FIG. S1: \footnotesize  Orbital densities for \textit{trans}-Pt, computed with 4c-CAM-B3LYP. Numbers below the orbital densities are $\alpha$- and $\beta$-occupations, respectively.}
\end{figure}

\begin{figure}[H]
\includegraphics[width=0.8\textwidth]{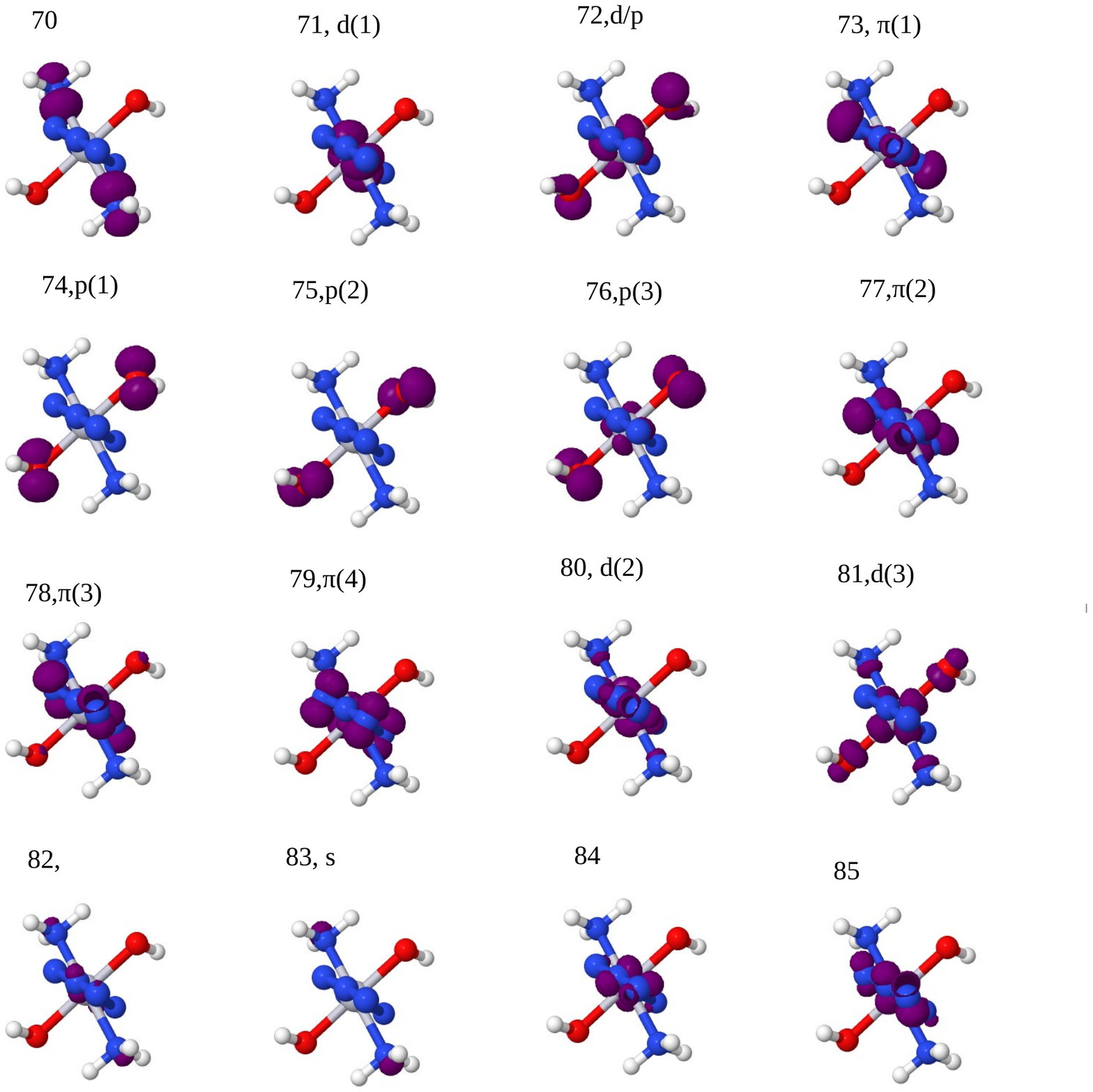}
\caption*{ FIG. S2: \footnotesize  Orbital densities for \textit{trans}-Pt, computed with NR-CAM-B3LYP. Numbers below the orbital densities are $\alpha$- and $\beta$-occupations, respectively. }
\end{figure}

\begin{figure}[H]
\includegraphics[width=0.8\textwidth]{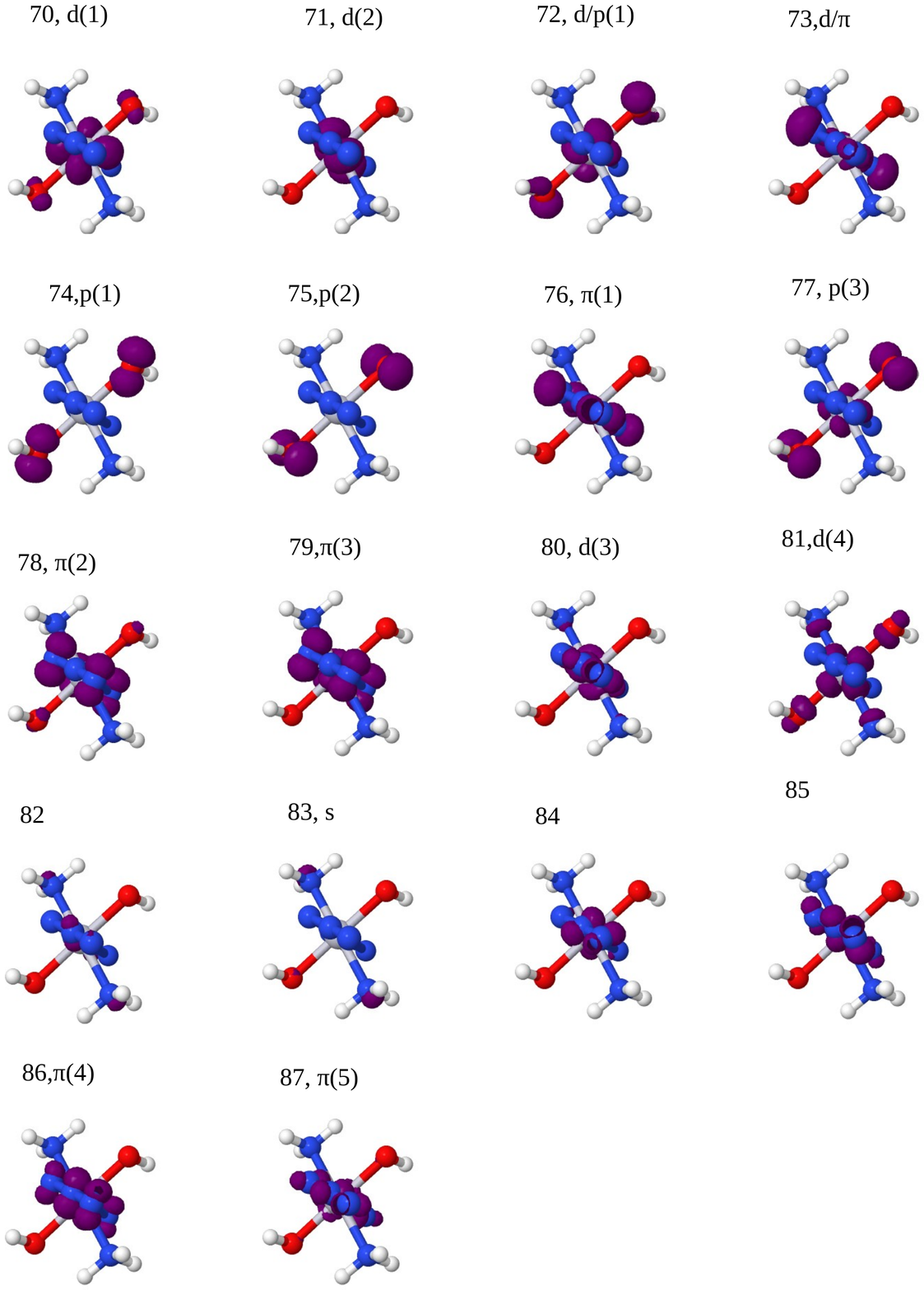}
\caption*{FIG. S3: \footnotesize  Orbitals densities for \textit{trans}-Pt, computed with SR-CAM-B3LYP. Numbers below the orbital densities are $\alpha$- and $\beta$-occupations, respectively. }
\end{figure}

\begin{figure}[H]
\includegraphics[width=0.8\textwidth]{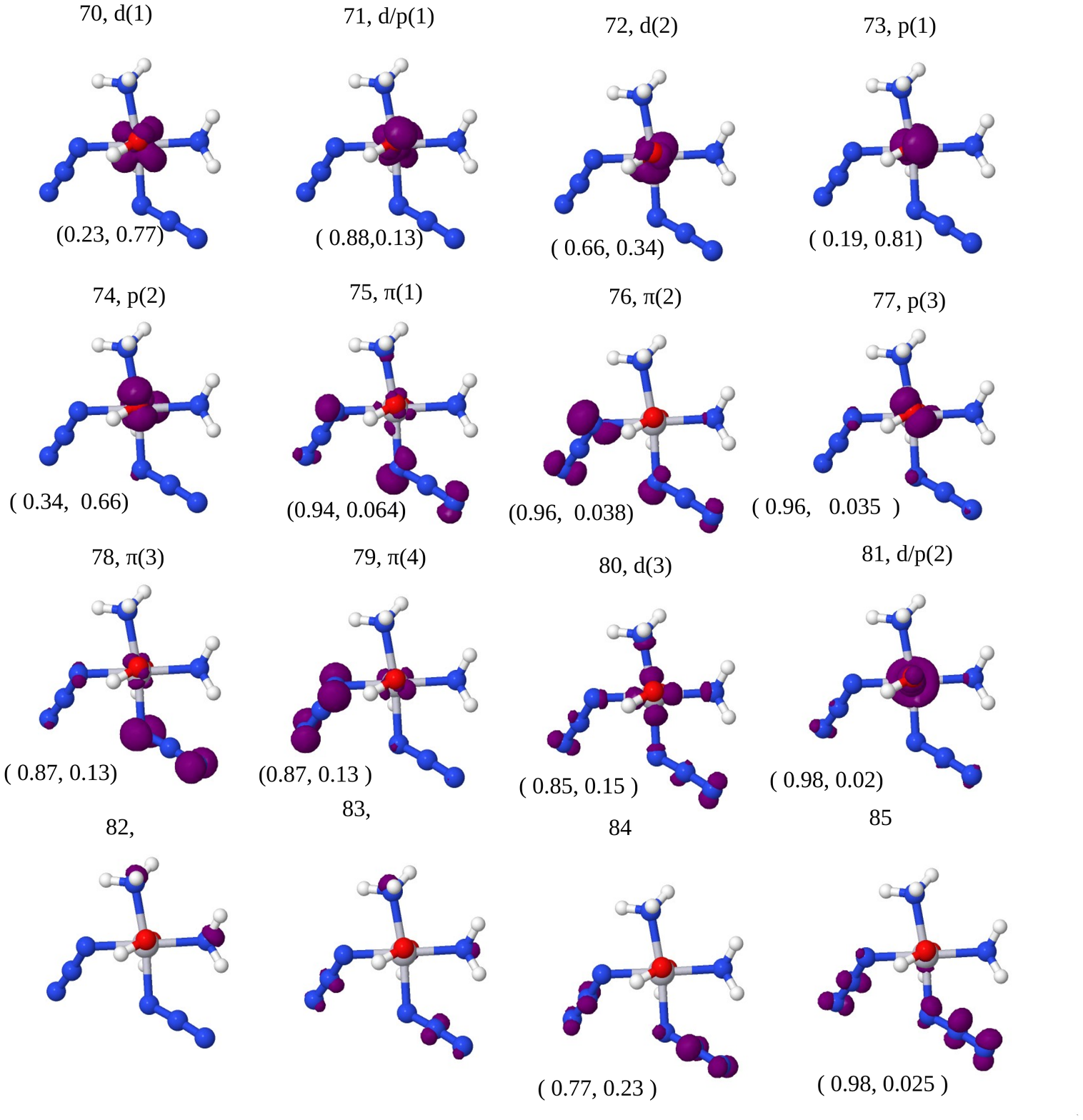}
\caption*{FIG. S4: \footnotesize  Orbital densities for \textit{cis}-Pt, computed with 4c-CAM-B3LYP. Numbers below the orbital densities are $\alpha$- and $\beta$-occupations, respectively. }
\end{figure}

\begin{figure}[H]
\includegraphics[width=0.8\textwidth]{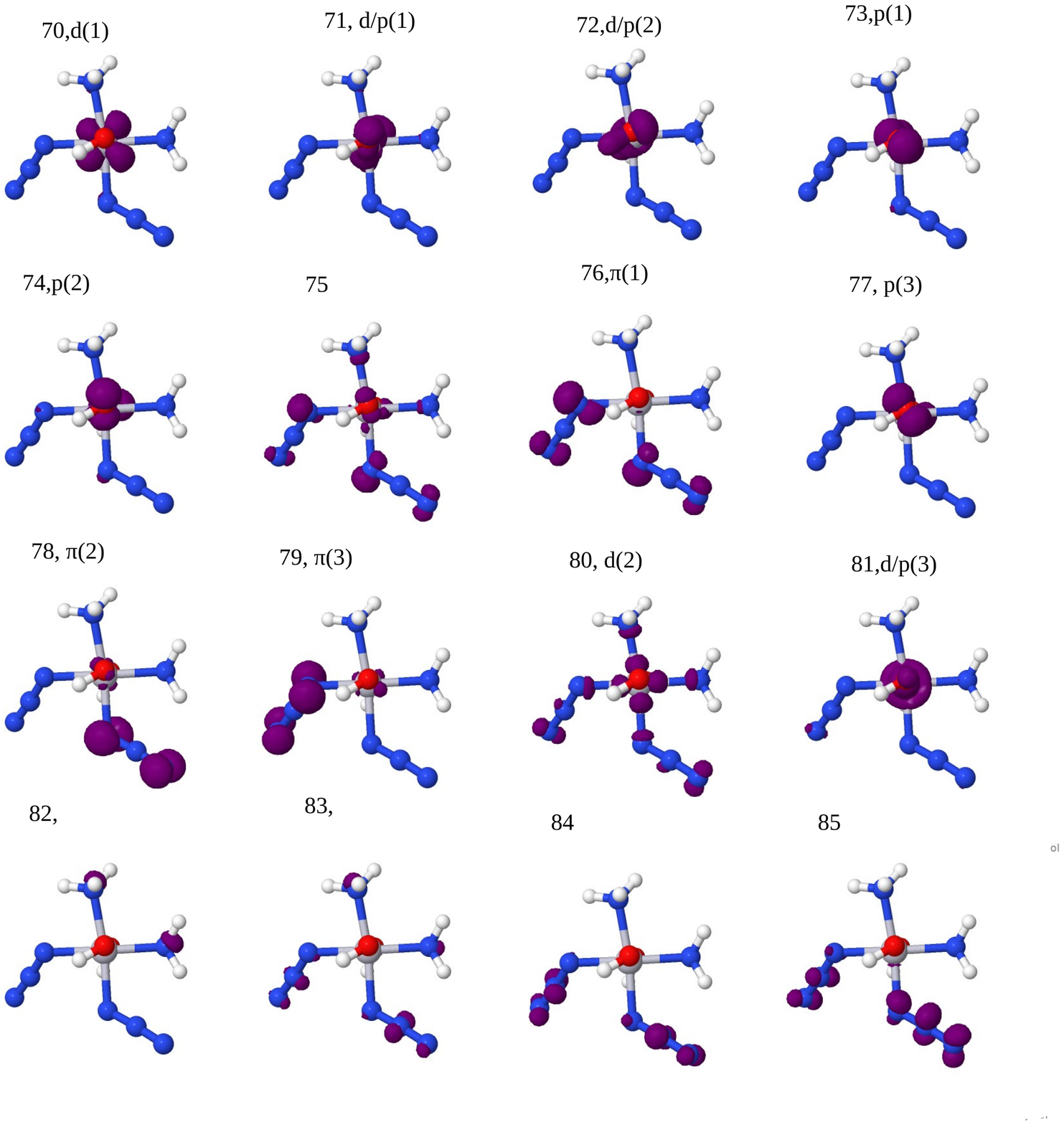}
\caption*{FIG. S5: \footnotesize  Orbital densities for \textit{cis}-Pt, computed with NR-CAM-B3LYP. Numbers below the orbital densities are $\alpha$- and $\beta$-occupations, respectively. }
\end{figure}

\begin{figure}[H]
\includegraphics[width=0.8\textwidth]{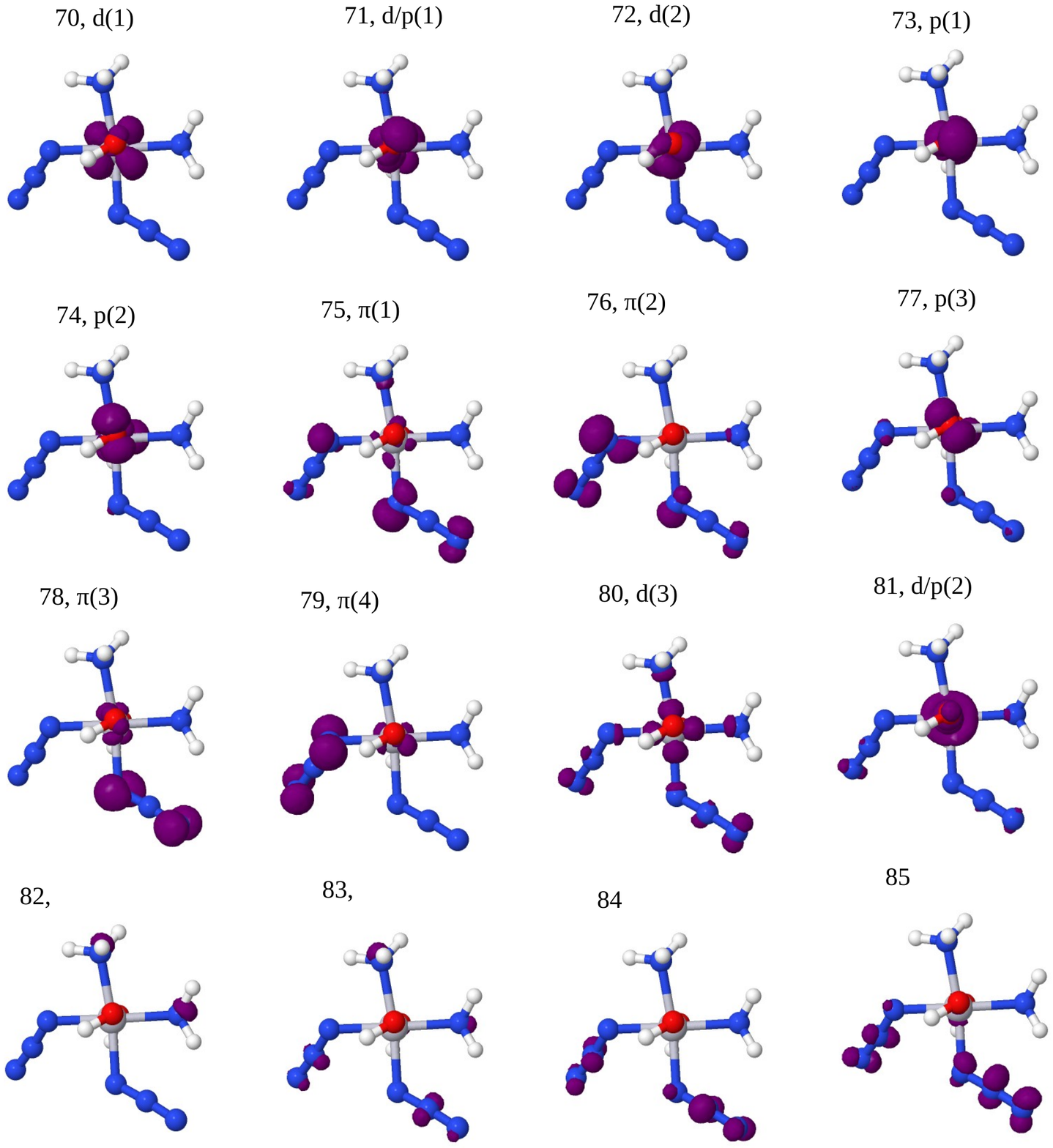}
\caption*{FIG. S6: \footnotesize  Orbital densities for \textit{cis}-Pt, computed with SR-CAM-B3LYP. Numbers below the orbital densities are $\alpha$- and $\beta$-occupations, respectively. }
\end{figure}

\section*{Orbitals for B3LYP calculations}

\begin{figure}[H]
\includegraphics[width=0.8\textwidth]{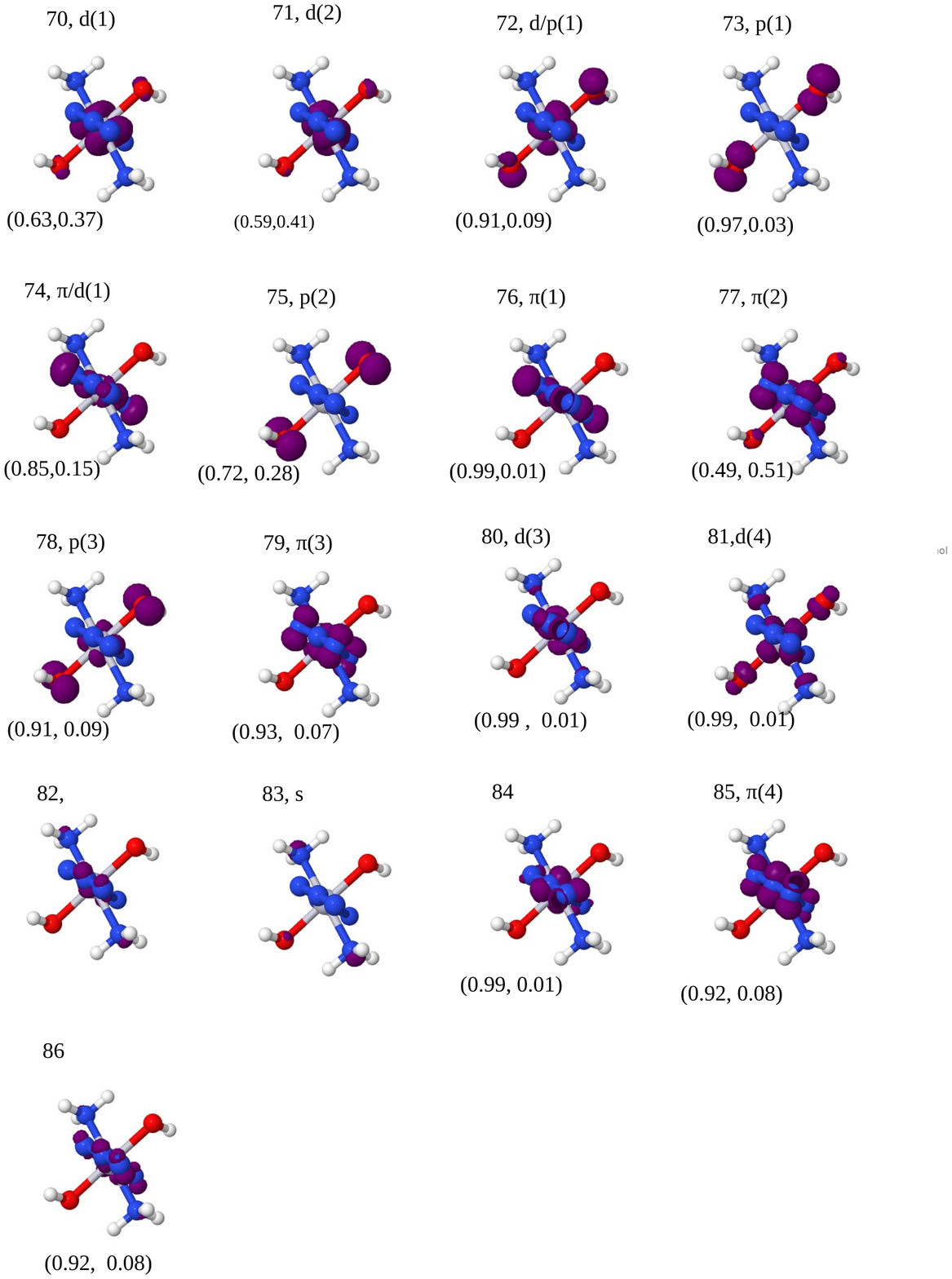}
\caption*{FIG. S7: \footnotesize Orbital densities for \textit{trans}-Pt, computed with 4c-B3LYP. Numbers below the orbital densities are $\alpha$- and $\beta$-occupations, respectively. }
\end{figure}

\begin{figure}[H]
\includegraphics[width=0.8\textwidth]{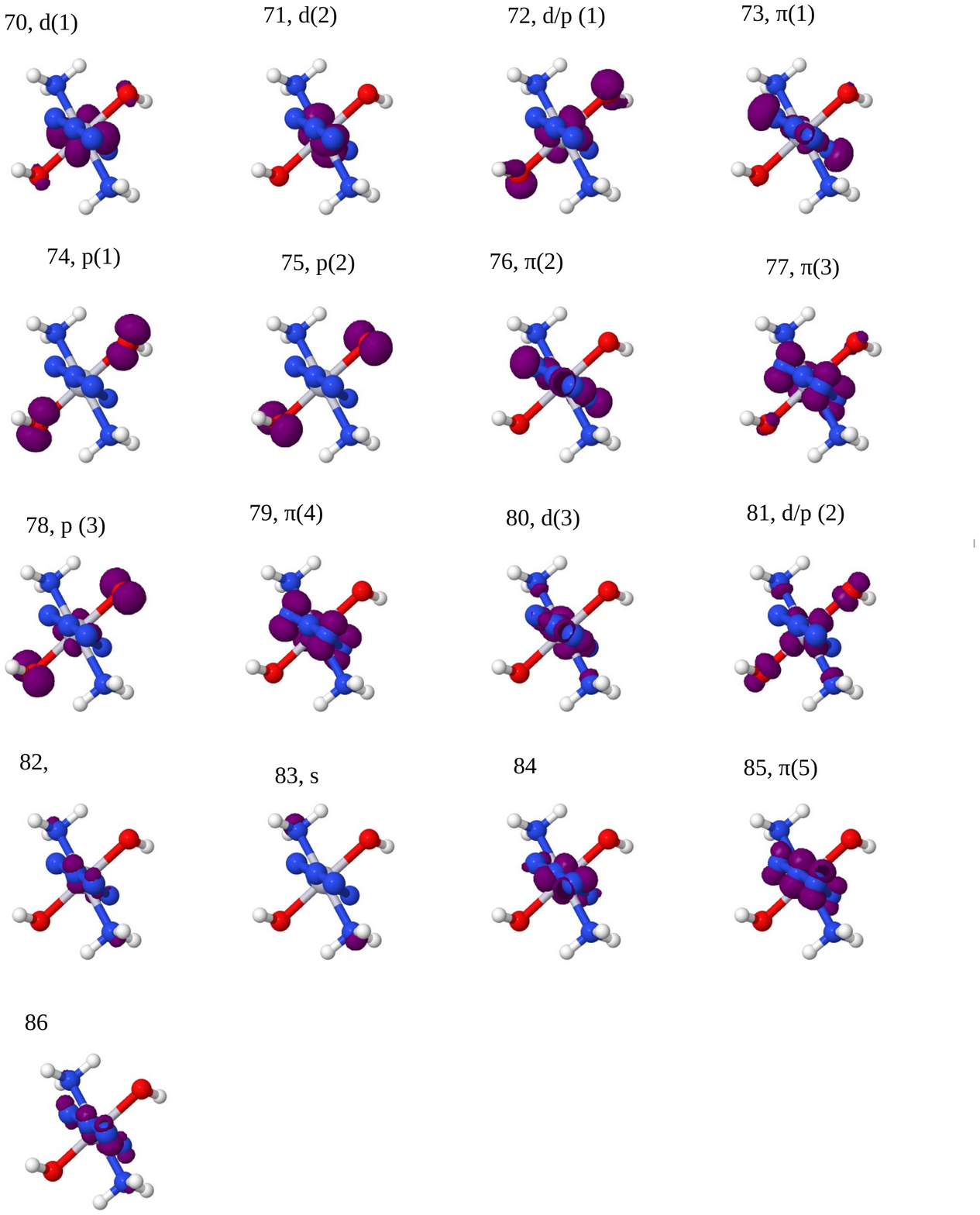}
\caption*{FIG. S8: \footnotesize Orbital densities for \textit{trans}-Pt, computed with NR-B3LYP. Numbers below the orbital densities are $\alpha$- and $\beta$-occupations, respectively. }
\end{figure}

\begin{figure}[H]
\includegraphics[width=0.8\textwidth]{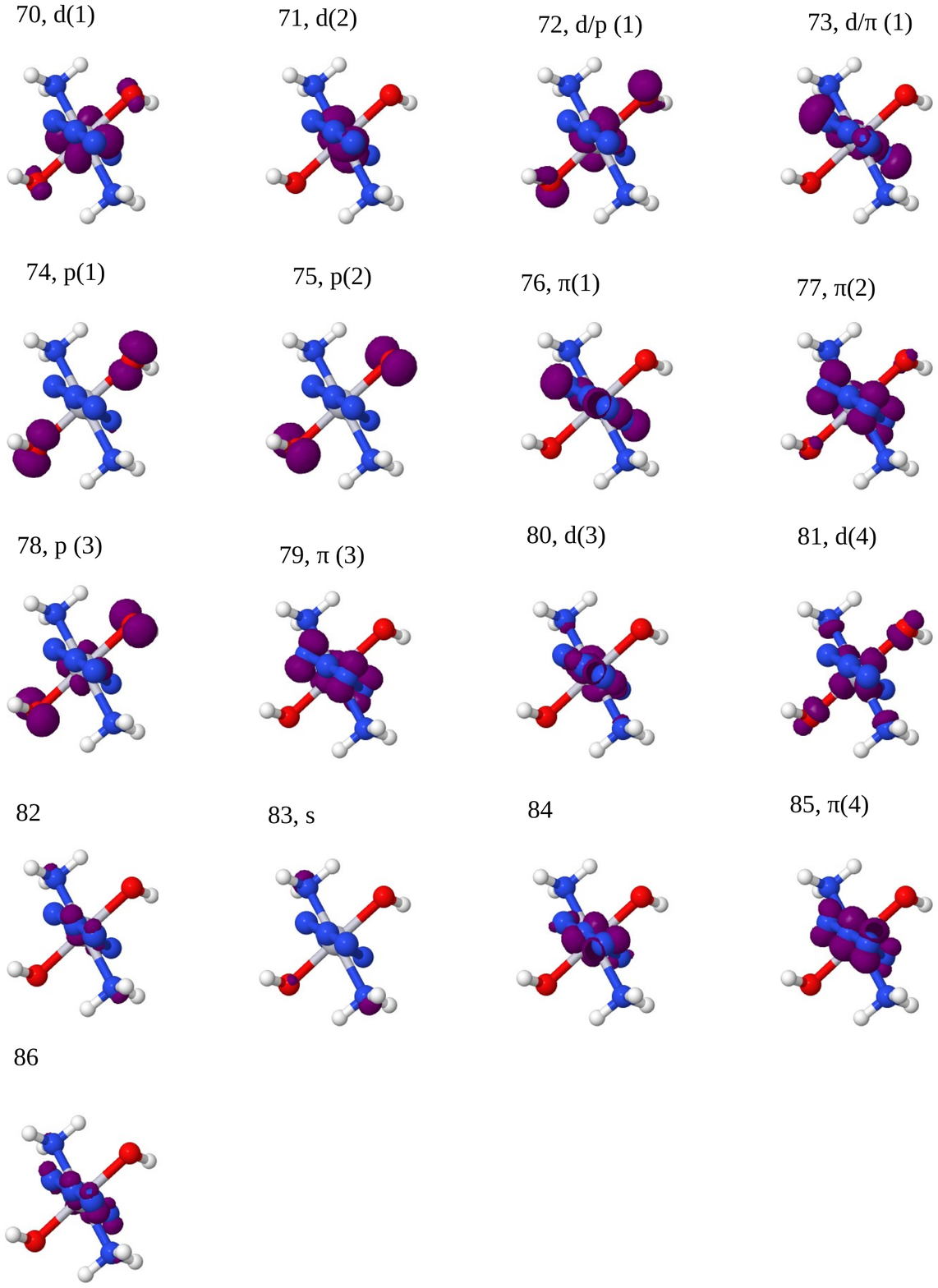}
\caption*{FIG. S9: \footnotesize Orbital densities for \textit{trans}-Pt, computed with SR-B3LYP. Numbers below the orbital densities are $\alpha$- and $\beta$-occupations, respectively. }
\end{figure}

\begin{figure}[H]
\includegraphics[width=0.8\textwidth]{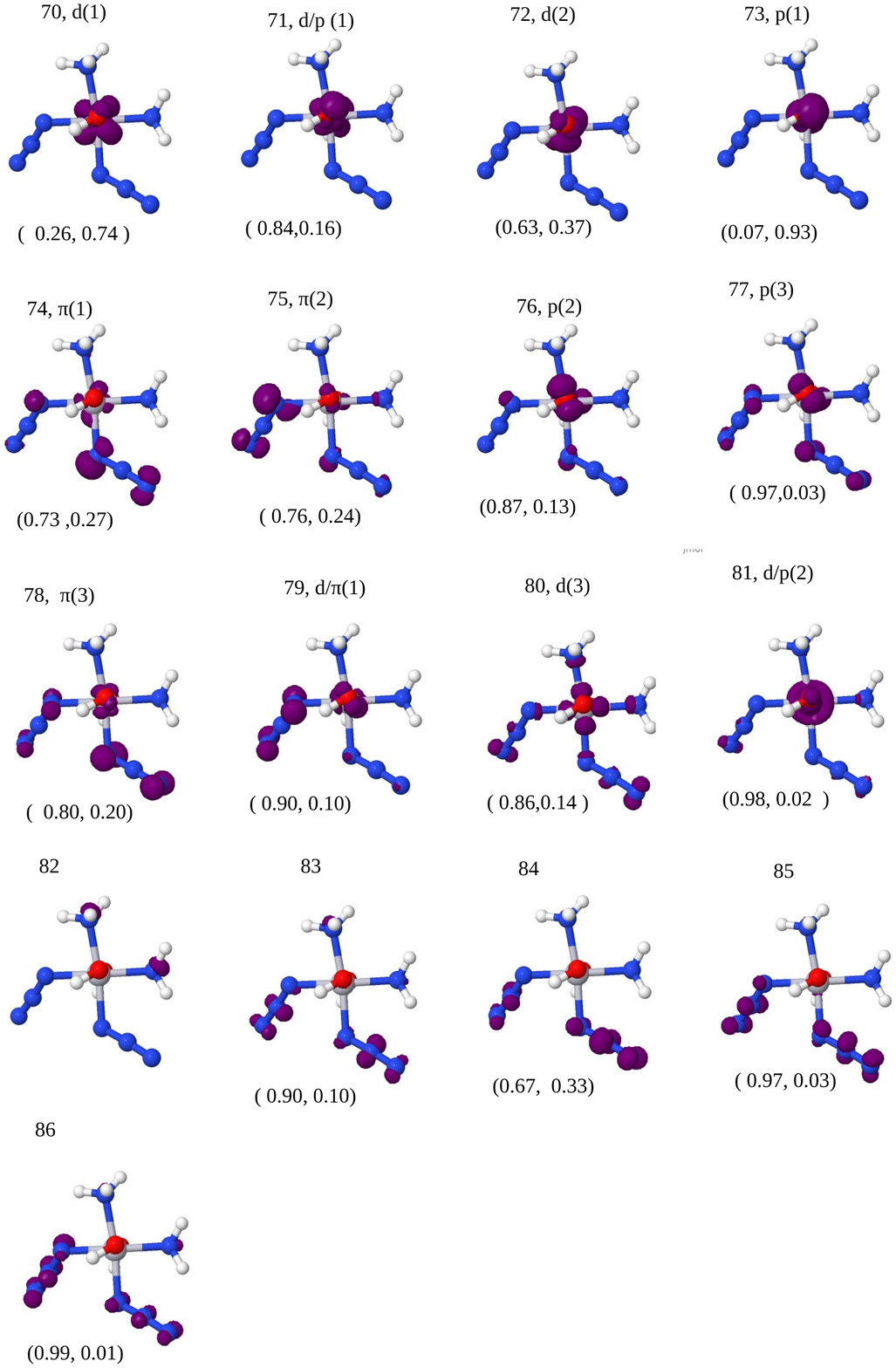}
\caption*{FIG. S10: \footnotesize Orbital densities for \textit{cis}-Pt, computed with 4c-B3LYP. Numbers below the orbital densities are $\alpha$- and $\beta$-occupations, respectively. }
\end{figure}

\begin{figure}[H]
\includegraphics[width=0.8\textwidth]{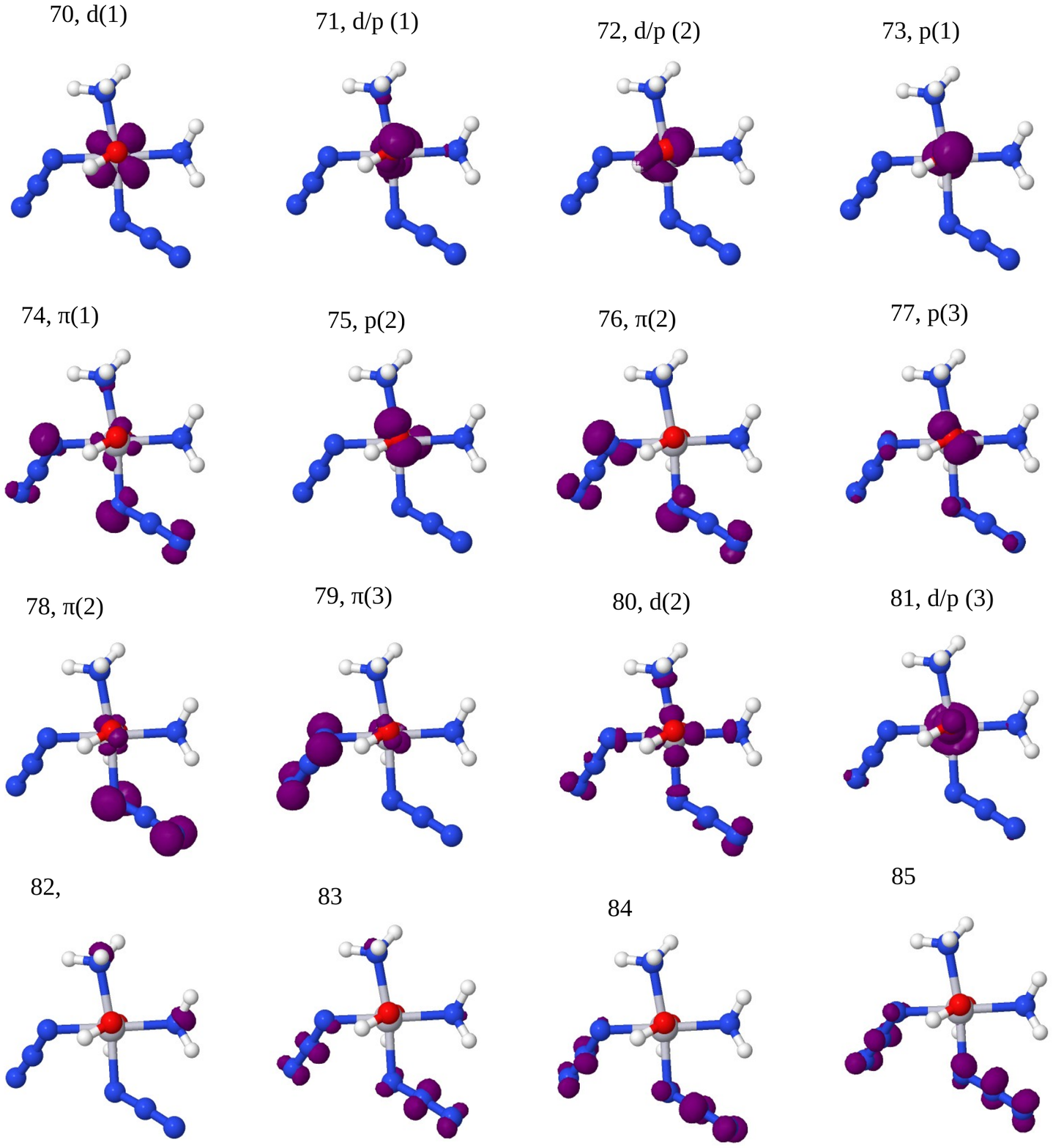}
\caption*{FIG. S11: \footnotesize Orbital densities for \textit{cis}-Pt, computed with NR-B3LYP. Numbers below the orbital densities are $\alpha$- and $\beta$-occupations, respectively. }
\end{figure}

\begin{figure}[H]
\includegraphics[width=0.8\textwidth]{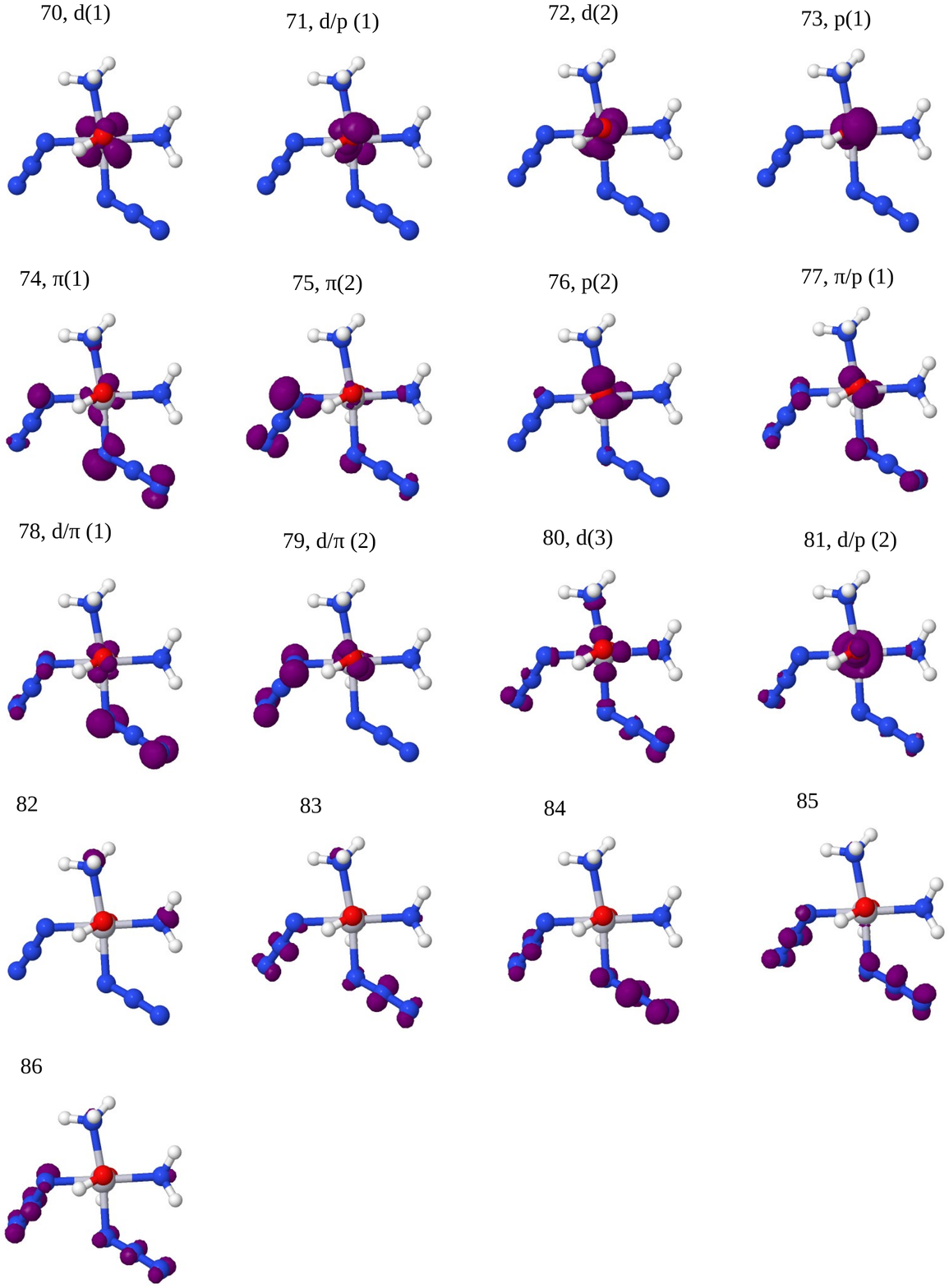}
\caption*{FIG. S12: \footnotesize Orbital densities for \textit{cis}-Pt, computed with SR-B3LYP. Numbers below the orbital densities are $\alpha$- and $\beta$-occupations, respectively. }
\end{figure}